\documentclass[11pt,showpacs,preprintnumbers,superscriptaddress,amsmath,amssymb,nofootinbib]{revtex4}
\usepackage{graphicx}% Include figure files
\usepackage{dcolumn}% Align table columns on decimal point
\usepackage{bm}% bold math
\usepackage{amssymb}
\usepackage{amsmath}
\usepackage{epsfig}    
\usepackage{color}
\usepackage{slashed}
\usepackage{hhline}
\def\be{\begin{equation}}
\def\ee{\end{equation}}
\newcommand{\bea}{\begin{eqnarray}}
\newcommand{\eea}{\end{eqnarray}}
\newcommand{\nn}{\nonumber}

\begin{document}

%{\begin{flushright}{APCTP Pre2023 - 0XX}\end{flushright}}

%%%%%%%%%
\title{Quasi two-zero texture  in Type-II seesaw at fixed points from modular $A_4$ symmetry}

%\preprint{KYUSHU-HET-268}

\author{Takaaki Nomura}
\email{nomura@scu.edu.cn}
\affiliation{College of Physics, Sichuan University, Chengdu 610065, China}

\author{Hiroshi Okada}
\email{hiroshi3okada@htu.edu.cn}
\affiliation{Department of Physics, Henan Normal University, Xinxiang 453007, China}

\date{\today}

\begin{abstract}
{
We study a quasi two-zero neutrino texture based on a type-II seesaw model with modular $A_4$ symmetry to evade the cosmological bound on the sum of neutrino masses while keeping some predictability in the neutrino sector.
Working on three fixed points for modulus, we discuss predictions of the model and show the allowed points satisfying the cosmological bound on neutrino mass from both CMB and CMB+BAO data.
}
%$A_4$
%%%%%%%%%%%%%%%%%%%%%%%%%%
 %
 \end{abstract}
\maketitle
\newpage

\section{Introduction}
Neutrino masses and their mixing patterns could be determined by a framework of beyond the standard model (BSM) physics. One attractive possibility to restrict neutrino mass and mixing is to realize two-zero textures in the neutrino mass matrix with the diagonal charged-lepton mass matrix. Currently it is found that only seven textures~\cite{Fritzsch:2011qv} satisfy the latest neutrino oscillation data~\cite{Esteban:2020cvm} as shown below:
\begin{align}
&\left[\begin{array}{ccc}
0 & 0 & \times \\
0 & \times & \times \\
\times & \times & \times \\
\end{array}\right],\quad 
%%%
\left[\begin{array}{ccc}
0 & \times & 0 \\
\times & \times & \times \\
0 & \times & \times \\
\end{array}\right],\quad
%%%
\left[\begin{array}{ccc}
\times & \times & 0 \\
\times & 0 & \times \\
0 & \times & \times \\
\end{array}\right],\quad
%%%
\left[\begin{array}{ccc}
\times & 0 & \times \\
0 & \times & \times \\
\times & \times & 0 \\
\end{array}\right], \nonumber \\
%%% %%% %%%
&\left[\begin{array}{ccc}
\times & 0 & \times \\
0 & 0 & \times \\
\times & \times & \times \\
\end{array}\right],\quad
%%%
\left[\begin{array}{ccc}
\times & \times & 0 \\
\times & \times & \times \\
0 & \times & 0 \\
\end{array}\right],\quad
%%%
\left[\begin{array}{ccc}
\times & \times & \times \\
\times & 0 & \times \\
\times & \times & 0 \\
\end{array}\right].
\end{align}
There are several ways to lead these textures introducing a symmetry regarding lepton flavor.
For example, a gauged $U(1)_{L_\mu-L_\tau}$ symmetry is a promising candidate to get some of these textures in addition to explanation of muon anomalous magnetic dipole moment~\cite{Ma:2001md, Baek:2001kca, Nagao:2022osm, Lou:2024fvw}; 
{In fact local $U(1)_{L_{i} - L_{j}} (i,j = e, \mu, \tau)$ symmetry and combination of them are useful to restrict neutrino mass structure~\cite{Asai:2018ocx, Asai:2019ciz, Nomura:2023vmh}.
Two-zero texture scenarios are attractive since they provide us predictions for neutrino mixing angles, masses, CP violating phases and neutrinoless double beta decay. 
However, these textures possess a shortcoming that they cannot satisfy the upper bounds on the sum of neutrino masses $\sum D_{\nu}$ derived from analysis of cosmic microwave background (CMB) and baryon acoustic oscillation (BAO) based on standard $\Lambda$CDM cosmology; the upper limit is 120 meV from CMB with Planck data~\cite{Vagnozzi:2017ovm, Planck:2018vyg} and recently more stringent limit of $\sum D_{\nu} < 72$ meV is derived by combining Planck CMB data and BAO data from Dark Energy Spectroscopic Instrument (DESI)~\cite{DESI:2024mwx}.
%%%
~\footnote{{
Note that the upper bound on sum of neutrino masses is controversial issue. Depending on experimental bounds, more relaxed bound would be imposed~\cite{Naredo-Tuero:2024sgf}.}
%For example, the new Planck likelihood provides the upper bound to be $\sum D_{\nu} < 200$ meV~\cite{Naredo-Tuero:2024sgf}.
}
%%%
%
Thus we would like to realize a scenario in which neutrino mass structure has quasi two-zero texture to obtain predictions for neutrino observables while satisfying constraints on $\sum D_{\nu}$ at the same time.
{The term 'quasi' implies, in this paper, that the charged-lepton mass matrix is not diagonal anymore. Thus, we expect that a non-trivial charged-lepton mixing matrix would contribute to  $\sum D_{\nu}$ as well as the neutrino oscillation data. 
}

{
In order to realize the quasi two-zero textures, we apply modular $A_4$ flavor symmetry for the type-II seesaw model~\cite{Lazarides:1980nt,Mohapatra:1980yp,Ma:1998dx,Schechter:1980gr,Cheng:1980qt,Bilenky:1980cx,Hirsch:2008gh}.} Here we consider $A_4$ since it is known as the minimal non-Abelian group with irreducible triplet representation~\cite{Feruglio:2017spp, Criado:2018thu, Kobayashi:2018scp, Okada:2018yrn, Nomura:2019jxj, Okada:2019uoy, deAnda:2018ecu, Novichkov:2018yse, Nomura:2019yft, Okada:2019mjf,Ding:2019zxk, Nomura:2019lnr,Kobayashi:2019xvz,Asaka:2019vev,Zhang:2019ngf, Gui-JunDing:2019wap,Kobayashi:2019gtp,Nomura:2019xsb, Wang:2019xbo,Okada:2020dmb,Okada:2020rjb, Behera:2020lpd, Behera:2020sfe, Nomura:2020opk, Nomura:2020cog, Asaka:2020tmo, Okada:2020ukr, Nagao:2020snm, Okada:2020brs,Kang:2022psa, Qu:2024rns, Ding:2024fsf, Ding:2023htn, Nomura:2023usj, Kobayashi:2023qzt, Petcov:2024vph, Kobayashi:2023zzc}.~\footnote{See for a different approach to realize texture zeros in quark and lepton mass matrices adopting the modular symmetries~\cite{Lu:2019vgm,Kikuchi:2022svo,Ding:2022aoe,Devi:2023vpe}.} 
We obtain the third type of two-zero texture in Eq.(1) by assigning $A_4$ singlets for three left-handed leptons, while the charged-lepton mass matrix is given by a non-trivial but predictable mass matrix by assigning the right-handed charged-lepton fields to be $A_4$ triplet. 
{The minimum assignments are summarized in Appendix A where the number of free parameters except modulus $\tau$ is only three in order to reproduce the mass eigenvalues of charged-leptons.
\footnote{Note that $B'_2$ marginally satisfies the cosmological (CMB) bound but there is no solution in case of IH.}
However, all these simplest models do not satisfy the cosmological bound $\sum D_\nu\le$120 meV. Thus, we move to a next minimum scenario. Since the next minimum one has enough free parameters,}
%Furthermore, 
we work on three fixed points of $\tau$; $i$, $\omega$ and $i\infty$ that preserve $Z_2$, $Z_3$, and $Z_2$ symmetry, respectively.
These fixed points are statistically favored in the flux compactification of
Type IIB string theory~\cite{Kobayashi:2021pav} and important in phenomenological points of view due to enhancement of predictions.

This paper is organized as follows.
In Sec. \ref{sec:II}, 
we show the field contents and their assignments, and derive two-zero textures in the neutrino mass matrix and specific mass matrix for the charged-lepton. We then analytically formulate their mixings and mass eigenvalues. 
In Sec. \ref{sec:III}, we give numerical analyses in each fixed points with normal hierarchy and inverted one and show some predictions for each case. 
%And then, we briefly mention  a possibility of testing models at collider experiments.
 % A
 Finally, we summarize and conclude in Sec. \ref{sec:IV}.

\section{Model setup}
\label{sec:II}

\begin{table}[t!]
\begin{tabular}{|c||c|c|c|c|c|c|}\hline\hline  
& ~$\hat L$~ & ~$\hat {\overline{\ell}}$~ & ~$\hat H_{u}$~ & ~$\hat H_{d}$~ & ~$\hat \Delta_{u}$~ & ~$\hat \Delta_{d}$~  \\\hline\hline 
%%%
$SU(2)_L$   & $\bm{2}$  & $\bm{1}$  & $\bm{2}$ & $\bm{2}$ & $\bm{3}$  & $\bm{3}$     \\\hline 
$U(1)_Y$    & $-\frac12$  & $+1$ & $+\frac12$ & $-\frac12$ & $+1$  & $-1$  \\\hline
$A_4$   & $\{\bm{1}\}$  & $ \bm{3}$  & $\bm{1}$ & $\bm{1}$ & $\bm{1}$  & $\bm{1}$        \\\hline 
$-k_I$    & $-5$  & $-1$ & $0$ & $0$ & $0$ & $0$     \\\hline
\end{tabular}
\caption{Charge assignments of the SM lepton and new superfields
under $SU(2)_L\otimes U(1)_Y \otimes A_4$ where $-k_I$ is the modular weight. Here {$\{ \bm{1} \} =\{1, 1'', 1'\}$} indicates assignment of $A_4$ singlets.    }\label{tab:1}
\end{table}

The Type-II seesaw mechanism requires an isospin triplet scalar field $\Delta$ to generate neutrino mass where we adopt a supersymmetric scenario
to forbid infinitely possible terms. Therefore we introduce $\Delta_{u}$ and $\Delta_{d}$ having $+1$ and $-1$ hypercharge, respectively, to cancel gauge anomalies.~$\footnote{In analysis, we do not consider superpartners assuming they are sufficiently heavy without affecting phenomenology.}$
The superfields are denoted by $\hat \psi$ where $\psi$ indicates any field. %where $\psi$ are fermions or bosons.
In the model we assign three types of $A_4$ singlets { $\{ \bm{1} \} =\{1, 1'', 1'\}$} to the left-handed leptons $L$ with $-5$ modular weight, and $A_4$ triplet to the right-handed charged-leptons $\bar \ell$ with $-1$ modular weight. The other fields are trivial $A_4$ representations with zero modular weight. Relevant fields and their assignments are summarized in  Table~\ref{tab:1}.
The renormalizable superpotential which is invariant under the modular $A_4$ is found as
\begin{align}
W_\ell = & 
 [Y^{(6)}_{{\bm 3},{\bm 3}'} \hat{\overline{\ell}} \hat L_\ell] \hat H_d 
+ { [Y^{(10)}_{1,1'} \hat{L} \hat \Delta_u \hat L]}
+ \mu_H \hat H_u \hat H_d
+ \mu_\Delta {\rm Tr}(\hat \Delta_u \hat \Delta_d)
+ \lambda_u \hat H_u \hat \Delta_d \hat H_u
+ \lambda_d \hat H_d \hat \Delta_u \hat H_d,
\label{eq:SP}
\end{align}
where $Y_{A}^{(k_Y)}$ indicates a modular form with $A_4$ representation $A$ and modular weight $k_Y$, and $[\cdots]$ represents $A_4$ singlets constructed by the $A_4$ representations associated with superfields and modular forms. Note that superpotential is invariant under modular $A_4$ symmetry since sum of modular weights is zero for each term.

\subsection{Charged-lepton mass matrix}
The mass matrix of charged-lepton comes from the first term in Eq.(\ref{eq:SP}).
After spontaneous electroweak symmetry breaking, the mass matrix is given by
\begin{align}
m_\ell = \frac{v_d}{\sqrt2}
 \left(\begin{array}{ccc} y_1 + p y'_1 & y_2 + q y'_2 & y_3 + r y'_3 \\
 y_3 + p y'_3 & y_1 + q y'_1 & y_2 + r y'_2 \\
  y_2 + p y'_2 & y_3 + q y'_3 & y_1 + r y'_1 \end{array} \right)
  %%%
   \left(\begin{array}{ccc} a_e & 0 & 0 \\
0 & b_e & 0 \\
0 & 0 & c_e \end{array} \right),
\label{eq:cgd-lep}
\end{align}
where $v_d$ is the vacuum expectation value (VEV) of $H_d$, and $\{a_e, b_e, c_e\}$ are real parameters without loss of generality
%which are used to fit the mass eigenvalues of charged-leptons.
 while $\{p,q,r\}$ are complex parameters.
Furthermore,  $Y^{(6)}_{\bm{3}} {= \left(Y_1^{(6)},Y_2^{(6)},Y_3^{(6)}\right)} \equiv (y_1,y_2,y_3 )$ and  $Y^{(6)}_{{\bm 3}'} {= \left(Y'^{(6)}_1,Y'^{(6)}_2,Y'^{(6)}_3\right)}  \equiv (y'_1,y'_2,y'_3)$.
The mass eigenvalues for the charged-leptons are obtained via mixings matrices $V_R, V_L$ as diag.$(m_e,m_\mu,m_\tau)\equiv V_R^\dag m_\ell V_L$.
Therefore, $V_L^\dag m_\ell^\dag m_\ell V_L ={\rm diag.}(|m_e|^2,|m_\mu|^2,|m_\tau|^2)$.
%%%
We numerically determine the free parameters $a_e,b_e,c_e$ in order to fit the three observed charged-lepton masses, applying the following three relations:
\begin{align}
&{\rm Tr}[m_\ell^\dag m_\ell] = |m_e|^2 + |m_\mu|^2 + |m_\tau|^2,\quad
 {\rm Det}[m_\ell^\dag m_\ell] = |m_e|^2  |m_\mu|^2  |m_\tau|^2,\nn\\
&({\rm Tr}[m_\ell^\dag m_\ell])^2 -{\rm Tr}[(m_e^\dag m_e)^2] =2( |m_e|^2  |m_\mu|^2 + |m_\mu|^2  |m_\tau|^2+ |m_e|^2  |m_\tau|^2 ).\label{eq:l-cond}
\end{align}
In the following, we estimate qualitative behavior of their mass eigenvalues and mixing matrix on the three fixed points.

\subsubsection{$\tau=i$}
In the case of $\tau=i$, we can simplify the modular forms, at the leading order, as follows;
$Y^{(6)}_3\sim (-3+2\sqrt3, -9+5\sqrt3, 12-7\sqrt3)$ and  $Y^{(6)}_{3'}\sim (-12+7\sqrt3, 3-2\sqrt3, 9-5\sqrt3)$.
They lead us to the one massless eigenvalue. Therefore, the electron mass is induced at sub-leading order.
The eigenvector in the massless eigenvalue is given by
\begin{align}
\frac1{A}
\left[
b_e c_e(1+q+qr), a_e c_e(1+r+p r), a_e b_e(1+p+pq)
\right]^T,
\end{align}
where $A\equiv 1/{\sqrt{b_e^2c_e^2|1+q+qr|^2+a_e^2b_e^2|1+p+pq|^2+a_e^2c_e^2|1+r+pr|^2}}$.
%%%
%%
%
Here, in some limiting cases, we find analytical forms of the first column of $V_L$.\\

The first case is to assume $a_e \ll b_e,c_e$ and it is given by
\begin{align}
i)\hspace{0.5cm} a_e \ll b_e,c_e,\quad
\left[1, 
\frac{a_e(1+r+pr)}{b_e(1+q+ q r)}, \frac{a_e(1+p+pq)}{c_e(1+q+ q r)}
\right]^T +{\cal O}(a_e^2).
\end{align}
Then, the electron mass is found at the next leading order in presuming the following hierarchies among parameters of $(p,q,r)$:
\begin{align}
& |p|,|q| \ll |r|,\quad
m_e^2\approx 2 v_d^2 (45-26\sqrt3) a_e^2(2\epsilon_1-\epsilon_2) (r^2-|r|^2),\\
 %4  (45-26\sqrt3) a_e^2(2\epsilon_1-\epsilon_2) (r^2-|r|^2),\\
%
& |p|,|r| \ll |q|,\quad
m_e^2\approx \frac{v_d^2}{2} a_e^2\left[ (288-166\sqrt3)(\epsilon_1+\epsilon_1^*) -(-963+556\sqrt3)(\epsilon_2+\epsilon_2^*) \right] \frac{|q|^2}{(1+q)^2},\\
& |q|,|r| \ll |p|,\quad
m_e^2\approx\frac{v_d^2}{2}  a_e^2 \left[ (648-374\sqrt3)(\epsilon_1+\epsilon_1^*) -(-783+556\sqrt3)(\epsilon_2+\epsilon_2^*) \right] p^2,
\end{align}
where relations among modular forms in $|\epsilon_{1,2}| \ll1$ are found in Appendix~\ref{apdxA}.

The second case is to assume $b_e \ll a_e,c_e$ and it is given by
\begin{align}
ii)\hspace{0.5cm} b_e \ll a_e,c_e,\quad
\left[
\frac{b_e(1+q+qr)}{a_e(1+r+ p r)}, 1, \frac{b_e(1+p+pq)}{c_e(1+r+ p r)}
\right]^T +{\cal O}(b_e^2).
\end{align}
Then, the electron mass is found at the next leading order in presuming the following hierarchies among parameters of $(p,q,r)$:
\begin{align}
& |p|,|q| \ll |r|,\quad
m_e^2\approx  \frac{v_d^2}{2}  b_e^2\left[ (288-166\sqrt3)(\epsilon_1+\epsilon_1^*)- (-963+556\sqrt3)(\epsilon_2 + \epsilon_2 ^*)\right] \frac{|r|^2}{(1+r)^2},\\
& |p|,|r| \ll |q|,\quad
m_e^2\approx \frac{v_d^2}{2}  b_e^2\left[ (648-374\sqrt3)(\epsilon_1+\epsilon_1^*) -(-783+452\sqrt3)(\epsilon_2+\epsilon_2^*) \right] q^2, \\
& |q|,|r| \ll |p|,\quad
m_e^2\approx 2 v_d^2 (45-26\sqrt3) b_e^2(2 \epsilon_1- \epsilon_2) (p^2-|p|^2).
%4 (45-26\sqrt3) b_e^2(2 \epsilon_1- \epsilon_2) (p^2-|p|^2).
\end{align}
%where $|\epsilon_{1,2}| \ll1$ are found in Appendix~\ref{apdx}.

%

The third case is to assume $c_e \ll a_e,b_e$ and it is given by
\begin{align}
iii)\hspace{0.5cm} c_e \ll a_e,b_e,\quad
\left[
\frac{c_e(1+q+qr)}{a_e(1+p+ p q)}, \frac{c_e(1+r+pr)}{b_e(1+p+ p q)}, 1
\right]^T +{\cal O}(c_e^2).
\end{align}
Then, the electron mass is found at the next leading order in presuming the following hierarchies among parameters of $(p,q,r)$:
\begin{align}
& |p|,|q| \ll |r|,\quad
m_e^2\approx \frac{v_d^2}{2} c_e^2\left[ (648-374\sqrt3)(\epsilon_1+\epsilon_1^*) -(-783+452\sqrt3)(\epsilon_2+\epsilon_2^*) \right] r^2\\
& |p|,|r| \ll |q|,\quad
m_e^2\approx 2 v_d^2 (45-26\sqrt3) c_e^2(2 \epsilon_1- \epsilon_2) (q^2-|q|^2), \\
%4 (45-26\sqrt3) c_e^2(2 \epsilon_1- \epsilon_2) (q^2-|q|^2), \\
%
& |q|,|r| \ll |p|,\quad
m_e^2\approx \frac{v_d^2}{2} c_e^2\left[ (288-166\sqrt3)(\epsilon_1+\epsilon_1^*)- (-963+556\sqrt3)(\epsilon_2 + \epsilon_2 ^*)\right] \frac{|p|^2}{(1+p)^2},
\end{align}
They indicate that we have some specific structure of $V_L$ when parameters $\{a_e, b_e, c_e, |p|, |q|, |r| \}$ are hierarchical. 
In fact such hierarchical points are obtained in our numerical analysis.

\subsubsection{$\tau=\omega$}
In the case of $\tau=\omega$, we can simplify the modular forms, at the leading order, such that
$Y^{(6)}_3\sim (0,0,0)$ and  $Y^{(6)}_{3'}\sim (-1, 2\omega, 2\omega^2)$.
They lead us to the unit mixing matrix for $V_L$ at the leading order.  Therefore, the neutrino oscillation data is determined only by the structure of the neutrino matrix at the leading order. 
The nonzero mass eigenvalues are given the leading order by
\begin{align}
 \left(\begin{array}{ccc} m_e^2 & 0 & 0 \\
0 & m_\mu^2 & 0 \\
0 & 0 & m_\tau^2 \end{array} \right)
\sim
 \frac{v_d^2}{2} \left(\begin{array}{ccc} a_e^2|p|^2 & 0 & 0 \\
0 & b_e^2|q|^2 & 0 \\
0 & 0 & c_e^2|r|^2 \end{array} \right).
\end{align}
In this case deviation from two-zero texture originates from difference of $\tau$ from the fixed point $\omega$.
The off-diagonal components of  $V_L$ are found at the next leading which is given by
\begin{align}
V_L&\approx
\left(\begin{array}{ccc} 
1 &\epsilon_{12} &\epsilon_{13}\\
-\epsilon^*_{12} & 1 &\epsilon_{23} \\
-\epsilon^*_{13} & -\epsilon^*_{23} & 1 \end{array} \right),\\
%%%
\epsilon_{12}\approx -31.8938 i a_e b_e q\omega(1+\frac12 p^*)\epsilon^*,\
\epsilon_{13}& \approx -31.8938 i a_e c_e p^*\omega^*(1+\frac12 r)\epsilon,\
\epsilon_{23} \approx  31.8938 i b_e c_e r \omega (1+\frac12 q^*)\epsilon^*,
\end{align}
 where modular forms with $|\epsilon|\ll1$ is given in Appendix~\ref{apdxB}.

\subsubsection{$\tau=i\infty$}
In the case of $\tau=i\infty$, we can simplify the modular forms, at the leading order, as
$Y^{(6)}_3\sim (1,0,0)$ and  $Y^{(6)}_{3'}\sim (0,0,0)$.
They also lead us to the unit mixing matrix for $V_L$ at the leading order.
Therefore, the neutrino oscillation data is determined only by the structure of the neutrino matrix at the leading order.
The nonzero mass eigenvalues are given the leading order by
\begin{align}
 \left(\begin{array}{ccc} m_e^2 & 0 & 0 \\
0 & m_\mu^2 & 0 \\
0 & 0 & m_\tau^2 \end{array} \right)
\sim
\frac{v_d^2}{2} \left(\begin{array}{ccc} a_e^2 & 0 & 0 \\
0 & b_e^2 & 0 \\
0 & 0 & c_e^2 \end{array} \right).
\end{align}
Again, we obtain deviation from two-zero texture due to difference of $\tau$ from fixed point $\tau = i \infty$.
The off-diagonal components of  $V_L$ are found at the next leading which is given by
\begin{align}
V_L&\approx
\left(\begin{array}{ccc} 
1 &\epsilon_{12} &\epsilon_{13}\\
-\epsilon^*_{12} & 1 &\epsilon_{23} \\
-\epsilon^*_{13} & -\epsilon^*_{23} & 1 \end{array} \right),\\
%%%
\epsilon_{12}\approx -6  a_e b_e {\bf p}^{1/3} \epsilon^{1/3} (1+2 q),\
\epsilon_{13}& \approx -6 a_e c_e {\bf p}^{1/3}\epsilon^{1/3} (1+2 p^*),\
\epsilon_{23} \approx  -6 b_e c_e {\bf p}^{1/3}\epsilon^{1/3} (1+2r),
\end{align}
  where modular forms with $ (|{\bf p}|,\ |\epsilon|) \ll1$ are given in Appendix~\ref{apdxC}.

\subsection{Neutrino mass matrix}
%In this subsection, 
The mass matrix of neutrino comes from the second term in Eq.(\ref{eq:SP}) {via type-II seesaw mechanism~\cite{Lazarides:1980nt,Mohapatra:1980yp,Ma:1998dx,Schechter:1980gr,Cheng:1980qt,Bilenky:1980cx,Hirsch:2008gh}.}
After spontaneous electroweak symmetry breaking with $\Delta_u$ VEV, the mass matrix is given by
\begin{align}
m_\nu = \frac{v_{\Delta_u}}{\sqrt2}
%\frac{v_{\Delta_u} |d_\nu|}{\sqrt2}
 \left(\begin{array}{ccc} a_\nu Y^{(10)}_{1} & b_\nu Y^{(10)}_{1'} & 0 \\
b_\nu Y^{(10)}_{1'} & 0 & c_\nu Y^{(10)}_{1} \\
0 & c_\nu Y^{(10)}_{1} &Y^{(10)}_{1'} \end{array} \right),
\label{eq:cgd-lep2}
\end{align}
where $\{a_\nu,b_\nu,c_\nu \}$ are complex free parameters.
Here $v_{\Delta_u}$ is VEV of $\Delta_u$ that is required to be $v_\Delta \lesssim {\cal O}(1)$ GeV to satisfy the constraint on the $\rho$ parameter~\cite{ParticleDataGroup:2020ssz}.
{The two-zero texture of the matrix is realized by the nature of modular $A_4$ symmetry. For weight 10 modular form, only $A_4$ singlet $1$ and $1'$ exist. Thus we obtain the two-zero texture by assigning singlet $A_4$ representations to $L$ as { $\{ \bm{1} \} =\{1, 1'', 1'\}$}.}
The mass eigenvalues for the active neutrinos $D_\nu = \{D_{\nu_1}, D_{\nu_2}, D_{\nu_3} \}$ are obtained via mixing matrix $V_\nu$ as $D_\nu \equiv V_\nu^T m_\nu V_\nu$.
Therefore, $V_\nu^\dag m_\nu^\dag m_\nu V_\nu ={\rm diag.}(|D_{\nu_1}|^2,|D_{\nu_2}|^2,|D_{\nu_3}|^2)$ where $D_{\nu_{1,2,3}}$ are the neutrino mass eigenvalues.
Pontecorvo-Maki-Nakagawa-Sakata (PMNS) mixing matrix $U(\equiv U_{PMNS})$ is defined by $U=V_L^\dag V_\nu$.
Then, the above texture is diagonalized in terms of $U$ and $V_L$ as follows:
\begin{align}
m_\nu=V^*_L U^*  
\left[\begin{array}{ccc}
D_{\nu_1} & 0 & 0 \\
0 & D_{\nu_2} & 0 \\
0 & 0 & D_{\nu_3} \\
\end{array}\right]
U^\dag V^\dag_L.
\end{align}
The matrix $U$ can be decomposed by $V  P$.
Here $V$ is the three by three unitary matrix consisting of three mixing angles $\theta_{12,23,13}$ and one Dirac CP-phase $\delta_{CP}$:
\begin{align}
V= 
\left[\begin{array}{ccc}
1 & 0 & 0 \\
0 & c_{23} & s_{23} \\
0 & -s_{23} & c_{23} \\
\end{array}\right]
\left[\begin{array}{ccc}
c_{13} & 0 & s_{13}e^{-i\delta_{CP}} \\
0 & 1 & 0 \\
-s_{13}e^{i\delta_{CP}} &0 & c_{13} \\
\end{array}\right]
\left[\begin{array}{ccc}
c_{12} & s_{12} &0 \\
-s_{12} & c_{12} & 0 \\
0 &0 & 1 \\
\end{array}\right]
,
\end{align}
where $s_{ij},\ c_{ij}$ are respectively  short-hand notations for $\sin\theta_{ij},\ \cos\theta_{ij}$ ($i j=12,23,13$).  
$P$ is the phase matrix given by $\rm{diag.}(1,e^{i\alpha/2},e^{i\beta/2})$ with $\{ \alpha,\ \beta \}$ being Majorana phases.
%{\color{red}where these Majorana phases are introduced in order to remove phases from the neutrino mass eigenvalues.}

Now we rewrite the neutrino mass matrix in terms of complex mass eigenvalues and mixings as follows:
 \begin{align}
m_\nu &=V^*_L V^*   
{
\left[\begin{array}{ccc}
1 & 0 & 0 \\
0 &e^{-i\alpha/2}  & 0 \\
0 & 0 & e^{-i\beta/2} \\
\end{array}\right]
}
\left[\begin{array}{ccc}
D_{\nu_1} & 0 & 0 \\
0 & D_{\nu_2} & 0 \\
0 & 0 & D_{\nu_3} \\
\end{array}\right]
\left[\begin{array}{ccc}
1 & 0 & 0 \\
0 & e^{-i\alpha/2}& 0 \\
0 & 0 &  e^{{\color{red}-}i\beta/2} \\
\end{array}\right]
V^\dag V^\dag_L \nonumber \\
& \equiv
V^*_L V^* 
\left[\begin{array}{ccc}
\lambda_1 & 0 & 0 \\
0 & \lambda_2 & 0 \\
0 & 0 & \lambda_3 \\
\end{array}\right]
V^\dag V^\dag_L,
\end{align}
 where $(\lambda_1,\lambda_2,\lambda_3)\equiv (D_{\nu_1}, D_{\nu_2} e^{-i\alpha}, D_{\nu_3} e^{-i\beta})$.
  {Here $D_{\nu_{1,2,3}}$ are real mass eigenvalues while $\lambda_{2,3}$ are complex ones ($\lambda_1$ coincides with $D_{{\nu_1}}$)}.
 Here, we redefine $U'\equiv V^*_L V^* $. Therefore, we find
  \begin{align}
m_\nu=
U' 
\left[\begin{array}{ccc}
\lambda_1 & 0 & 0 \\
0 & \lambda_2 & 0 \\
0 & 0 & \lambda_3 \\
\end{array}\right]
U'^T.
\end{align}
 Then, applying the above neutrino mass matrix for  two-zero textures denoted by $(m_\nu)_{ab}=(m_\nu)_{cd}=0$ ($ab\neq cd$),
 one finds the following relations:
 \begin{align}
 \Lambda_{12}&\equiv \frac{\lambda_1}{\lambda_2}= 
 \frac{U'_{a3} U'_{b3} U'_{c2} U'_{d2} - U'_{a2} U'_{b2} U'_{c3} U'_{d3}}
 {U'_{a1} U'_{b1} U'_{c3} U'_{d3} -U'_{a3} U'_{b3} U'_{c1} U'_{d1}},\nn\\
 %%%
  \Lambda_{13}&\equiv \frac{\lambda_1}{\lambda_3}= 
 \frac{U'_{a3} U'_{b3} U'_{c2} U'_{d2} - U'_{a2} U'_{b2} U'_{c3} U'_{d3}}
 {U'_{a2} U'_{b2} U'_{c1} U'_{d1} - U'_{a1} U'_{b1} U'_{c2} U'_{d2}}.
% \Lambda_{23}&\equiv \frac{\lambda_2}{\lambda_3}=  \frac{U'_{a1} U'_{b1} U'_{c3} U'_{d3} -U'_{a3} U'_{b3} U'_{c1} U'_{d1}}{U'_{a2} U'_{b2} U'_{c1} U'_{d1} - U'_{a1} U'_{b1} U'_{c2} U'_{d2}}.
 \label{eq:relori}
 \end{align}
% in Nufit 5.2~\cite{nufit52}.
Furthermore, one finds the following real mass ratios and Majorana phases {applying the above equations and our definition $(\lambda_1,\lambda_2,\lambda_3)\equiv (D_{\nu_1}, D_{\nu_2} e^{-i\alpha}, D_{\nu_3} e^{-i\beta})$}: 
 \begin{align}
 \frac{D_{\nu_1}}{D_{\nu_2}}&=\left| \Lambda_{12} \right|,\
%\left|  \frac{V_{a3} V_{b3} V_{c2} V_{d2} - V_{a2} V_{b2} V_{c3} V_{d3}}{V_{a2} V_{b2} V_{c1} V_{d1} - V_{a1} V_{b1} V_{c2} V_{d2}}\right|,\\
 %%%
 \frac{D_{\nu_1}}{D_{\nu_3}} =\left| \Lambda_{13}\right|,\
%\left|  \frac{V_{a1} V_{b1} V_{c3} V_{d3} - V_{a3} V_{b3} V_{c1} V_{d1}}{V_{a2} V_{b2} V_{c1} V_{d1} - V_{a1} V_{b1} V_{c2} V_{d2}}\right|,\\
%%% %%%
\alpha =
{\rm arg}\left[ \Lambda_{13} \right],\
%\left[  \frac{V_{a3} V_{b3} V_{c2} V_{d2} - V_{a2} V_{b2} V_{c3} V_{d3}}{V_{a2} V_{b2} V_{c1} V_{d1} - V_{a1} V_{b1} V_{c2} V_{d2}}\right],\\
 %%%
\beta =
{\rm arg}\left[ \Lambda_{12} \right].
%\left[  \frac{V_{a1} V_{b1} V_{c3} V_{d3} - V_{a3} V_{b3} V_{c1} V_{d1}}{V_{a2} V_{b2} V_{c1} V_{d1} - V_{a1} V_{b1} V_{c2} V_{d2}}\right].
 \end{align}
 Two mass square differences $\Delta m^2_{\mathrm{sol}}\equiv D_{\nu_2}^2-D_{\nu_1}^2$ and $\Delta m^2_{\mathrm{atm}}$
are given by
\begin{align}
& \Delta m^2_{\mathrm{sol}} = D_{\nu_1}^2 (\left| \Lambda_{12} \right|^{-2}-1),\\
& ({\bf NH}):~ \Delta m^2_{\mathrm{atm}} = D_{\nu_1}^2 (\left| \Lambda_{13} \right|^{-2}-1),\quad
 ({\bf IH}):~ \Delta m^2_{\mathrm{atm}} = D_{\nu_1}^2 (\left| \Lambda_{12} \right|^{-2}-\left| \Lambda_{13} \right|^{-2}). \label{eq:atm}
 \end{align}
In addition, the effective mass for the neutrinoless double beta decay is given by
\begin{align}
&\langle m_{ee}\rangle=\left|\sum_{i=1}^3 D_{\nu_i} U_{ei}^2\right|
=\left| D_{\nu_1} c^2_{12} c^2_{13}+D_{\nu_2} s^2_{12} c^2_{13}e^{i\alpha}
+D_{\nu_3} s^2_{13}e^{i(\beta - 2\delta_{CP})}\right|
\nn\\
&
=D_{\nu_1} \left|  c^2_{12} c^2_{13}+\left| \Lambda_{12} \right|^{-1} s^2_{12} c^2_{13}e^{i\alpha}
+\left| \Lambda_{13} \right|^{-1} s^2_{13}e^{i(\beta - 2\delta_{CP})}\right|.
\end{align}
Since concrete mixing of $V_L$ for $\tau=\omega$ and $i\infty$, we approximately estimate $\langle m_{ee}\rangle$.
At first, we show our $D_{\nu_1}$ from Eq.~(\ref{eq:atm}) depending on NH and IH.
\begin{align}
 ({\bf NH}):~D_{\nu_1} &=\sqrt{ \Delta m^2_{\mathrm{atm}}}  (\left| \Lambda_{13} \right|^{-2}-1)^{-1/2} \nn \\
&
\approx s^5_{23} \sqrt{\Delta m^2_{\mathrm{atm}}} \sqrt{-1+\frac{c_{23}^8}{s_{23}^8}}
\left[
\frac{s^3_{23}}{c^8_{23}-s^8_{23}} + \frac{c_{12} c_{23}^9 s_{23}}{2s_{12} (c^2_{23}-s^2_{23})^2(c^4_{23}+s^4_{23})^2} (\epsilon_{12}+\epsilon_{12}^*)\right.\nn\\
&\left. 
+ \frac{c_{23}^{7} s^2_{23}}{(c^8_{23}-s^8_{23})^2} (\epsilon_{23}+\epsilon_{23}^*)
+ \frac{c_{12} c_{23}^{10} }{2s_{12} (c^8_{23}-s^8_{23})^2} (\epsilon_{13}+\epsilon_{13}^*)
\right],  \label{eq:NHD1} \\
 ({\bf IH}):~ D_{\nu_1} &= \sqrt{\Delta m^2_{\mathrm{atm}}} (\left| \Lambda_{12} \right|^{-2}-\left| \Lambda_{13} \right|^{-2})^{-1/2} \nn \\
&
\approx \frac{s^5_{23}}{c^8_{23}-s^8_{23}} \sqrt{\Delta m^2_{\mathrm{atm}}} \sqrt{-1+\frac{c_{23}^8}{s_{23}^8}}
\left[
-s^3_{23} - \frac{c_{23}^5 s^2_{23}}{c^8_{23}-s^8_{23}} (\epsilon_{23}+\epsilon_{23}^*)\right.\nn\\
&\left. 
+ \frac{c_{23} s^9_{23}}{2c_{12} s_{12} (c^8_{23}-s^8_{23})} (\epsilon_{12}+\epsilon_{12}^*)
+ \frac{s^8_{23}c^2_{23}-c^{10}_{23}c^2_{12}}{2s_{12} c_{12}(c^8_{23}-s^8_{23})} (\epsilon_{13}+\epsilon_{13}^*)
\right].  \label{eq:IHD1} %
 \end{align}
On the other hand, the common part of $\langle m_{ee}\rangle$ is found as follows:
\begin{align}
&
 \left|  c^2_{12} c^2_{13}+\left| \Lambda_{12} \right|^{-1} s^2_{12} c^2_{13}e^{i\alpha}
+\left| \Lambda_{13} \right|^{-1} s^2_{13}e^{i(\beta - 2\delta_{CP})}\right|
\nn\\
&
=
\left| 
c^2_{12} c_{13}^2 +s^2_{12} s^2_{13}e^{i\alpha} + \left(\frac{c_{23}}{s_{23}}\right)^4 s^2_{13} e^{i(\beta-2\delta_{CP})}
-\frac{c_{12}c_{23}^5s_{13}^2}{2s_{12}s^6_{23}}e^{i(\beta - 2\delta_{CP})} (\epsilon_{12}+\epsilon_{12}^*)\right.\nn\\
&\left. 
-\frac{c_{23}s_{12} c_{13}^2}{2c_{12}s^2_{23}}e^{i\alpha} (\epsilon_{12}+\epsilon_{12}^*)
-\frac{c_{23}^3 s_{13}^2}{s^5_{23}}e^{i(\beta - 2\delta_{CP})} (\epsilon_{23}+\epsilon_{23}^*)\right.\nn\\
&\left. 
-\frac{1}{2c_{12} s_{12} s^2_{23}}\left(c^2_{12} c_{23}^6 e^{i(\beta-2\delta_{CP})}+ c_{23}^2 s_{12}^2 s_{23}^4 e^{i\alpha}\right) (\epsilon_{13}+\epsilon_{13}^*)
\right|.\label{eq:mee2}
\end{align}
Combining Eqs.~(\ref{eq:NHD1}), ~(\ref{eq:IHD1}), and ~(\ref{eq:mee2}), we find $\langle m_{ee}\rangle$ takes corrections from deviations from fixed points of $\tau=\omega,\ i\infty$. 
where its predicted value is constrained by the current KamLAND-Zen data and could be measured in future~\cite{KamLAND-Zen:2024eml}.
The upper bound is found as $\langle m_{ee}\rangle<(28-122)$ meV at 90 \% confidence level where the range of the bound comes from the use of different method estimating nuclear matrix elements. 

The sum of neutrino masses is given by
\begin{align}
&\sum_{i=1}^3 D_{\nu_i}
=D_{\nu_1}\left(1+ \left| \Lambda_{12} \right|^{-1} 
+\left| \Lambda_{13} \right|^{-1}\right).
\end{align}
%
%
%Since concrete mixing of $V_L$ for $\tau=\omega$ and $i\infty$, 
Similar way to the computation of  $\langle m_{ee}\rangle$,
we approximately estimate $\sum D_{\nu}$ as follows:
 \begin{align}
 \sum D_{\nu}&\approx
 \sqrt{ \Delta m^2_{\mathrm{atm}} }\left[
\frac{s_{23}^4}{c_{23}^4} +\frac{c_{12}  s_{23}^2}{2 c_{23}^3 s_{12}} (\epsilon_{12}+\epsilon_{12}^*) 
+\frac{c_{12} s_{23}}{2 c_{23}^2 s_{12}} (\epsilon_{13}+\epsilon_{13}^*)
+\frac{s_{23}^3 }{c_{23}^5} (\epsilon_{23}+\epsilon_{23}^*)\right],
\end{align}
where the above form is correct for both the cases of NH and IH, and we assume $s_{13}\ll s_{12},s_{23}$ that is expected by the experimental result. The first terms of above equations come from two-zero texture in Eq.(\ref{eq:cgd-lep2}).
While the other terms of above equations, which are proportional to $\epsilon_{ij}+\epsilon_{ij}^*$, originate from our new contribution from the modular $A_4$ symmetry. 
The sum of neutrino masses is constrained by the minimal cosmological model
$\Lambda$CDM $+\sum D_{\nu_i}$ that provides the upper bound on $\sum D_{\nu}\le$ 120 meV~\cite{Vagnozzi:2017ovm, Planck:2018vyg}, although it becomes weaker if the data are analyzed in the context of extended cosmological models~\cite{ParticleDataGroup:2014cgo}.
%%%
Recently, DESI and CMB data combination provides more stringent upper bound on the sum of neutrino masses~$\sum D_{\nu}\le$ 72 meV~\cite{DESI:2024mwx}. 
%%%
{ We note that $\sum D_{\nu} \simeq 172$ meV with IH when the two-zero texture is exact~\cite{Fritzsch:2011qv} and it is beyond the Planck CMB limit. 
Thus some deviation from the two-zero texture is necessary to satisfy the constraints on $\sum D_{\nu}$.}
The two observable $\langle m_{ee}\rangle$ and $\sum D_{\nu_i}$ are also taken into account in the numerical analysis.

\if0
% \begin{align} \epsilon_{13,23} \ll \epsilon_{12}:\ &\sum D_{\nu}\approx \frac{s_{12}s_{23}^4}{c_{23}^4 s_{12}} +\frac{c_{12} c_{23} s_{23}^2}{2 c_{23}^4 s_{12}} (\epsilon_{12}+\epsilon_{12}^*),\\
%
%\epsilon_{12,23} \ll \epsilon_{13}:\ &\sum D_{\nu}\approx.\frac{s_{23}^4}{c_{23}^4} +\frac{c_{12} s_{23} }{2 c_{23}^2 s_{12}} (\epsilon_{13}+\epsilon_{13}^*),\\
%
%\epsilon_{12,13} \ll \epsilon_{23}:\ &\sum D_{\nu}\approx\frac{s_{23}^4c_{23}}{c_{23}^5} +\frac{s_{23}^3 }{c_{23}^5} (\epsilon_{23}+\epsilon_{23}^*),\end{align}
\fi

\section{Numerical analysis and phenomenology}
\label{sec:III}
In this section, we carry out numerical analysis to fit the neutrino data and find some predictions of the models.
Then some implications for phenomenology are discussed.
We randomly select our input parameters within the following ranges 
\begin{align}
(|p|,|q|,|r|) \in [10^{-3},10^{3}]\ {\rm eV}.
\end{align}
Also, we allow $|\epsilon|=[10^{-5}-0.05]$ at each of the fixed point where $\tau=\tau_{fixed}+\epsilon$.
Note that Im[$\tau_{fixed}$]$=[2-10]$ in case of $\tau=i\infty$.
%In addition to the experimental values in Eqs.~(17,18),  we randomly select the value of $m_3$ within the range of
%\begin{align}m_3 \in [10^{-15},10^{-9}]\ {\rm eV}.\end{align}
{We apply Nufit5.2 experimental results within 3$\sigma$ intervals; three mixings ($s_{12},\ s_{13},\ s_{23}$), two mass squared differences ($\Delta m^2_{\mathrm{atm}},\ \Delta m^2_{\mathrm{sol}}$), and Dirac CP phase $\delta_{\rm CP}$, for our model.
Note, in Nufit, the definition of Majorana phases is given by  $\rm{diag.}(e^{ia},e^{ib},1)$ instead of our definition $\rm{diag.}(1,e^{i\alpha/2},e^{i\beta/2})$.
%$P$ is the phase matrix given by $\rm{diag.}(1,e^{i\alpha/2},e^{i\beta/2})$ with $\{ \alpha,\ \beta \}$ being Majorana phases.
%one has to replace our $\alpha,\beta$ into $2\beta,$
}

 %%%%%%%%%%%%%%%%%%%
\begin{figure}[tb]
\begin{center}
\includegraphics[width=50.0mm]{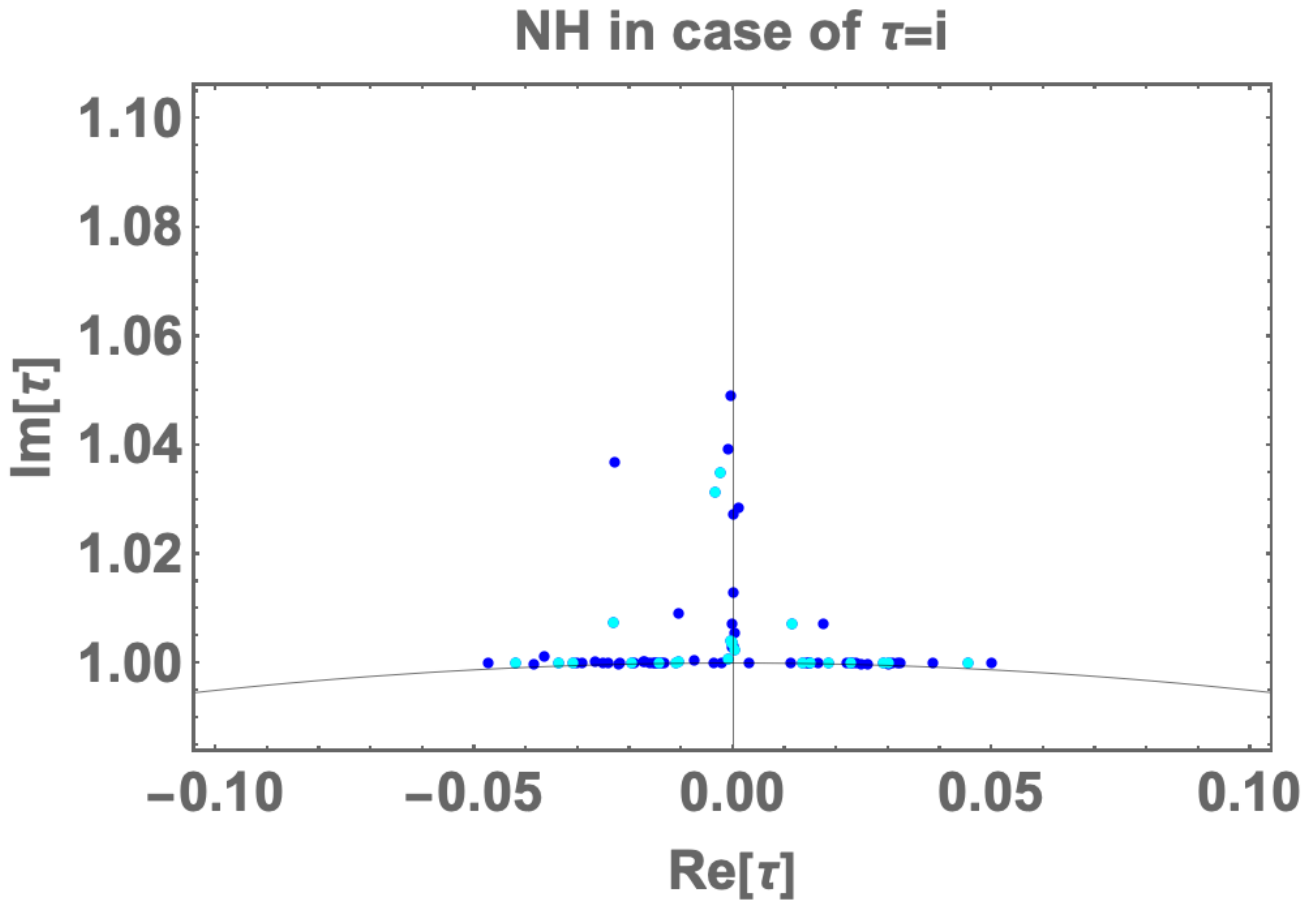} \quad
%%%
\includegraphics[width=50.0mm]{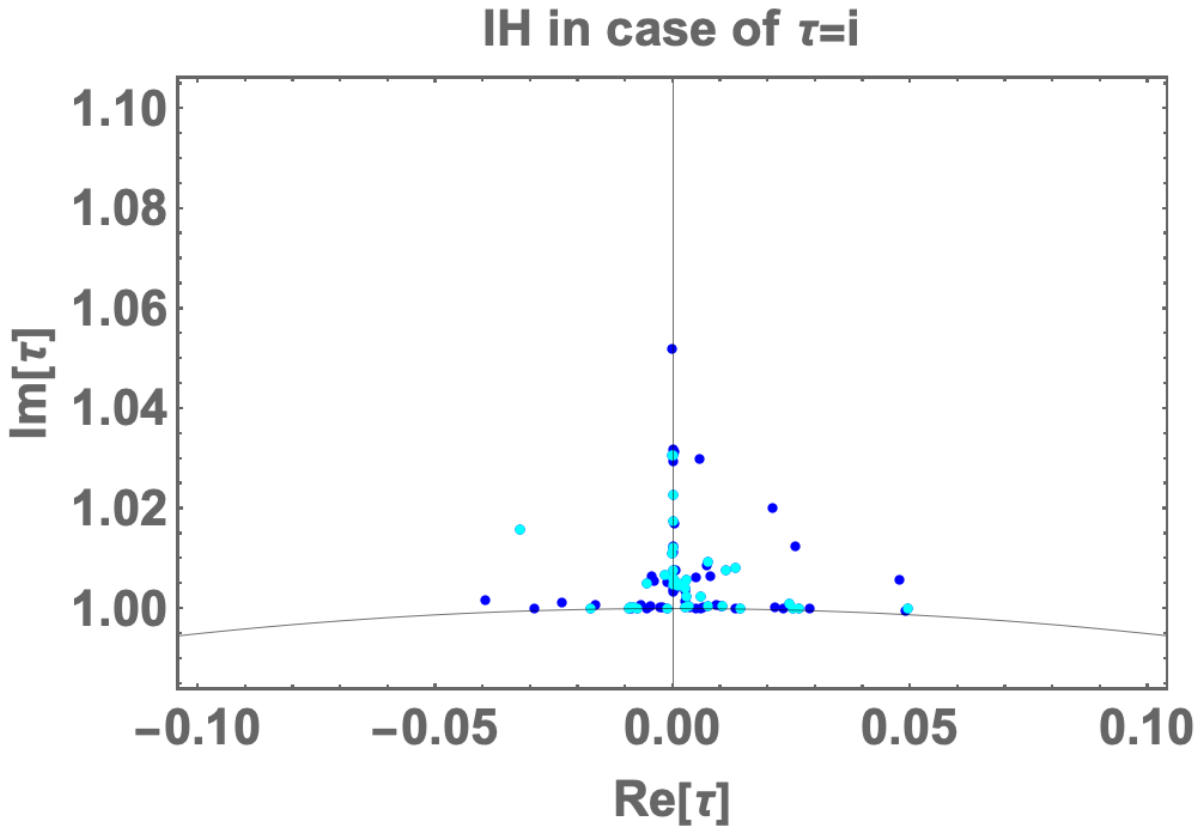} \quad
\caption{Numerical analyses at nearby $\tau=i$. The left figure represents the case of NH and the right one IH. Cyan points satisfy $\sum D_\nu\le$ 120 meV.}
  \label{fig:tau=i_1}
\end{center}\end{figure}
%%%%%%%%%%%%%%%%%%%   

 %%%%%%%%%%%%%%%%%%%
\begin{figure}[tb]
\begin{center}
\includegraphics[width=50.0mm]{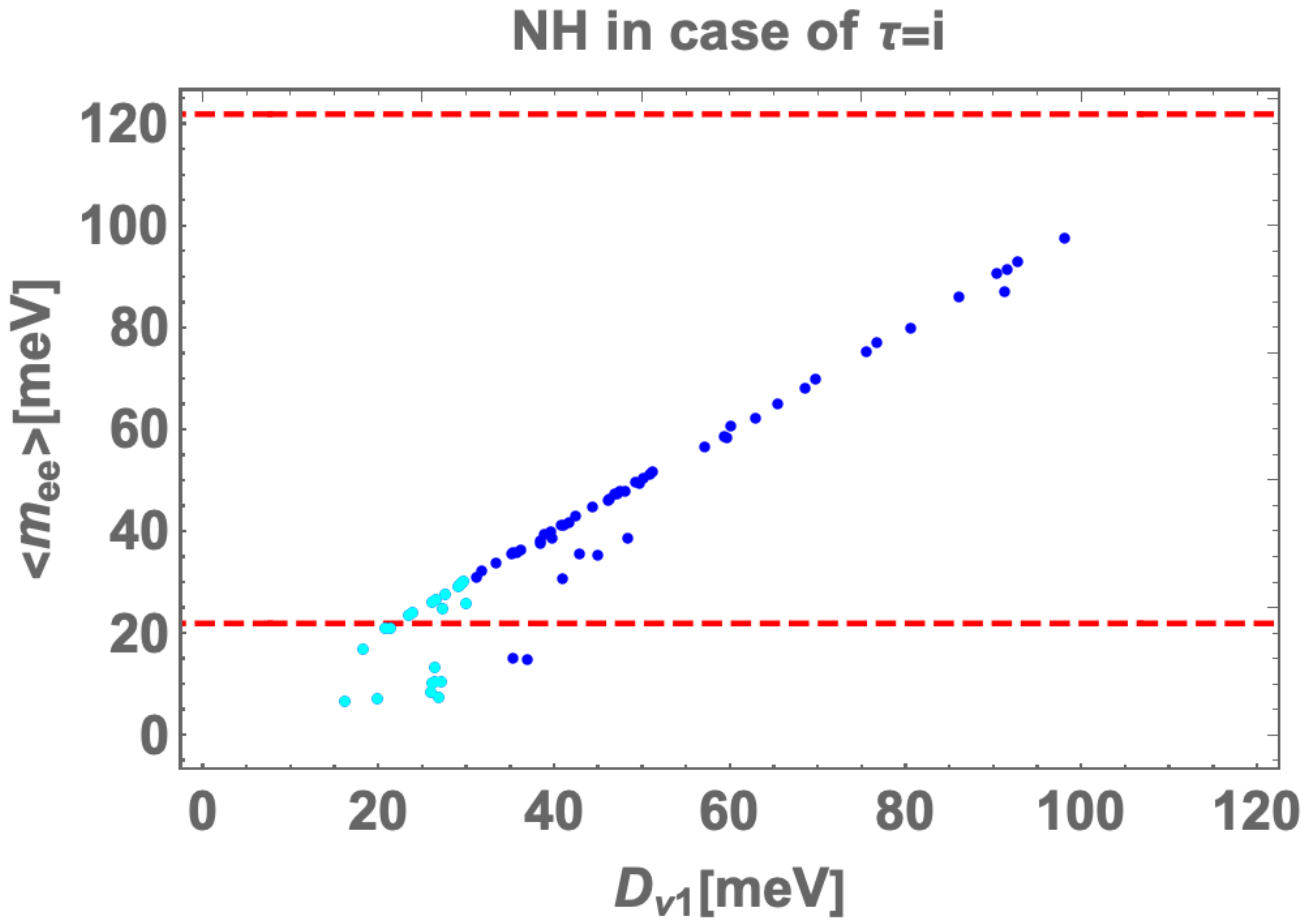} \quad
\includegraphics[width=50.0mm]{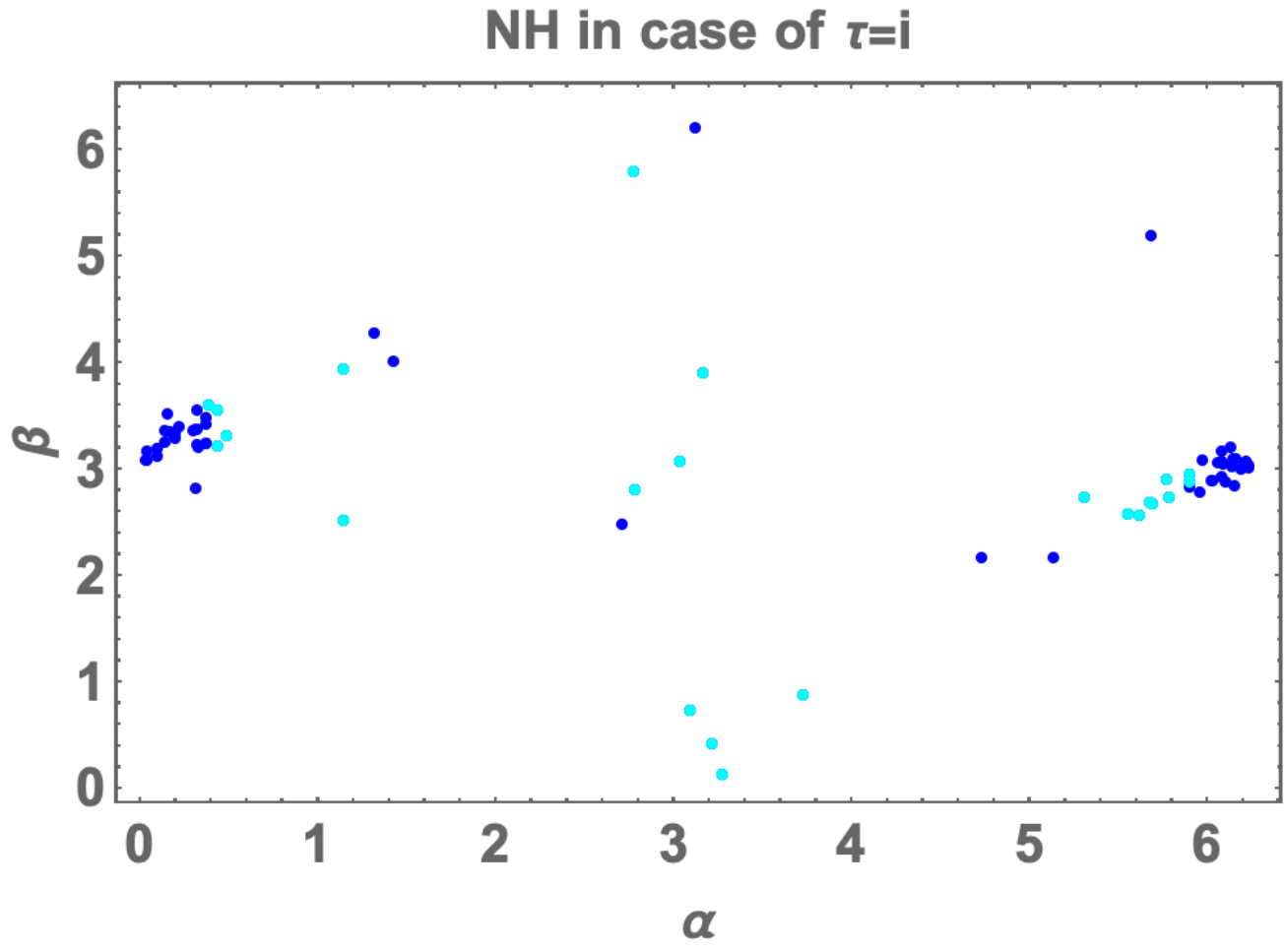}  \quad
\includegraphics[width=50.0mm]{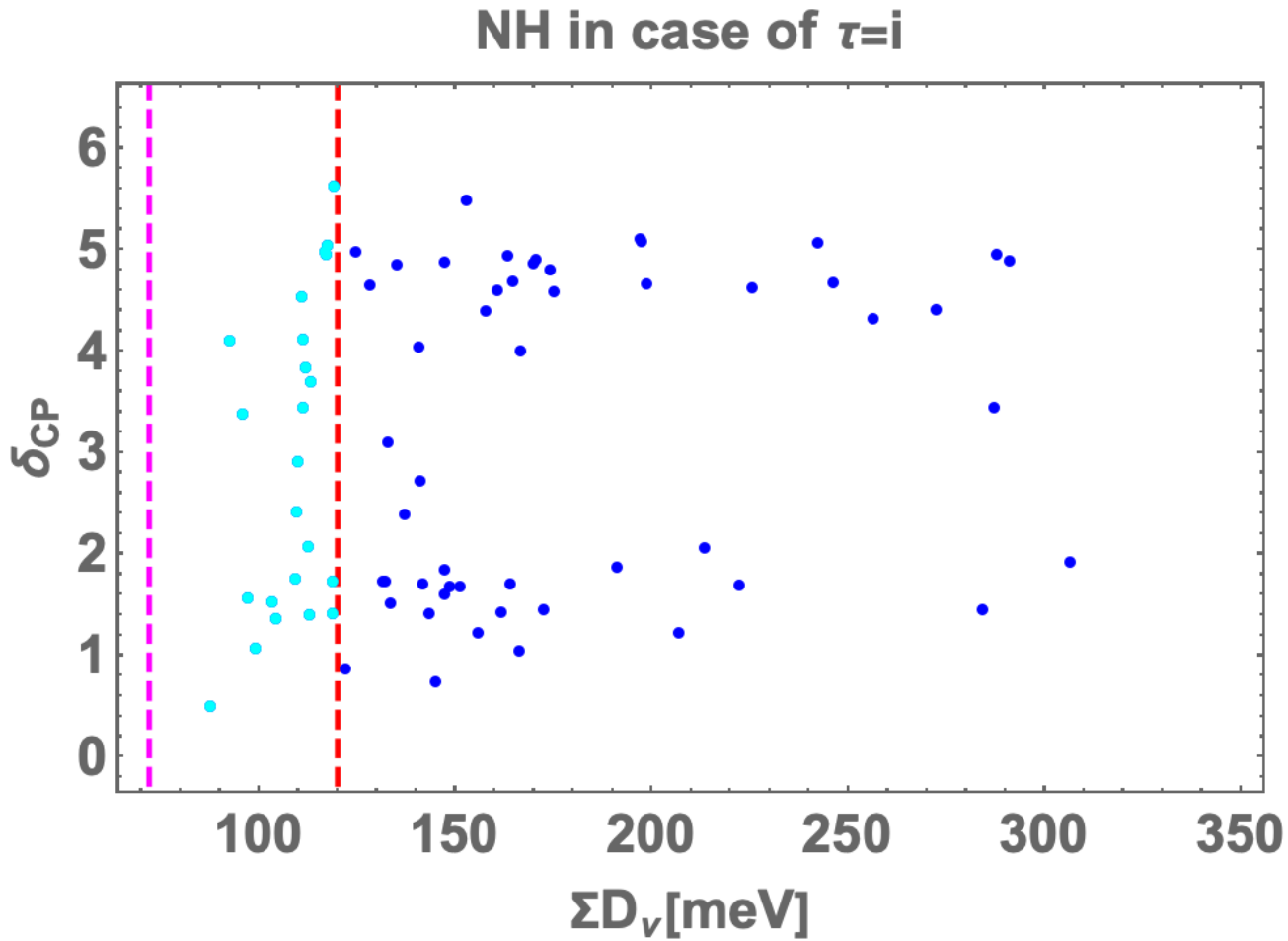} \\
%%%
\includegraphics[width=50.0mm]{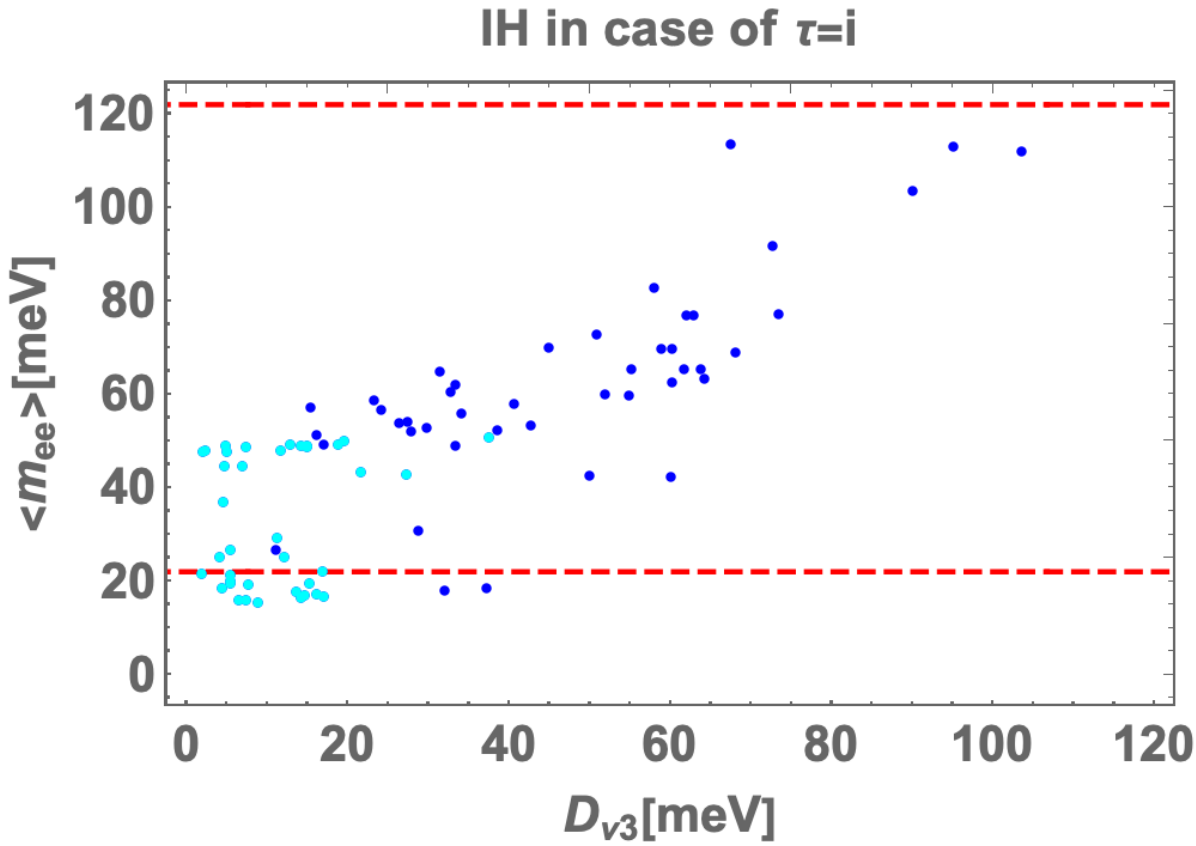} \quad
\includegraphics[width=50.0mm]{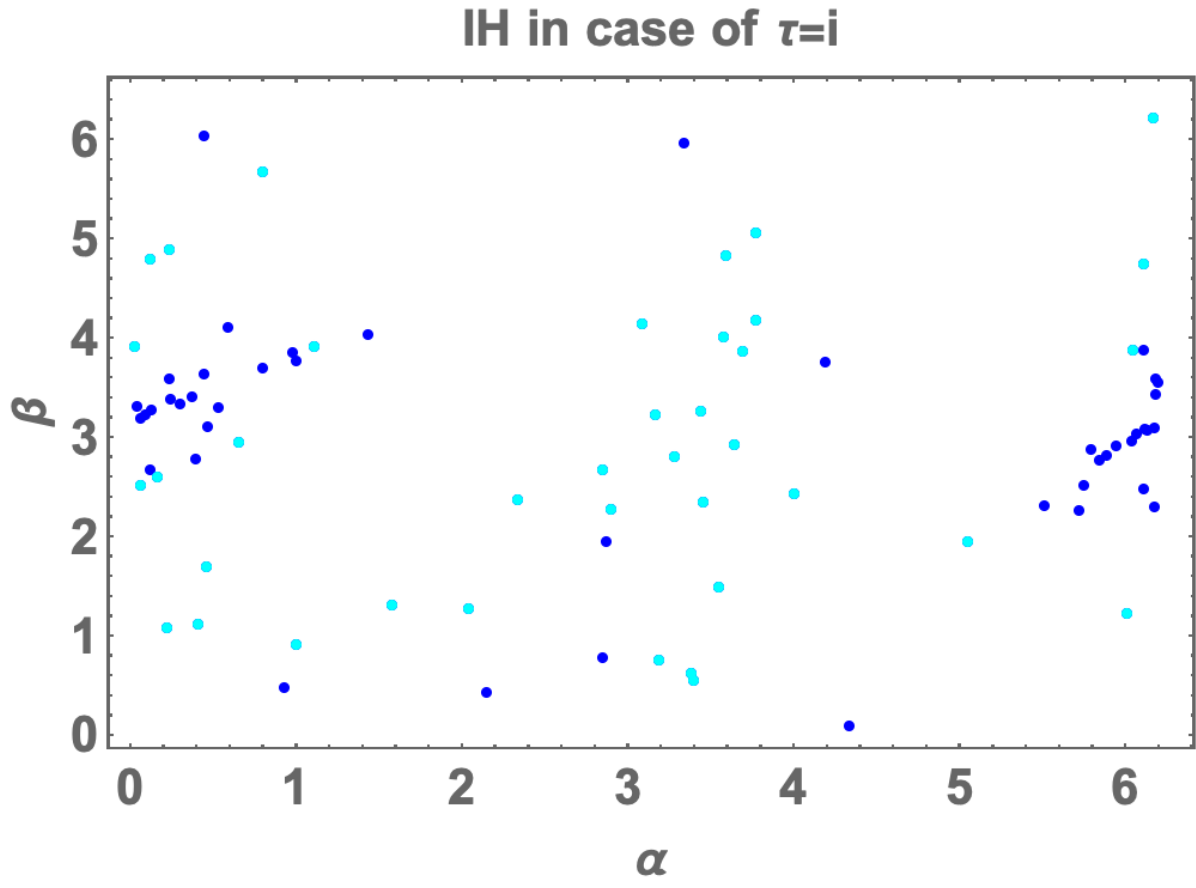}  \quad
\includegraphics[width=50.0mm]{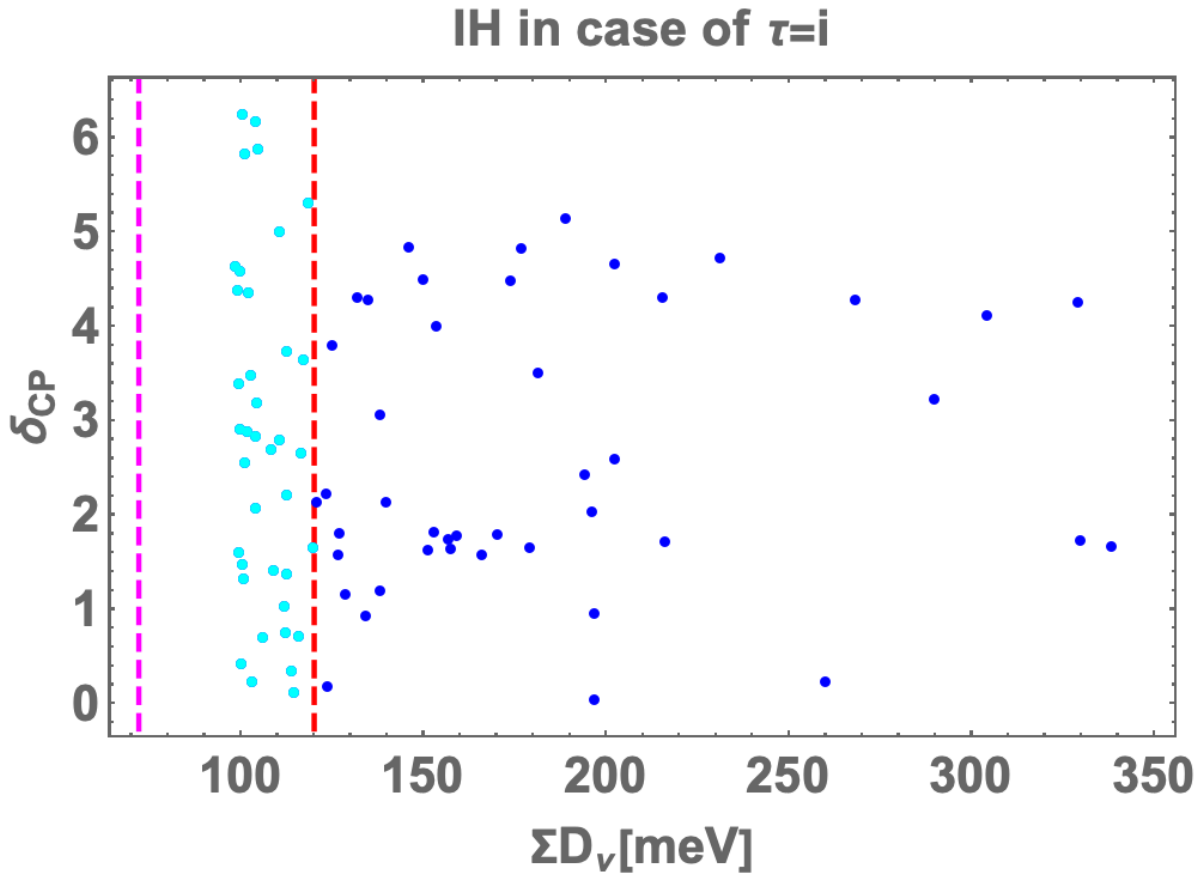} 
\caption{Numerical analyses at nearby $\tau=i$.
The up and down figures represent the NH and the IH cases, respectively. 
The left ones demonstrate neutrinoless double beta decay $\langle m_{ee}\rangle$ in terms of the lightest neutrino mass {where red dashed horizontal lines indicate the current upper bound from KamLAND-Zen with different estimation for nuclear matrix elements.}.
The center ones shows Majorana phases $\alpha$ and $\beta$.
The right ones depict  the Dirac CP phase in terms of the sum of the neutrino masses {where the red(magenta) dashed vertical lines indicate the upper bound on $\sum D_{\nu}$ from Planck CMB (+ DESI BAO) data}. }
  \label{fig:tau=i_2}
\end{center}\end{figure}
%%%%%%%%%%%%%%%%%%%   
%
\subsection{$\tau=i$}
In Fig.~\ref{fig:tau=i_1}, we figure out the allowed range of $\tau$ at nearby $\tau=i$
where the left figure represents the case of NH and the right one is for IH.
It implies that IH would have more points at nearby $\tau=i$.
Cyan points satisfy $\sum D_\nu\le$ 120 meV.
%%%
In Fig.~\ref{fig:tau=i_2}, we show predictions in the case of $\tau=i$.
The up and down figures represent the NH and the IH cases, respectively. 
The left ones demonstrate neutrinoless double beta decay $\langle m_{ee}\rangle$ in terms of the lightest neutrino masses.
The dotted red horizontal lines are bounds for the current KamLAND-Zen as discussed in the neutrino sector.
In both the cases, most of the points are within this bound and good testability would be expected. 
The center ones shows Majorana phases $\alpha$ and $\beta$.
The figures imply that points of NH case would be localized at nearby $\alpha=0$ and $\beta=\pi$. 
%although these tendency would be the same.
%
{At nearby this localized region, the smaller regions of $\langle m_{ee}\rangle$  and $\sum D_\nu$ are forbidden, therefore, $20$meV$\lesssim \langle m_{ee}\rangle$ and $100$~meV$\lesssim \sum D_\nu$. In case of IH, if we concentrate on this localized region,
small region of $\langle m_{ee}\rangle$ is restricted; $45$~meV$\lesssim \langle m_{ee}\rangle$.
Indeed we also have the allowed points that are away from the localized region where we can obtain smaller $\langle m_{ee}\rangle$ and $\sum D_\nu$ values.
}
The right ones depict  the Dirac CP phase  in terms of the sum of neutrino masses.
The vertical red dashed line shows the upper bound of cosmological limit $\sum D_\nu\le120$ meV while
the vertical magenta dashed line shows the upper bound of $\sum D_\nu\le72$ meV.
For both NH and IH case, we find the allowed region to satisfy $\sum D_\nu\le120$ meV due to the deviation from two-zero texture originated from the modular structure of charged-lepton mass matrix.

 %%%%%%%%%%%%%%%%%%%
\begin{figure}[tb]
\begin{center}
\includegraphics[width=50.0mm]{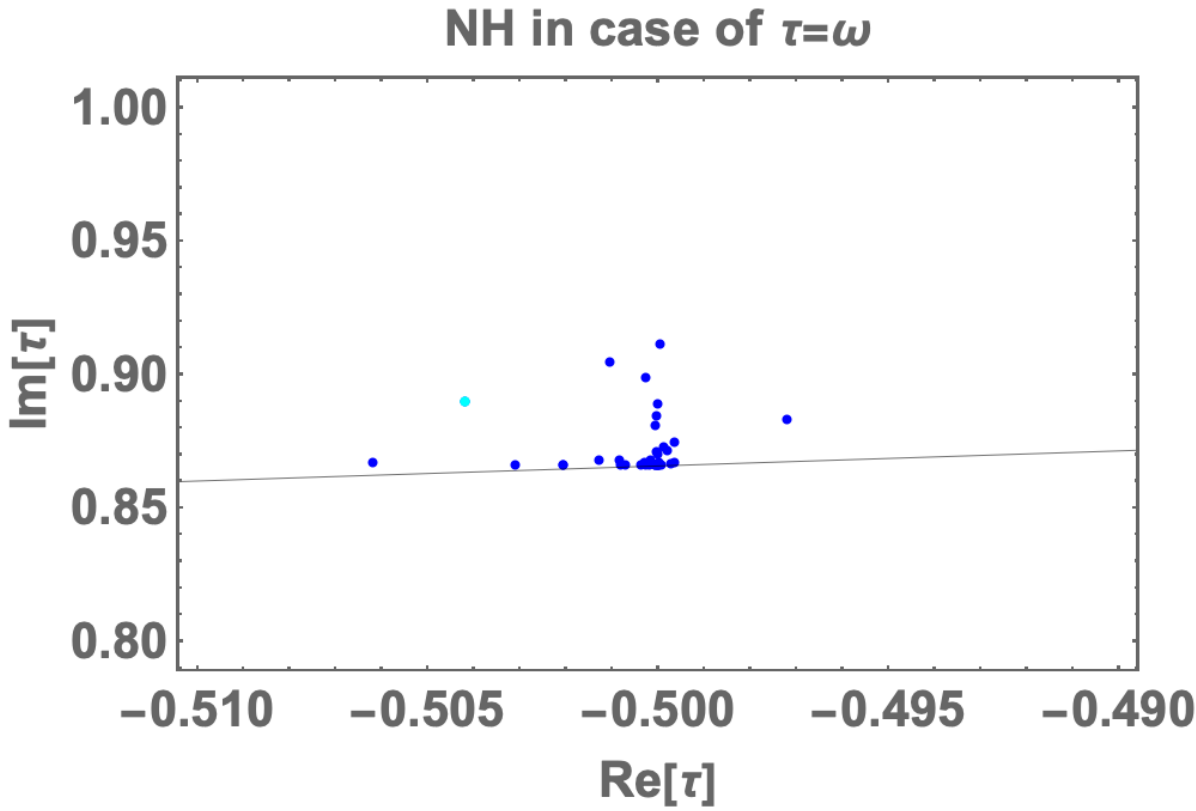} \quad
%%%
\includegraphics[width=50.0mm]{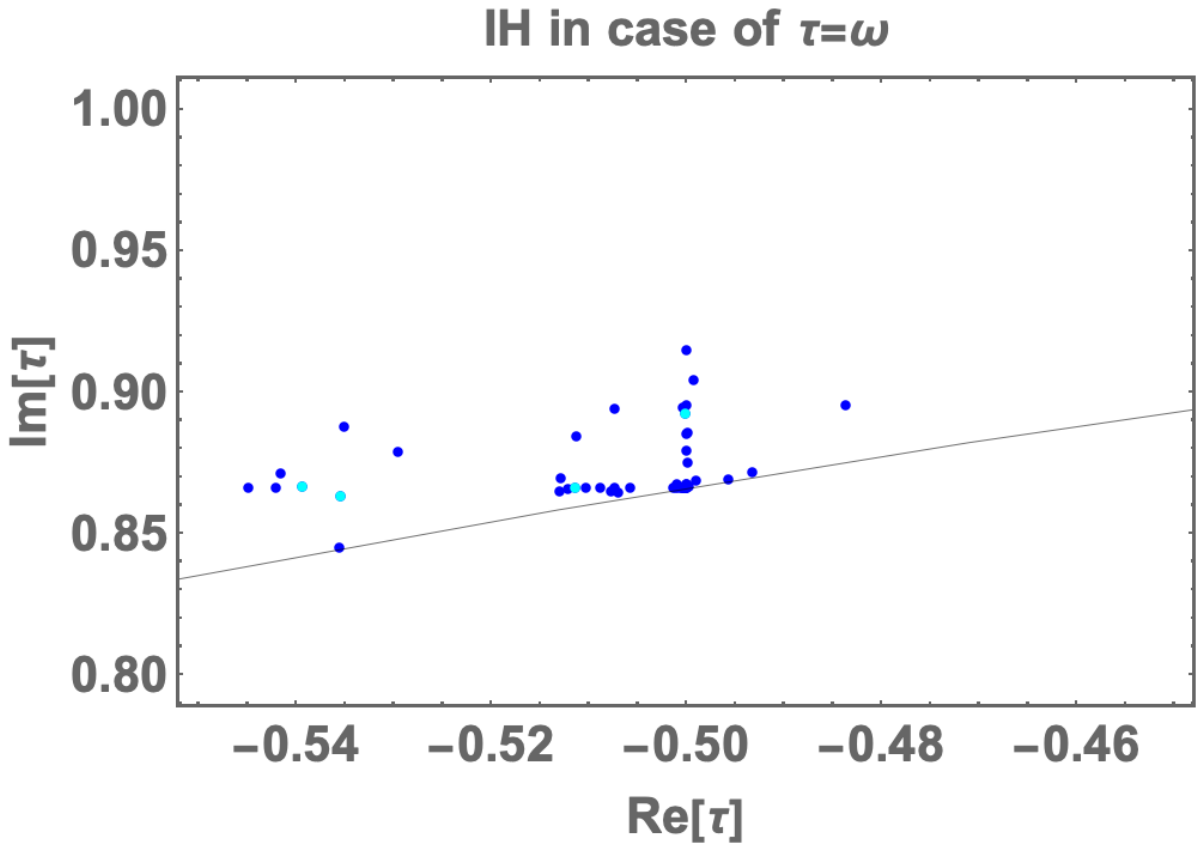} \quad
\caption{Numerical analyses at nearby $\tau=\omega$ where the legends of figures are the same as Fig.~\ref{fig:tau=i_1}.}
  \label{fig:tau=omega_1}
\end{center}\end{figure}
%%%%%%%%%%%%%%%%%%%   

 %%%%%%%%%%%%%%%%%%%
\begin{figure}[tb]
\begin{center}
\includegraphics[width=50.0mm]{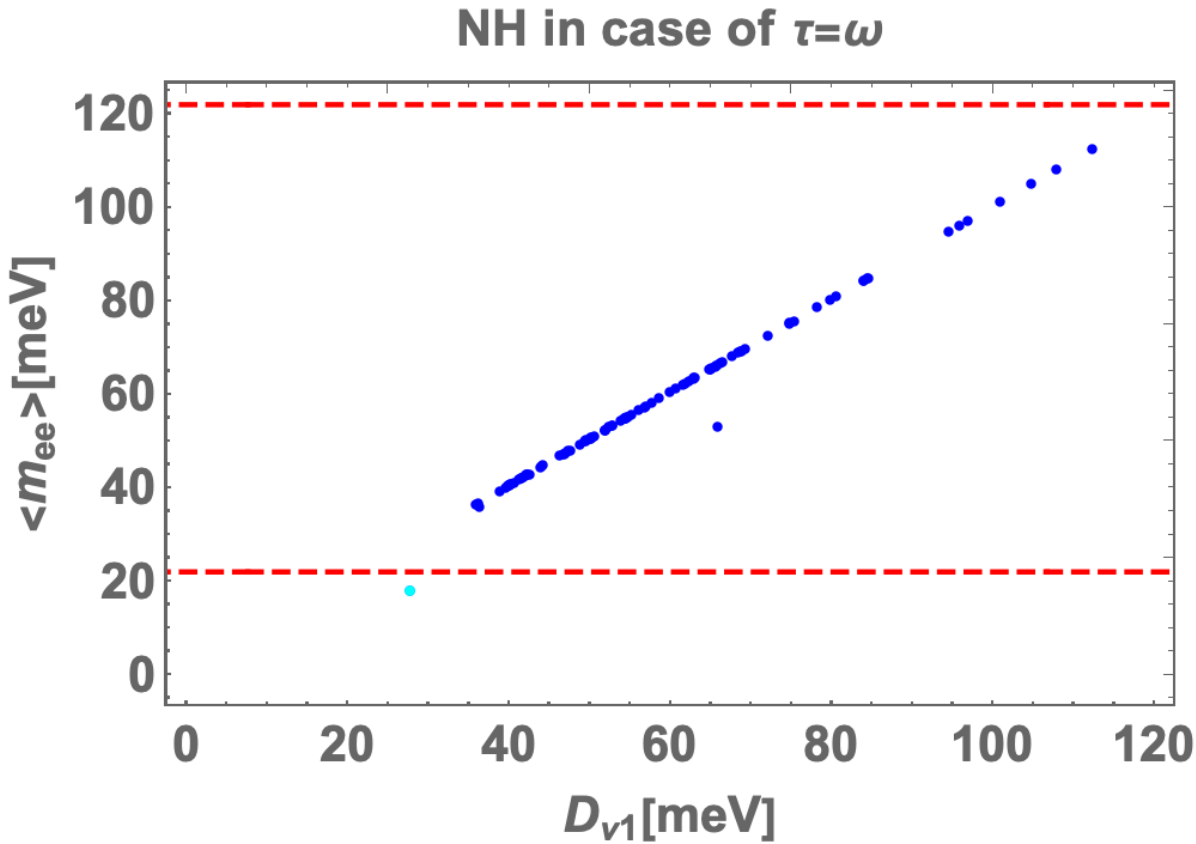} \quad
\includegraphics[width=50.0mm]{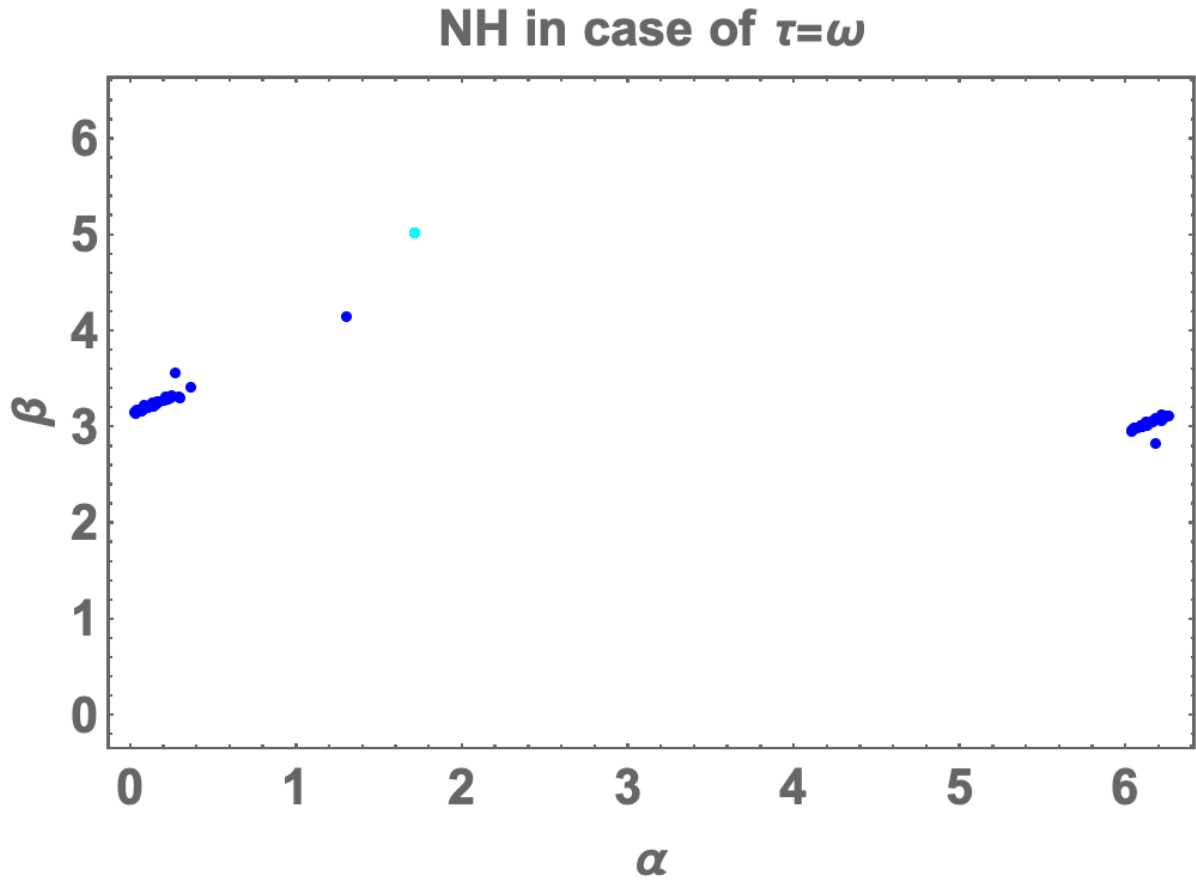}  \quad
\includegraphics[width=50.0mm]{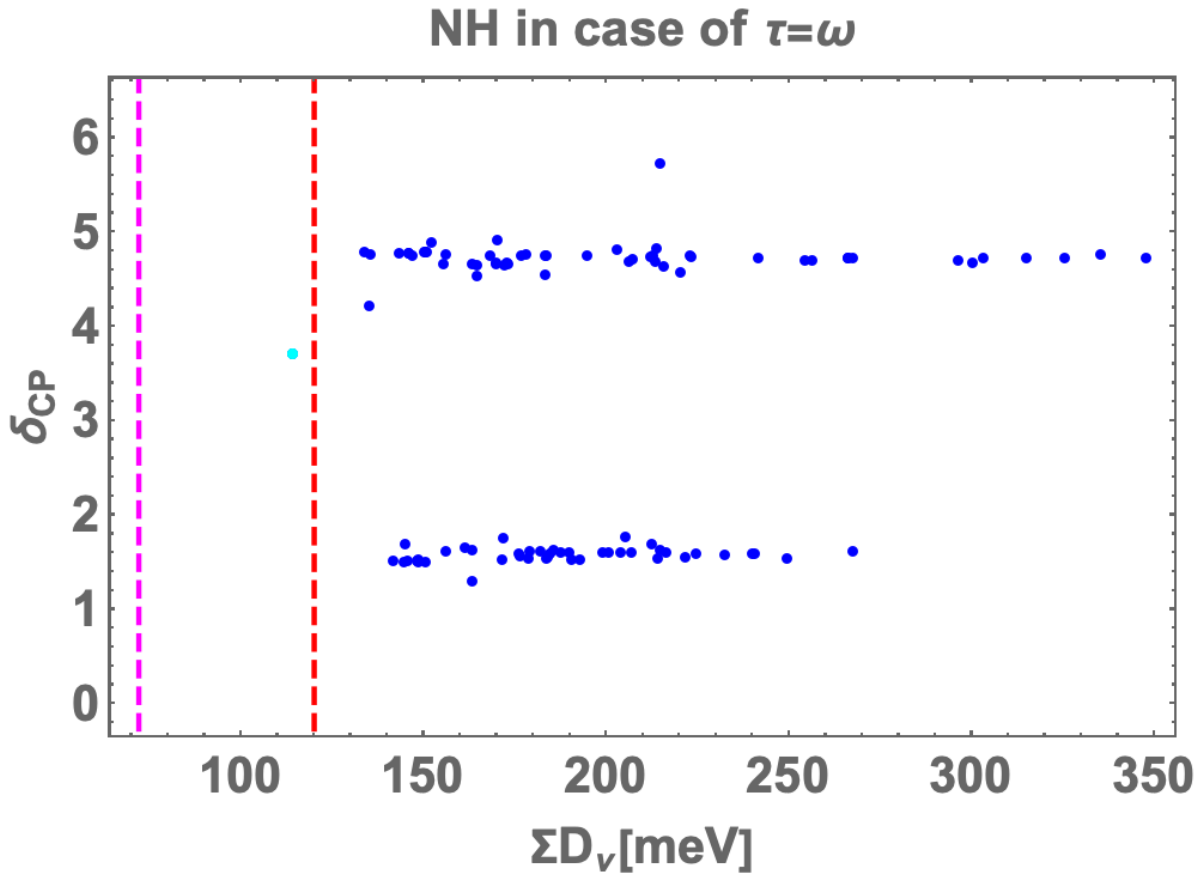} \\
%%%
\includegraphics[width=50.0mm]{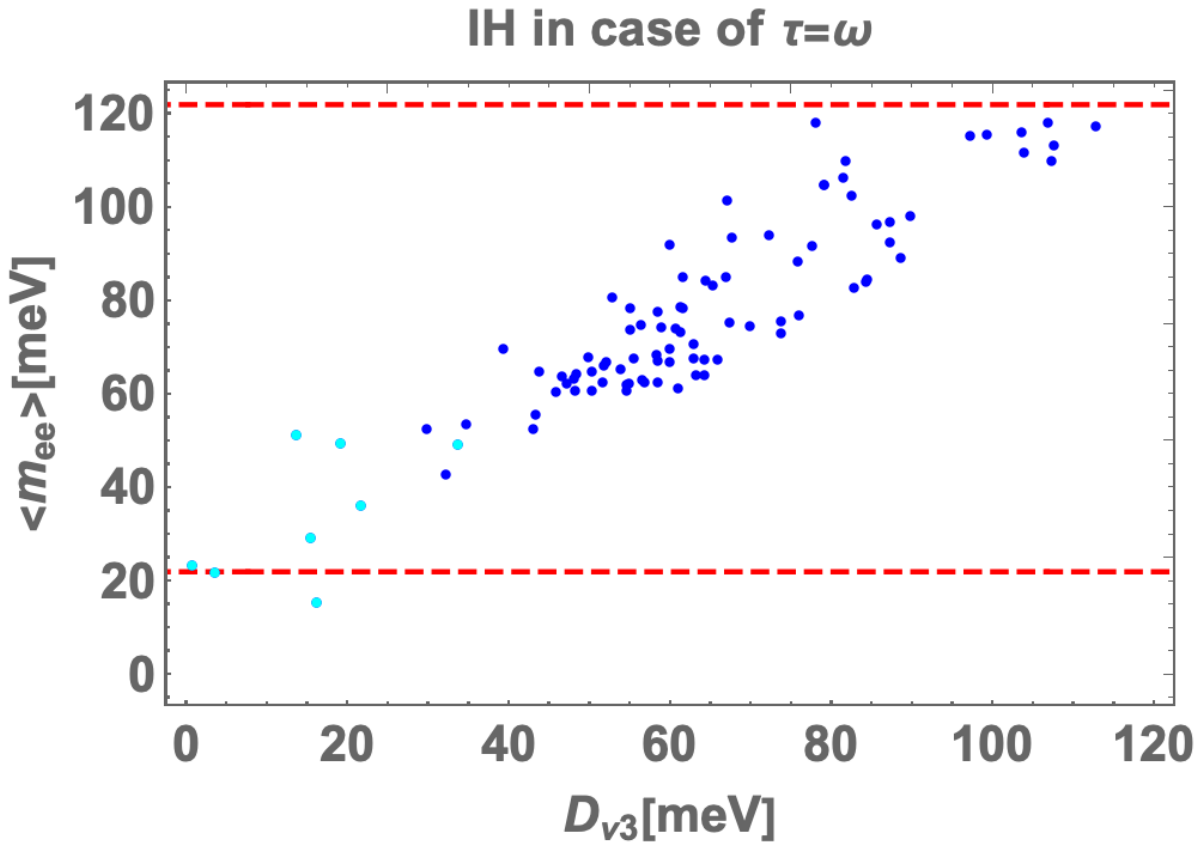} \quad
\includegraphics[width=50.0mm]{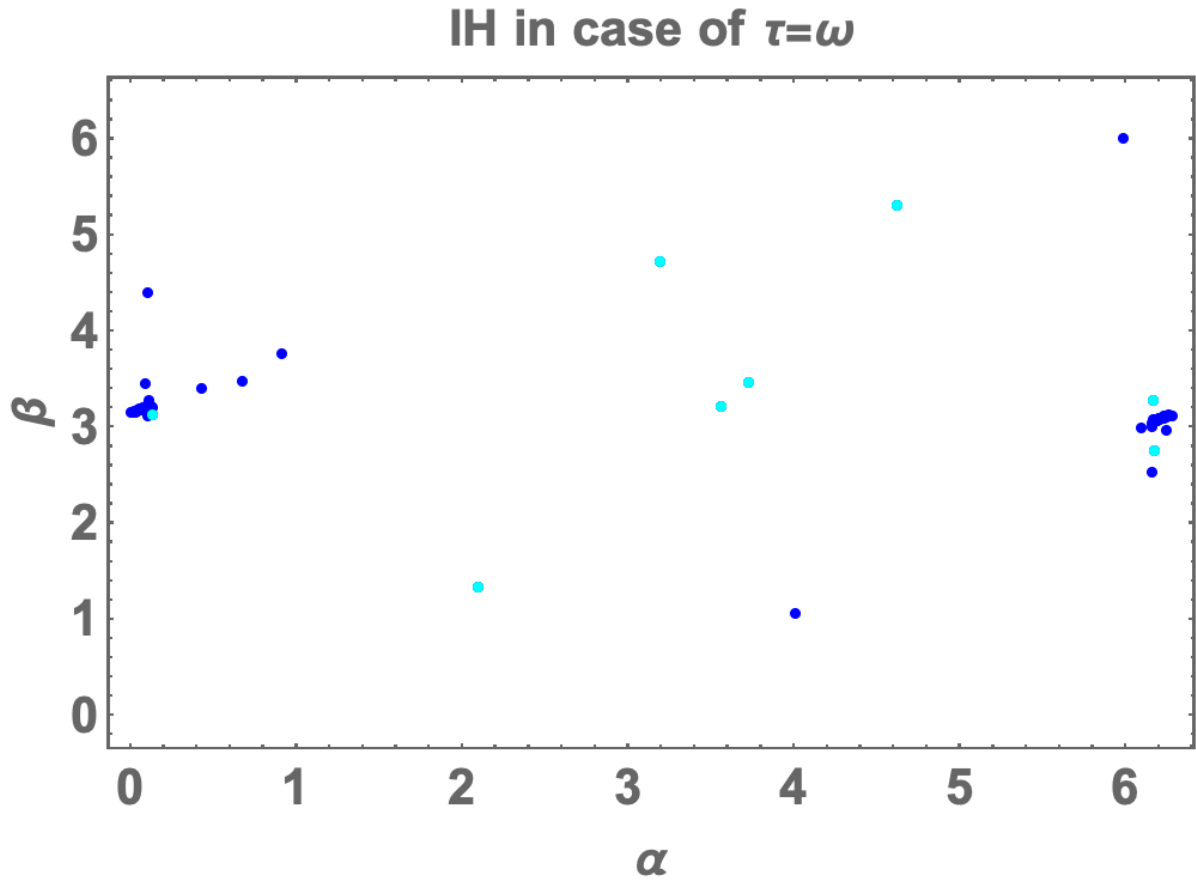}  \quad
\includegraphics[width=50.0mm]{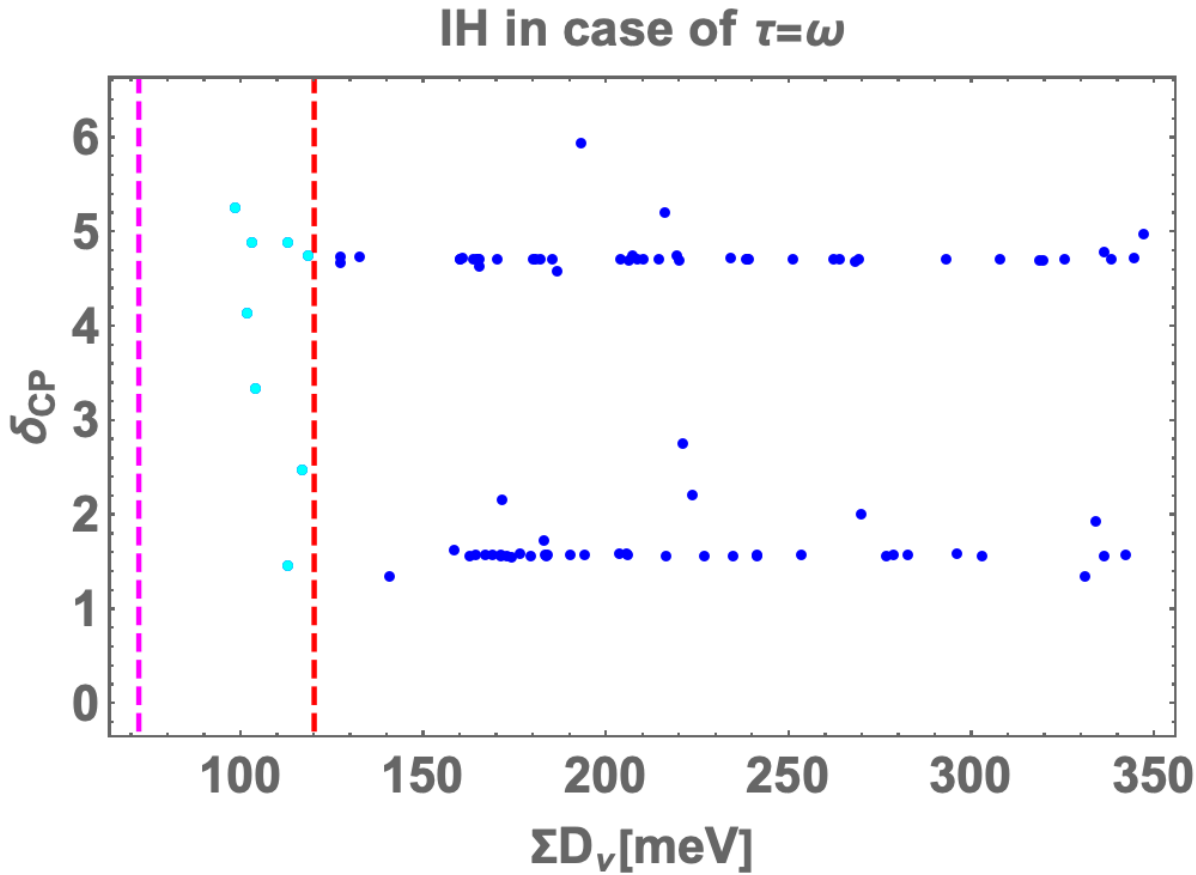} 
\caption{Numerical analyses at nearby $\tau=\omega$.
All the legends are the same as Figs.~\ref{fig:tau=i_2}. }
  \label{fig:tau=omega_2}
\end{center}\end{figure}
%%%%%%%%%%%%%%%%%%%   
%
\subsection{$\tau=\omega$}
In Fig.~\ref{fig:tau=omega_1}, we figure out the allowed range of $\tau$ at nearby $\tau=\omega$
where the left figure represents the case of NH and the right one is for IH.
These figures suggest that both the tendency are the same.
%%%
In Fig.~\ref{fig:tau=omega_2}, we show predictions in the case of $\tau=\omega$.
The up and down figures represent the NH and the IH cases, respectively. 
The left ones demonstrate neutrinoless double beta decay $\langle m_{ee}\rangle$ in terms of the lightest neutrino masses where 
the dotted red horizontal lines are bounds from the current KamLAND-Zen as given in the Fig.~\ref{fig:tau=i_2}.
In both the cases, most of the points are within the bound and good testability would be expected. 
The center ones show Majorana phases $\alpha$ and $\beta$, and
the figures imply that {both points of NH and IH cases} would be localized at nearby $\alpha=0$ and $\beta=\pi$.
% although these tendency would be the same
%
{
At nearby this localized region, the smaller regions of $\langle m_{ee}\rangle$  and $\sum D_\nu$ are forbidden, therefore, $35$~meV$\lesssim \langle m_{ee}\rangle$ and $130$~meV$\lesssim \sum D_\nu$. In case of IH, the tendency is the same as the case of NH; $30$~meV$\lesssim \langle m_{ee}\rangle$ and $110$~meV$\lesssim \sum D_\nu$. Note that only small number of points away from the localized region can realize smaller values of $\langle m_{ee}\rangle$ and $\sum D_\nu$.
}
The right ones represent the Dirac CP phase in terms of the sum of neutrino masses where
the vertical red (magenta) dashed line shows the upper bound of $\sum m_\nu$ the same as the Fig.~\ref{fig:tau=i_2}.
It would show that IH case is better than the NH one to satisfy the bound of $\sum m_\nu < 120$ meV.

 %%%%%%%%%%%%%%%%%%%
\begin{figure}[tb]
\begin{center}
\includegraphics[width=50.0mm]{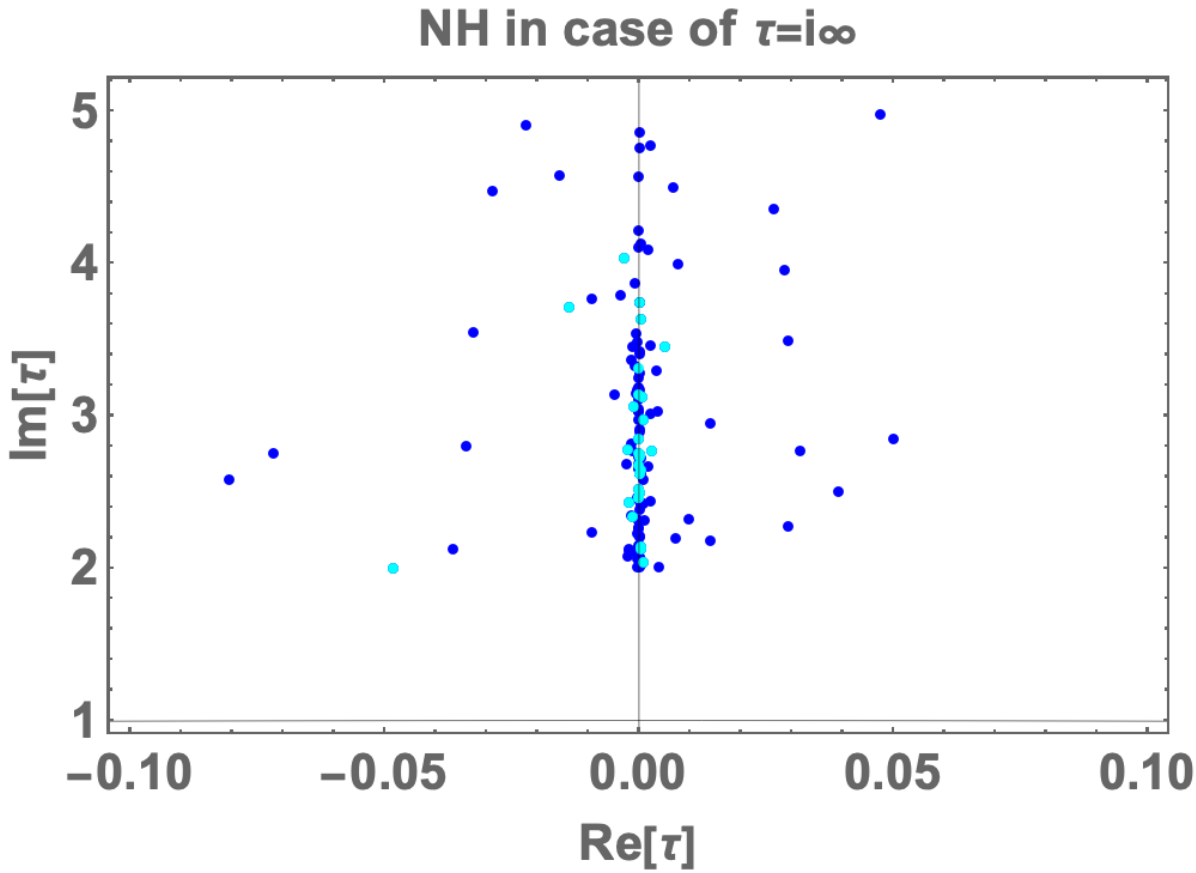} \quad
%%%
\includegraphics[width=50.0mm]{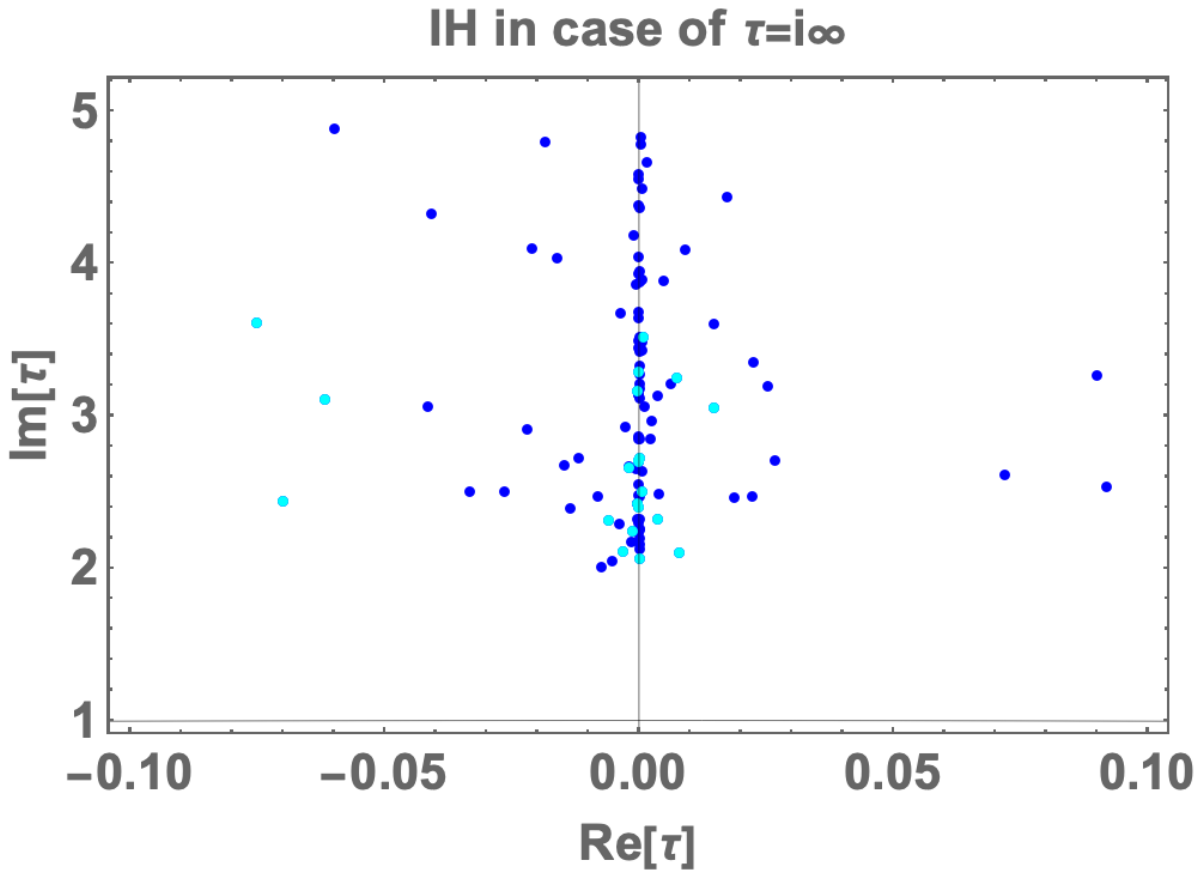} \quad
\caption{Numerical analyses at nearby $\tau=i\infty$ where the legends of figures are the same as Fig.~\ref{fig:tau=i_1}.}
  \label{fig:tau=infty_1}
\end{center}\end{figure}
%%%%%%%%%%%%%%%%%%%   

 %%%%%%%%%%%%%%%%%%%
\begin{figure}[tb]
\begin{center}
\includegraphics[width=50.0mm]{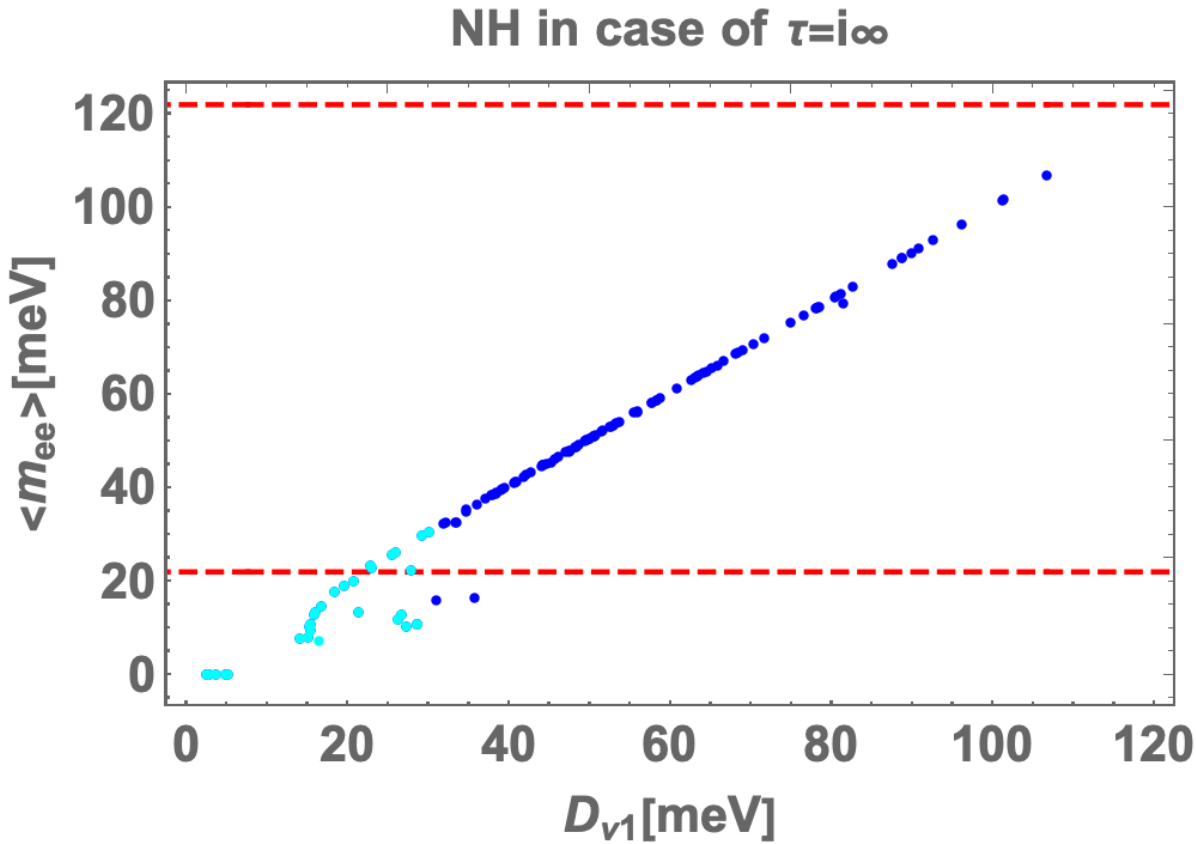} \quad
\includegraphics[width=50.0mm]{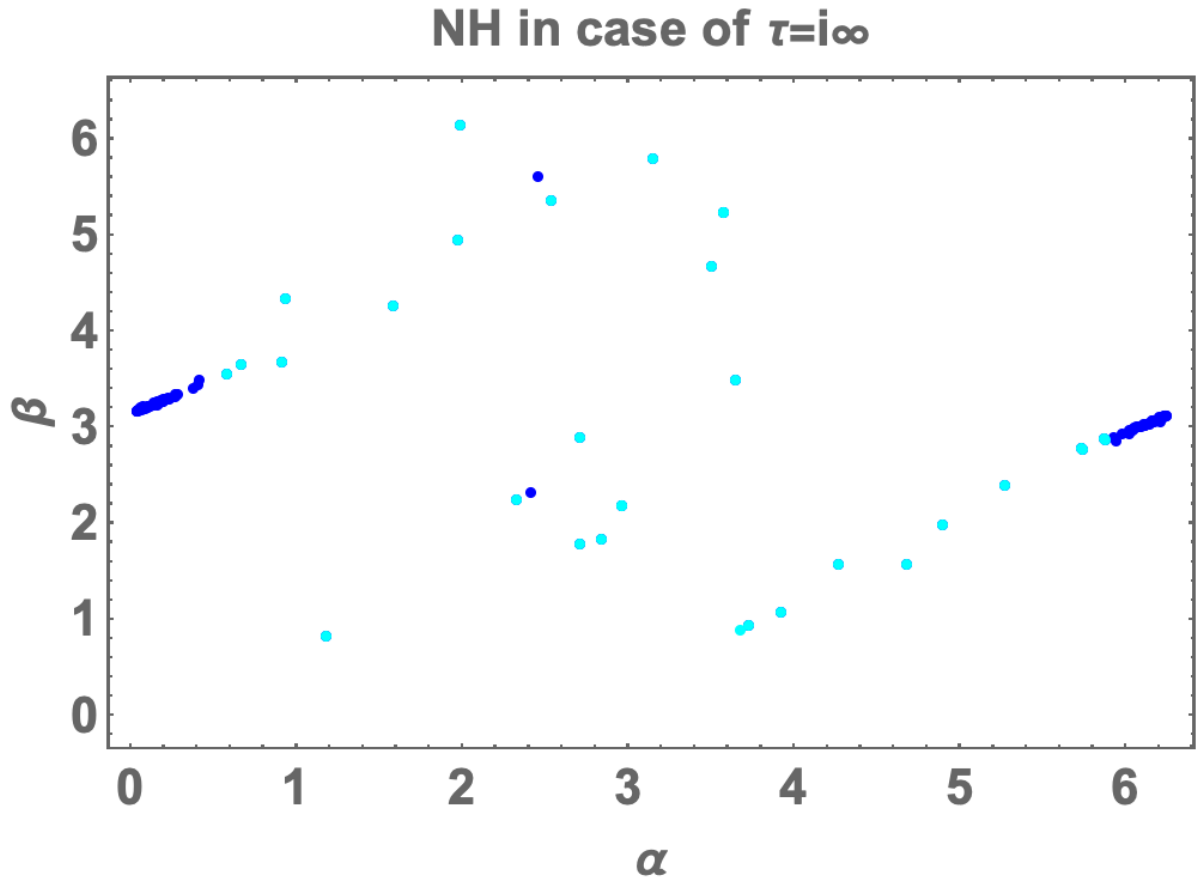}  \quad
\includegraphics[width=50.0mm]{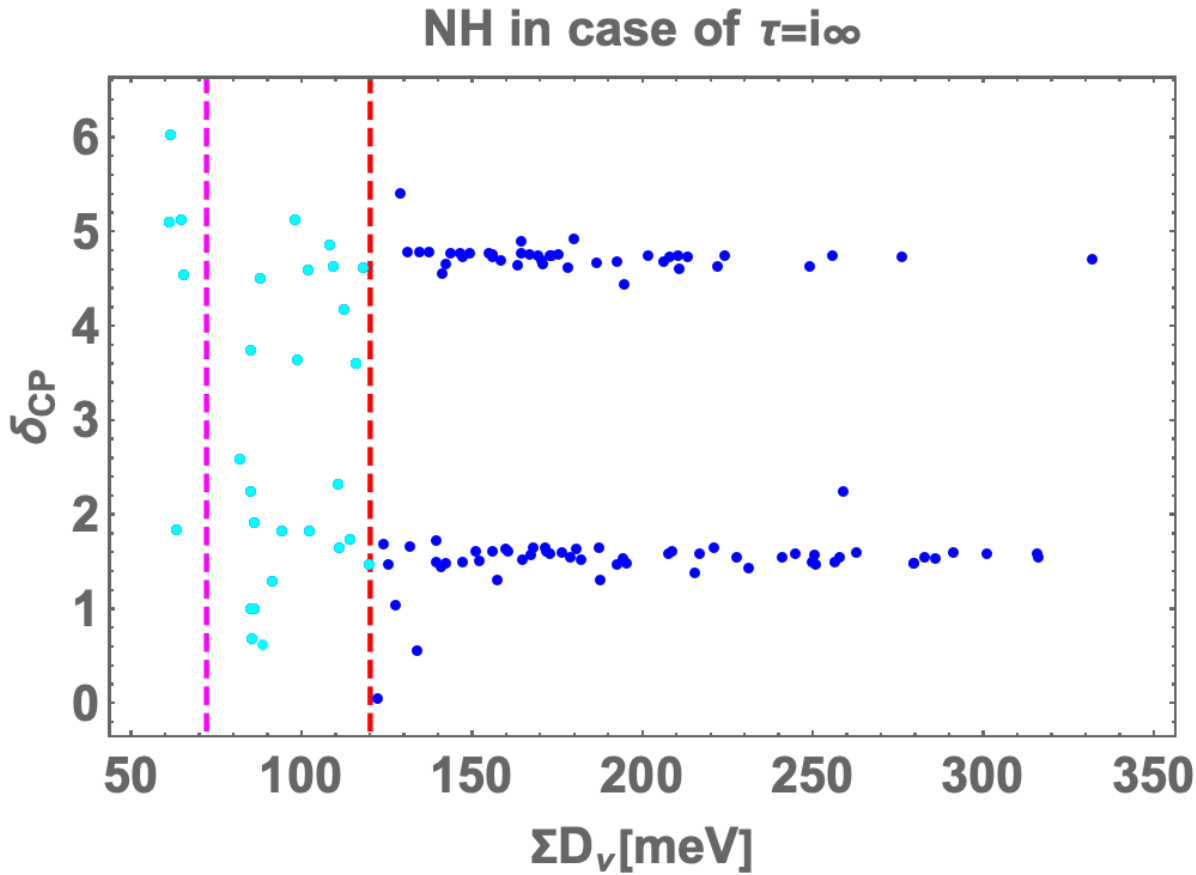} \\
%%%
\includegraphics[width=50.0mm]{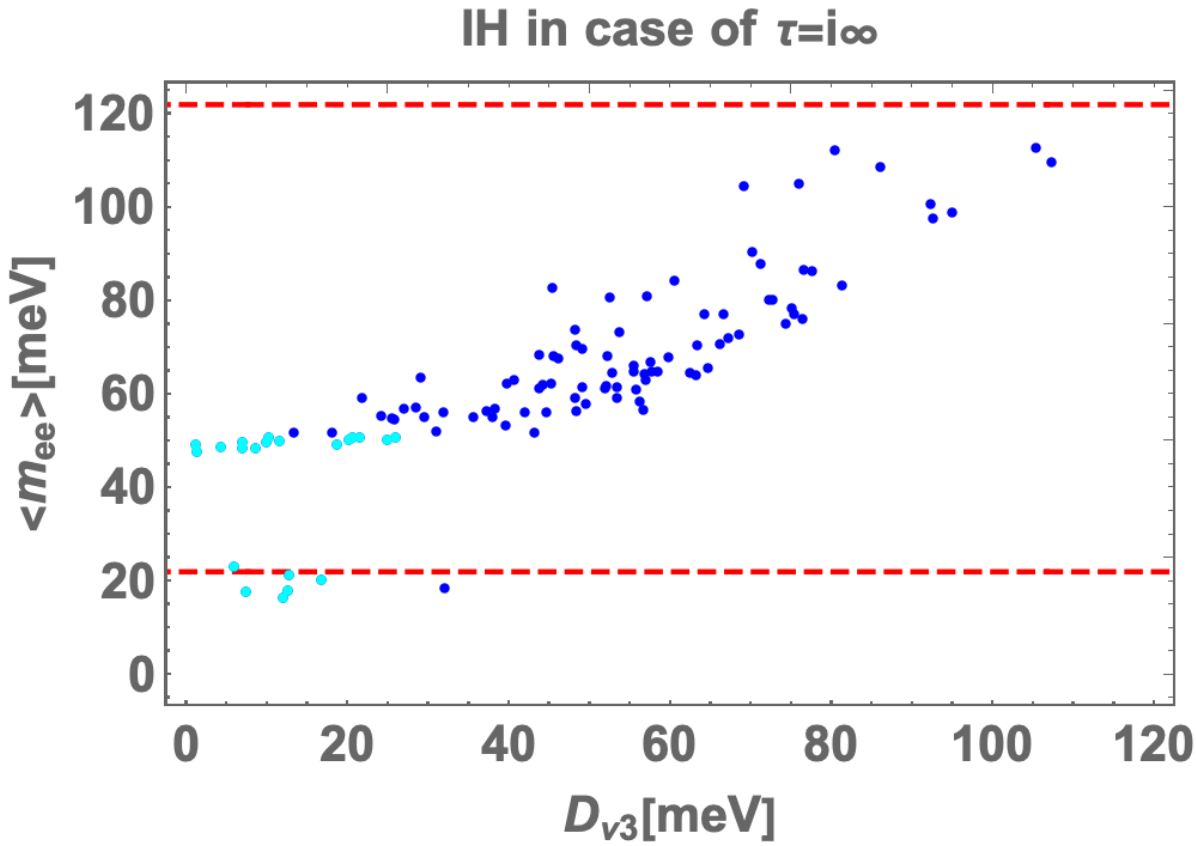} \quad
\includegraphics[width=50.0mm]{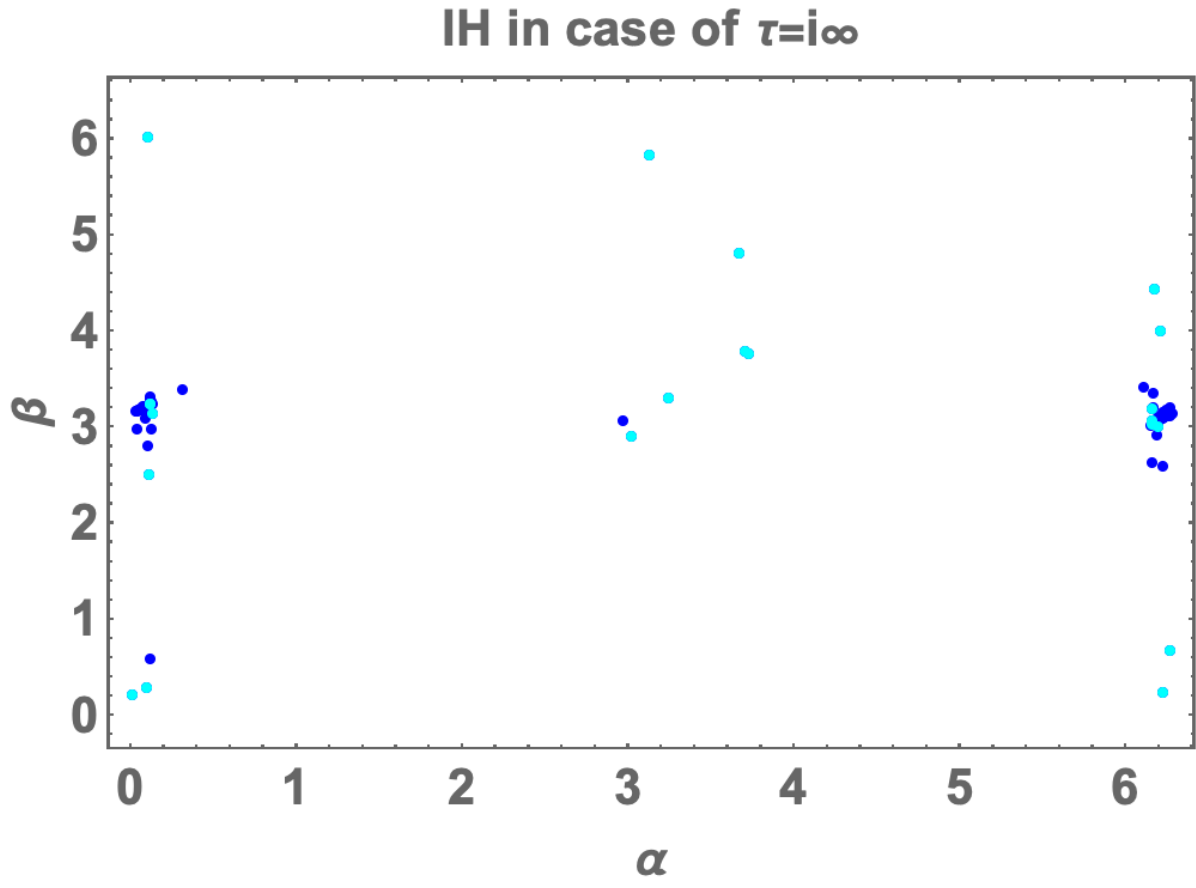}  \quad
\includegraphics[width=50.0mm]{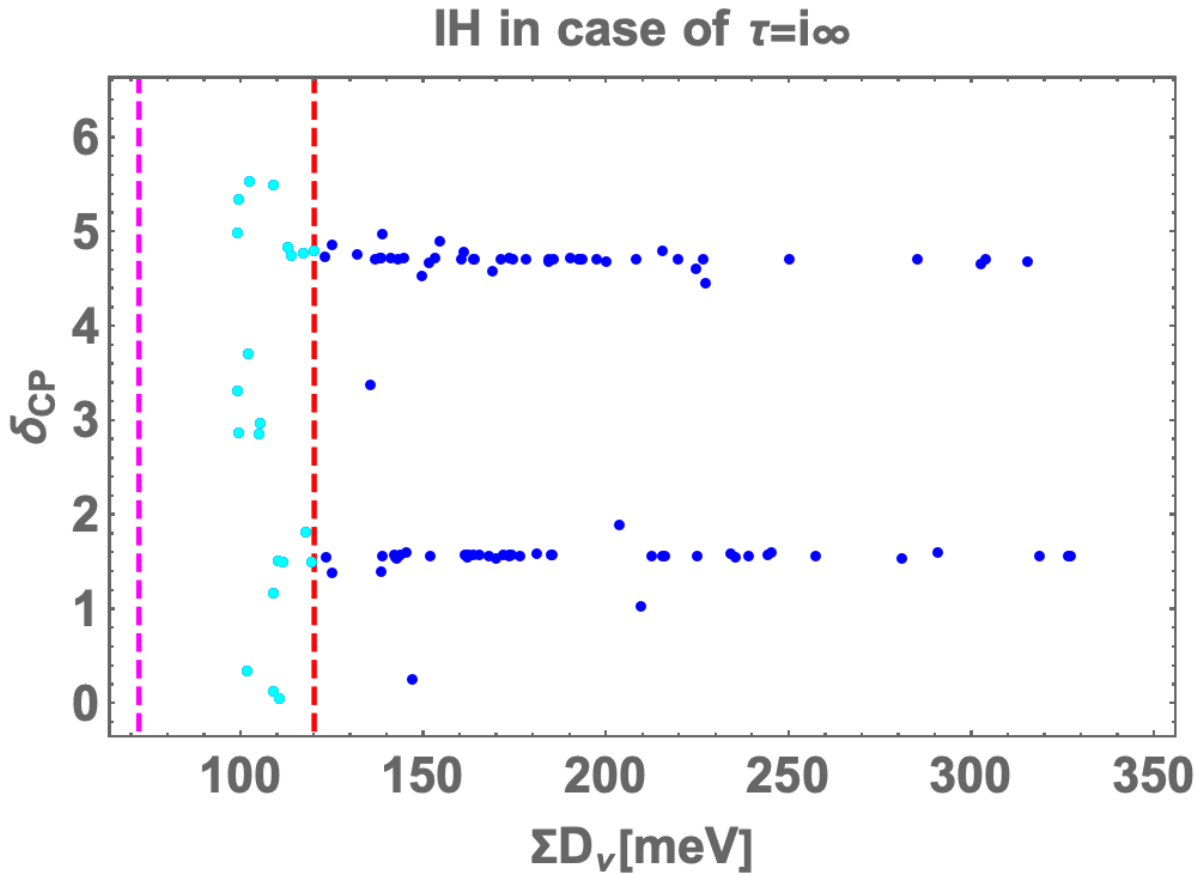} 
\caption{Numerical analyses at nearby $\tau=i\infty$.
All the legends are the same as Figs.~\ref{fig:tau=i_2}. }
  \label{fig:tau=infty_2}
\end{center}\end{figure}
%%%%%%%%%%%%%%%%%%%   
%
\subsection{$\tau=i\infty$}
In Fig.~\ref{fig:tau=infty_1}, we figure out the allowed range of $\tau$ at nearby $\tau=i\infty$
where the left figure represents the case of NH and the right one is for IH.
The NH case allows the range of Im[$\tau$]$=[2-4]$ while the IH case Im[$\tau$]$=[2-5]$.
%%%
In Fig.~\ref{fig:tau=infty_2}, we show predictions in the case of $\tau=i\infty$ where 
the up and down figures represent the NH and the IH cases, respectively. 
The left ones demonstrate neutrinoless double beta decay $\langle m_{ee}\rangle$ in terms of the lightest neutrino masses.
The dashed red horizontal lines are bounds from the current KamLAND-Zen as the Fig.~\ref{fig:tau=i_2}.
In both the cases, most of the points are within this bound and would be tested in near future. 
The center ones shows Majorana phases $\alpha$ and $\beta$ where 
the figures imply that both points would be localized at nearby $\alpha=0$ and $\beta=\pi$.
Furthermore, IH case would be more localized compared to the NH case at the other regions.
the figures imply that {both points of NH and IH cases} would be localized at nearby $\alpha=0$ and $\beta=\pi$.
{
At nearby this localized region, the smaller regions of $\langle m_{ee}\rangle$  and $\sum D_\nu$ are forbidden, therefore, $18$~meV$\lesssim \langle m_{ee}\rangle$ and $90$~meV$\lesssim \sum D_\nu$. In case of IH, the tendency is the same as the case of NH; $30$~meV$\lesssim \langle m_{ee}\rangle$ and $110$~meV$\lesssim \sum D_\nu$.
Note also that points away from the localized region can realize smaller values of $\langle m_{ee}\rangle$ and $\sum D_\nu$ as in the previous cases.
}
The right ones represent the Dirac CP phase in terms of the sum of neutrino masses where
the vertical red (magenta) dashed line shows the upper bound of $\sum m_\nu$ the same as the Fig.~\ref{fig:tau=i_2}.
It would show that NH case is better than the IH one to satisfy $\sum m_\nu < 120$ meV that is opposite behavior from the case of $\tau=\omega$.
Moreover, the only case of NH in $\tau=i\infty$ satisfies the bound of DESI and CMB $\sum D_\nu\le72$ meV.

%%%%%%%%%%%%%%%%%%%%
\section{Summary and discussion}
\label{sec:IV}
{
We have studied a model with quasi two-zero texture of neutrino mass matrix in order to satisfy the cosmological bounds on neutrino mass while keeping predictability of neutrino observables.
The quasi two-zero texture is realized by
applying a modular $A_4$ symmetry for the type-II seesaw model.
The two-zero texture of Yukawa couplings among triplet scalar and neutrino is obtained due to modular $A_4$ symmetry, and deviation from two-zero structure is originated from charged-lepton mass matrix which has non-trivial structure determined by modular $A_4$.

We have worked on three fixed points of $\tau$ that are favored by the flux compactification of type IIB string theory as well as phenomenological points of view.
We have demonstrated several predictions for each points and shown that CMB bound can be satisfied for all the fixed points while CMB+BAO bound can be satisfied only near the fixed point $\tau = i \infty$.
In addition the model would indicate specific decay patterns of doubly charged scalar boson from triplet via Yukawa coupling.
As the Yukawa coupling has the two-zero texture we expect decay of doubly charged scalar into $\mu \mu$ and $e \tau$ are suppressed.
Therefore the model could be tested in various ways such as neutrino observables, neutrinoless double beta decay, cosmological bound on sum of neutrino masses and collider physics associated with scalar boson from the triplet Higgs.
}

%%%%%%%%%%%%%%%%%%%%%%%%%%%%%%%%%%%
\section*{Acknowledgments}
%\vspace{0.3cm}
The work was supported by the Fundamental Research Funds for the Central Universities (T.~N.).
%%%%%%%%%%%%%%%%%%%%%%%%%%%%%%%%%%%

\appendix

\section{Minimal models }
\begin{table}[t!]
\begin{tabular}{|c||c|c|c|c|c|c|}\hline\hline  
& ~$\hat L$~ & ~$\hat {\overline{\ell}}$~ & ~$\hat H_{u}$~ & ~$\hat H_{d}$~ & ~$\hat \Delta_{u}$~ & ~$\hat \Delta_{d}$~  \\\hline\hline 
%%%
$SU(2)_L$   & $\bm{2}$  & $\bm{1}$  & $\bm{2}$ & $\bm{2}$ & $\bm{3}$  & $\bm{3}$     \\\hline 
$U(1)_Y$    & $-\frac12$  & $+1$ & $+\frac12$ & $-\frac12$ & $+1$  & $-1$  \\\hline
$A_4$   & $[\bm{1}]$  & $ \bm{3}$  & $\bm{1}$ & $\bm{1}$ & $\bm{1}$  & $\bm{1}$        \\\hline 
$-k_I$    & $-2$  & $0$ & $0$ & $0$ & $0$ & $0$     \\\hline
\end{tabular}
\caption{Charge assignments of the SM lepton and new superfields
under $SU(2)_L\otimes U(1)_Y \otimes A_4$ where $-k_I$ is the number of modular weight. Here $[\bm{1}]$ indicates assignment of $A_4$ singlets that depends on model in the main text of Appendix A.    }\label{tab:min1}
\end{table}

%%%%%%%%%%%%%%%%%%%%%%%%%%%%%%%%%%%%%%%%%%%%%%%%%%%%%%%%%%%
%\subsection{Minimu models }
%%%%%%%%%%%%%%%%%%%%%%%%%%%%%%%%%%%%%%%%%%%%%%%%%%%%%%%%%%%%%%%%% 

Here, we show neutrino observables in minimum scenarios in our lepton models.
The charge assignments are summarized in Tab.~\ref{tab:min1} where $[\bm{1}]$ indicates assignment of $A_4$ singlets that depends on model that we discuss later.
We have obtained $B_1$-like type and $B_2$-like type by appropriately assigning the singlet $A_4$ charges of left-handed leptons,
but $B_2$ does not have any solutions in case of IH.
Moreover, only the NH case of $B'_2$ has one point that allows the cosmological bound.

\subsection{$B_1$-like model}
$B_1$-like model is realized by assigning $[\bm{1}]=\{1,1'',1'\}$.
Similar way to the case of main text, the charged-lepton mass matrix is given by 
\begin{align}
m_\ell = \frac{v_d}{\sqrt2}
 \left(\begin{array}{ccc} f_1 & f_2 & f_3 \\
 f_3  & f_1 & f_2 \\
 f_2 & f_3 & f_1 \end{array} \right)
  %%%
   \left(\begin{array}{ccc} a_e & 0 & 0 \\
0 & b_e & 0 \\
0 & 0 & c_e \end{array} \right),
\label{eq:cgd-lep-b1}
\end{align}
where $(f_1,f_2,f_3)$ is $A_4$ triplet with modular $+2$ that is defined as $(Y_1(\tau),Y_2(\tau),Y_3(\tau))$ in Appendix B.
The neutrino mass matrix is given by
\begin{align}
m_\nu = \frac{v_{\Delta_u}}{\sqrt2}
%\frac{v_{\Delta_u} |d_\nu|}{\sqrt2}
 \left(\begin{array}{ccc} a_\nu & b_\nu  & 0 \\
b_\nu  & 0 & c_\nu  \\
0 & c_\nu  &d_\nu \end{array} \right),
\label{eq:cgd-lep2-b1}
\end{align}
where $a_\nu,b_\nu,c_\nu, d_\nu$ implicitly include $Y_1^{(4)}$ or $Y_{1'}^{(4)}$. 
%$a_\nu,b_\nu,c_\nu, d_\nu$ respectively include $Y_1^{(4)},Y_{1'}^{(4)},Y_{1}^{(4)},Y_{1'}^{(4)}$. 

 %%%%%%%%%%%%%%%%%%%
\begin{figure}[tb]
\begin{center}
\includegraphics[width=50.0mm]{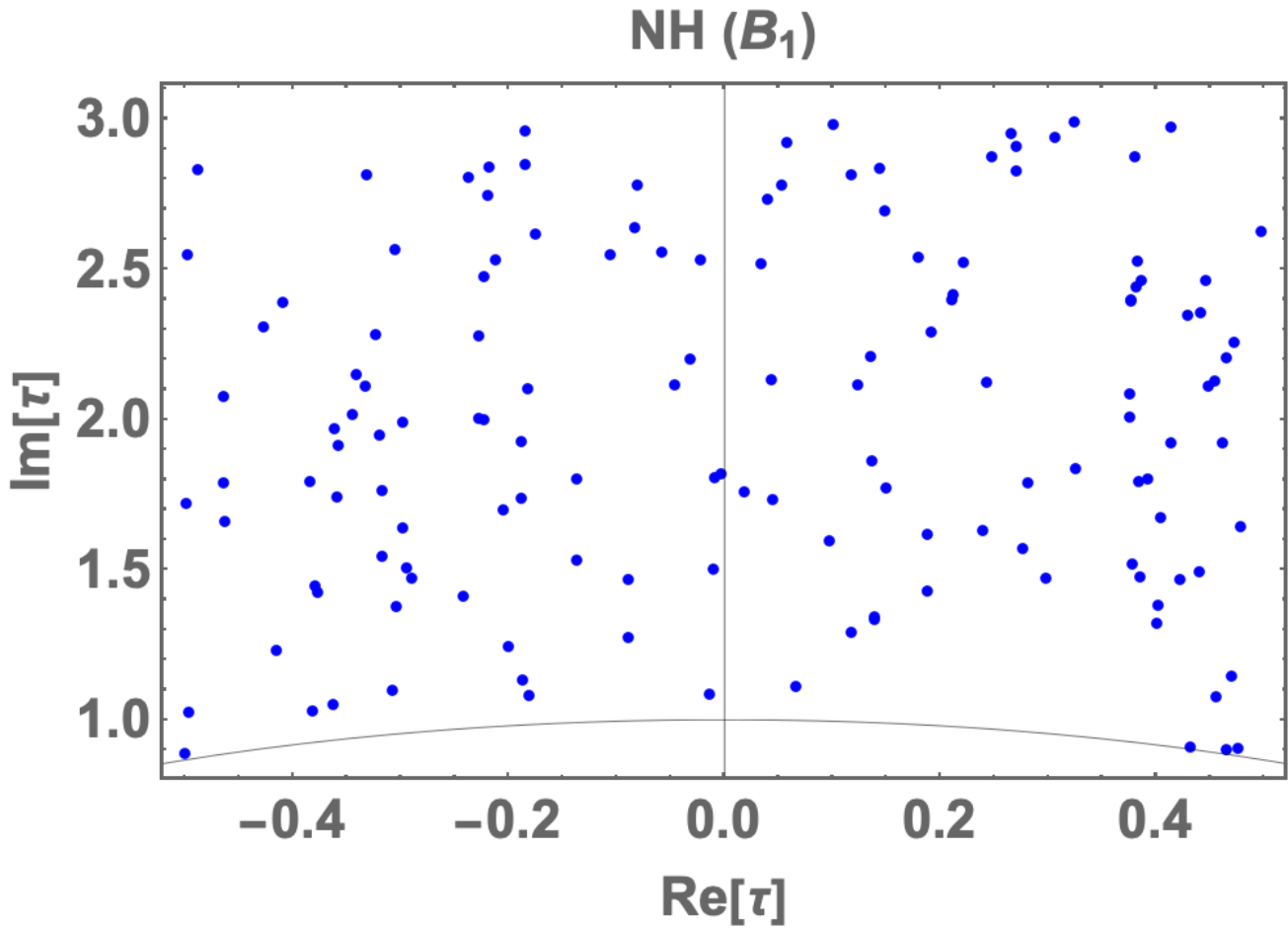} \quad
%%%
\includegraphics[width=50.0mm]{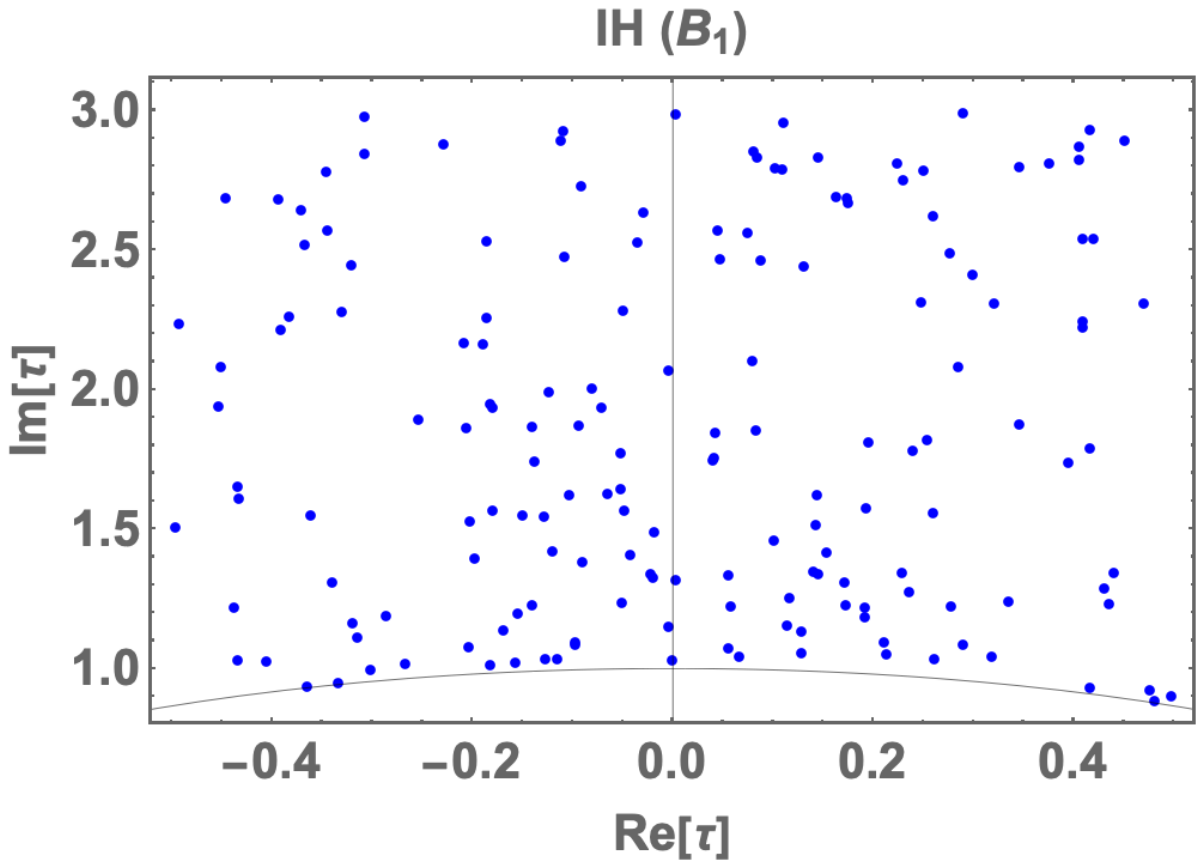} \quad
\caption{Numerical analysis of $\tau$ in case of $B_1$-like model, where the legends of figures are the same as Fig.~\ref{fig:tau=i_1}.}
  \label{fig:b1_1}
\end{center}\end{figure}
%%%%%%%%%%%%%%%%%%%   
%
In Fig.~\ref{fig:b1_1}, we figure out the allowed range of $\tau$
where the left figure represents the case of NH and the right one is for IH.
They tell us that all ranges are allowed in the fundamental region up to ${\rm Im}[\tau]=3$.

 %%%%%%%%%%%%%%%%%%%
\begin{figure}[tb]
\begin{center}
\includegraphics[width=50.0mm]{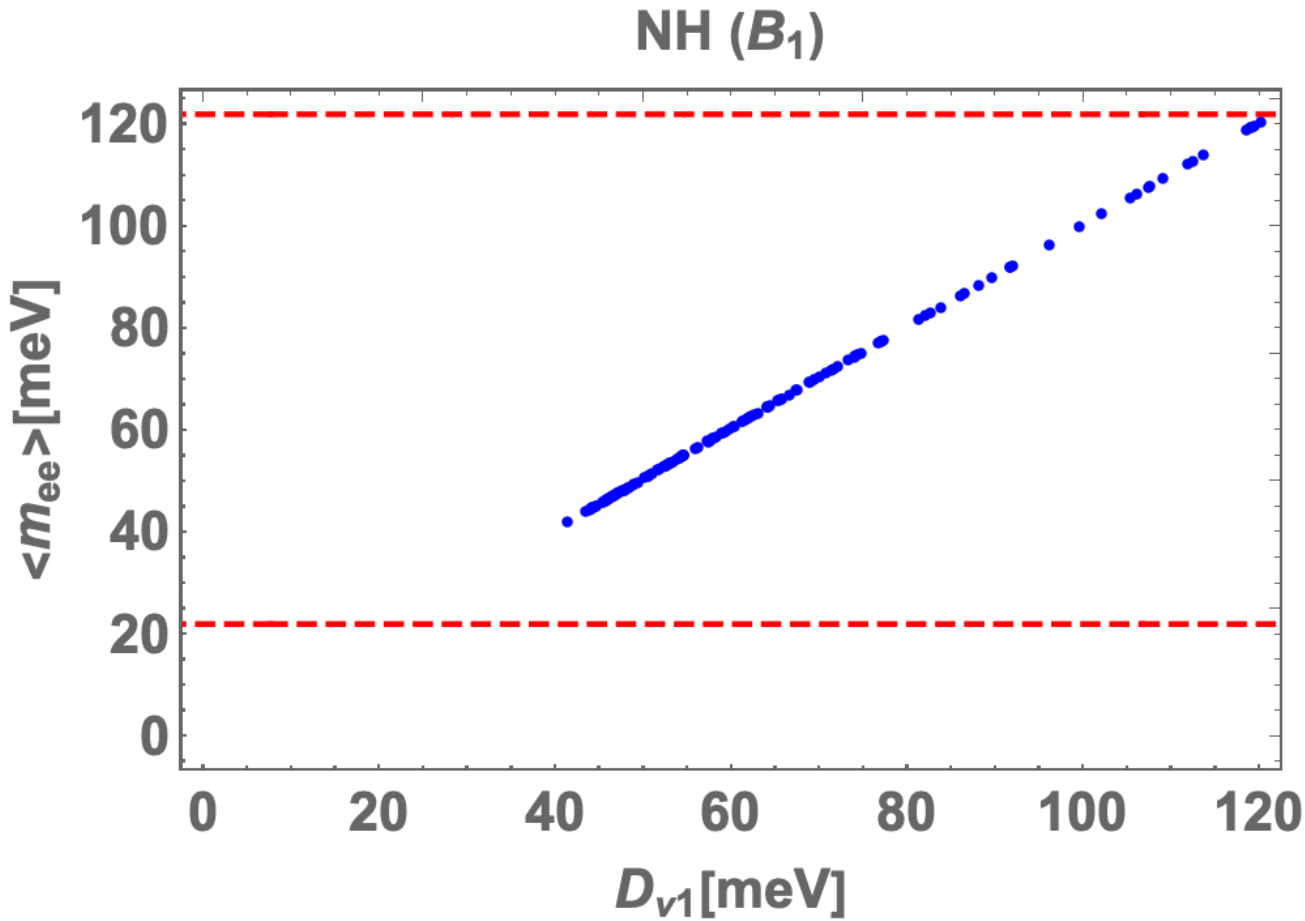} \quad
\includegraphics[width=50.0mm]{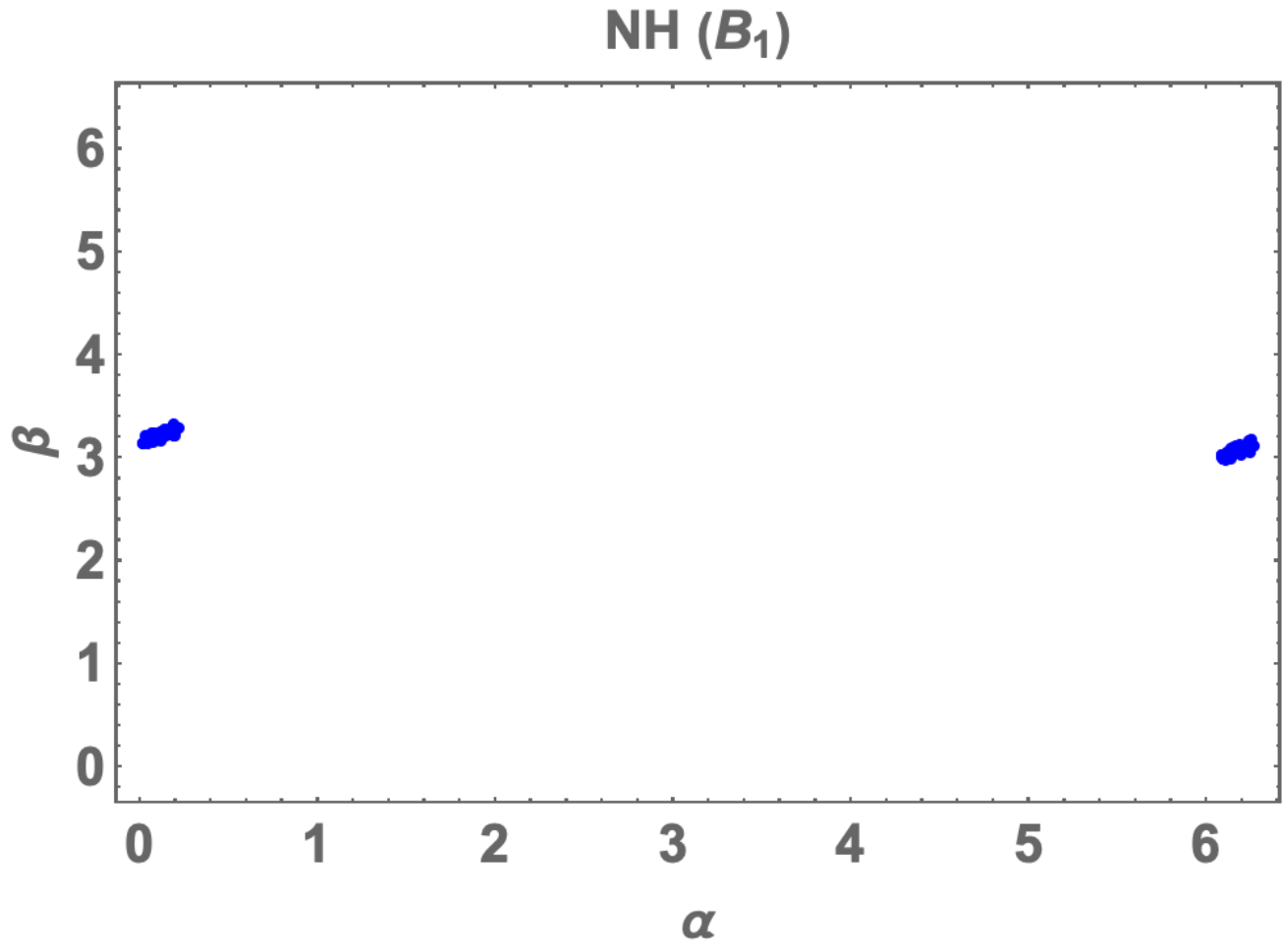}  \quad
\includegraphics[width=50.0mm]{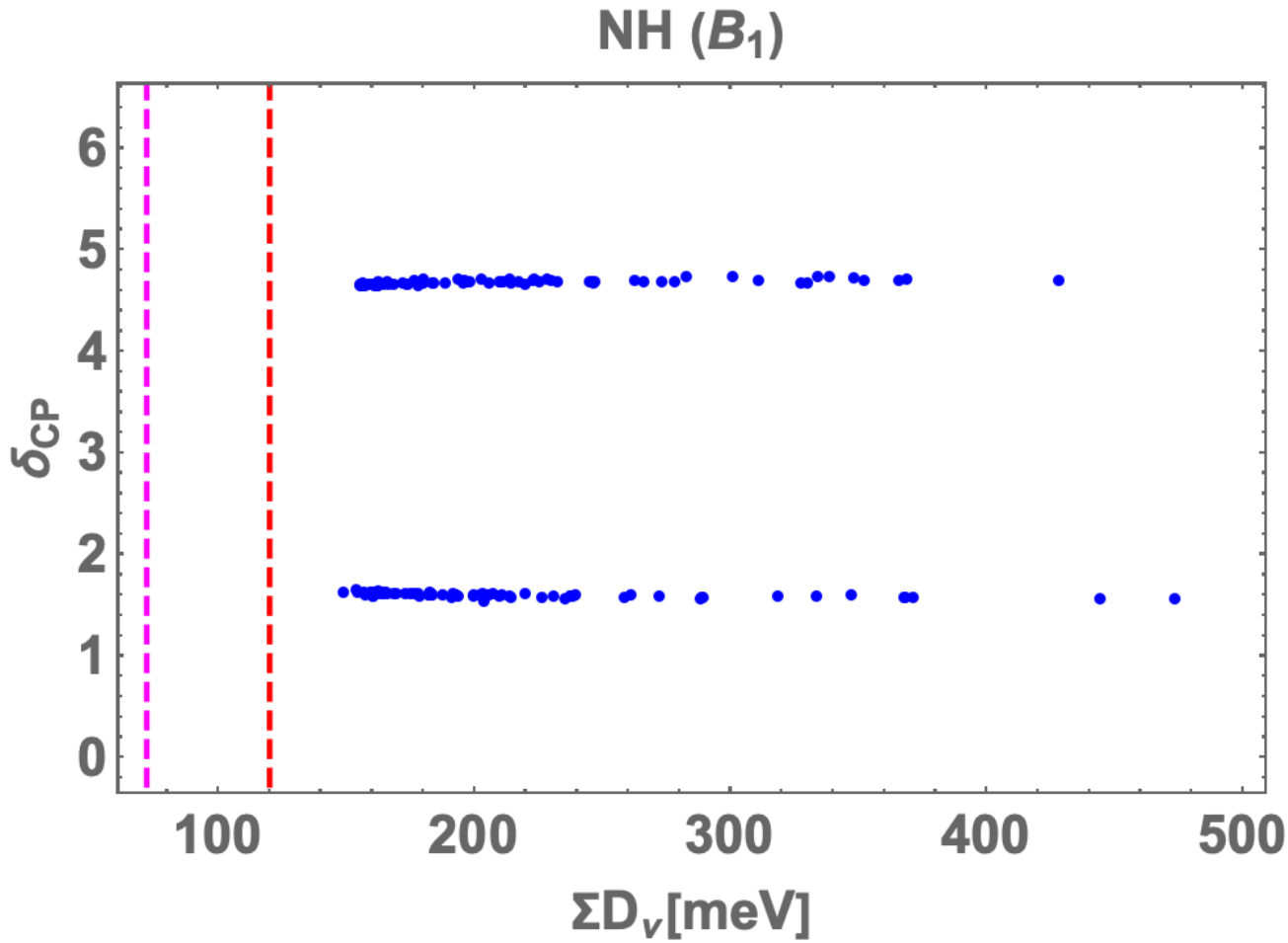} \\
%%%
\includegraphics[width=50.0mm]{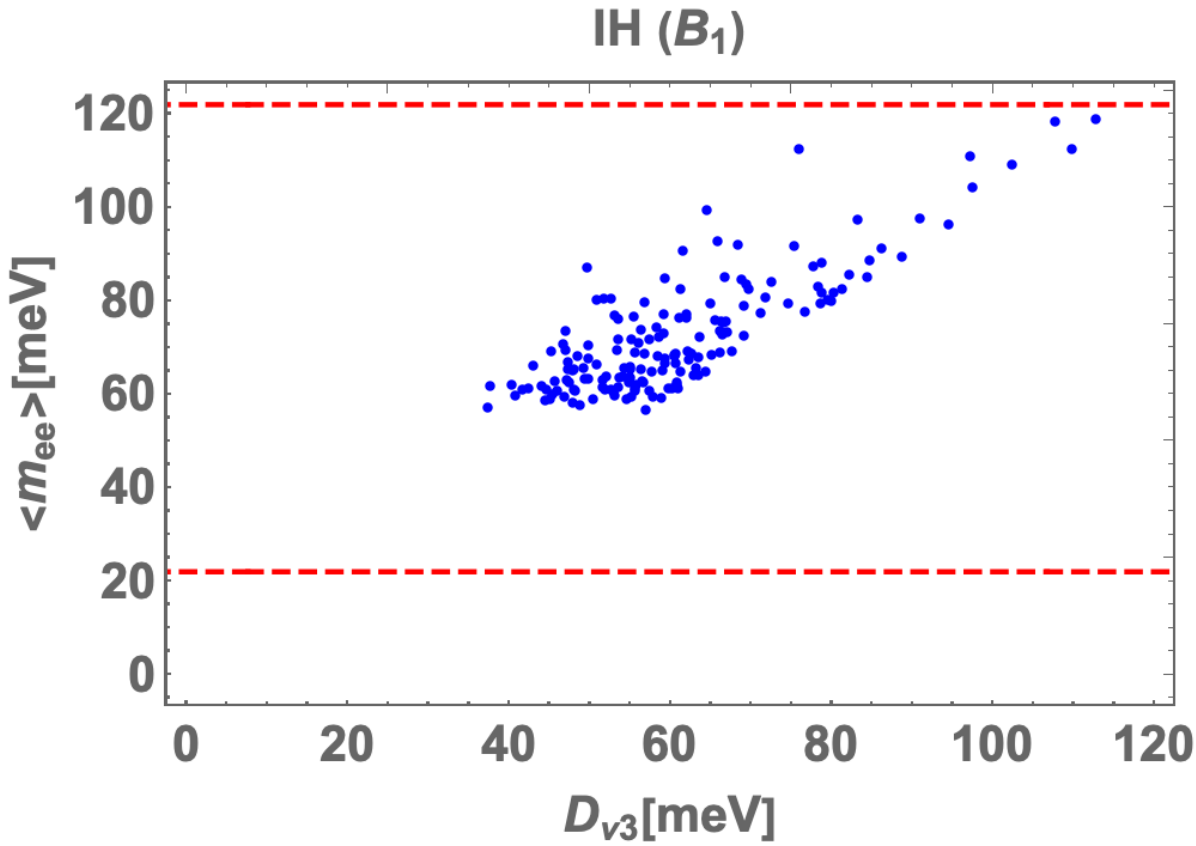} \quad
\includegraphics[width=50.0mm]{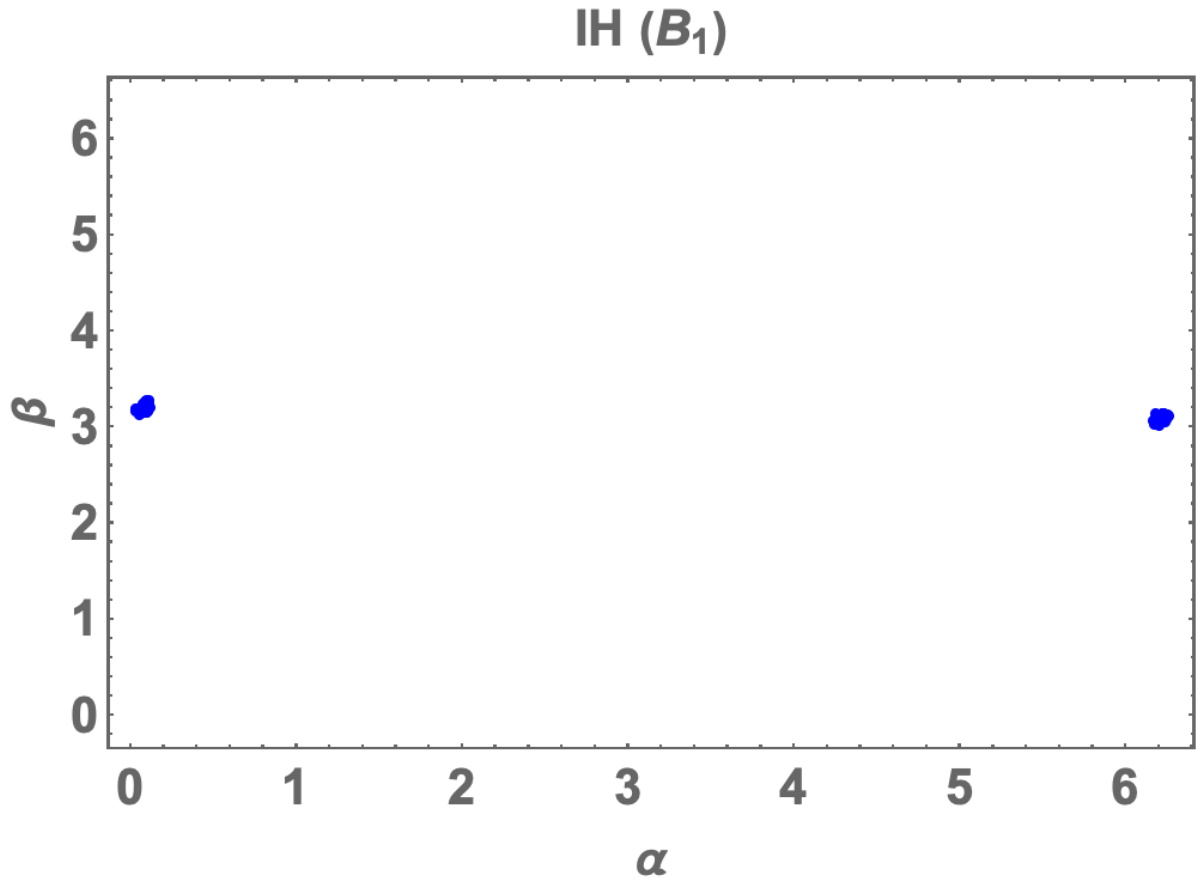}  \quad
\includegraphics[width=50.0mm]{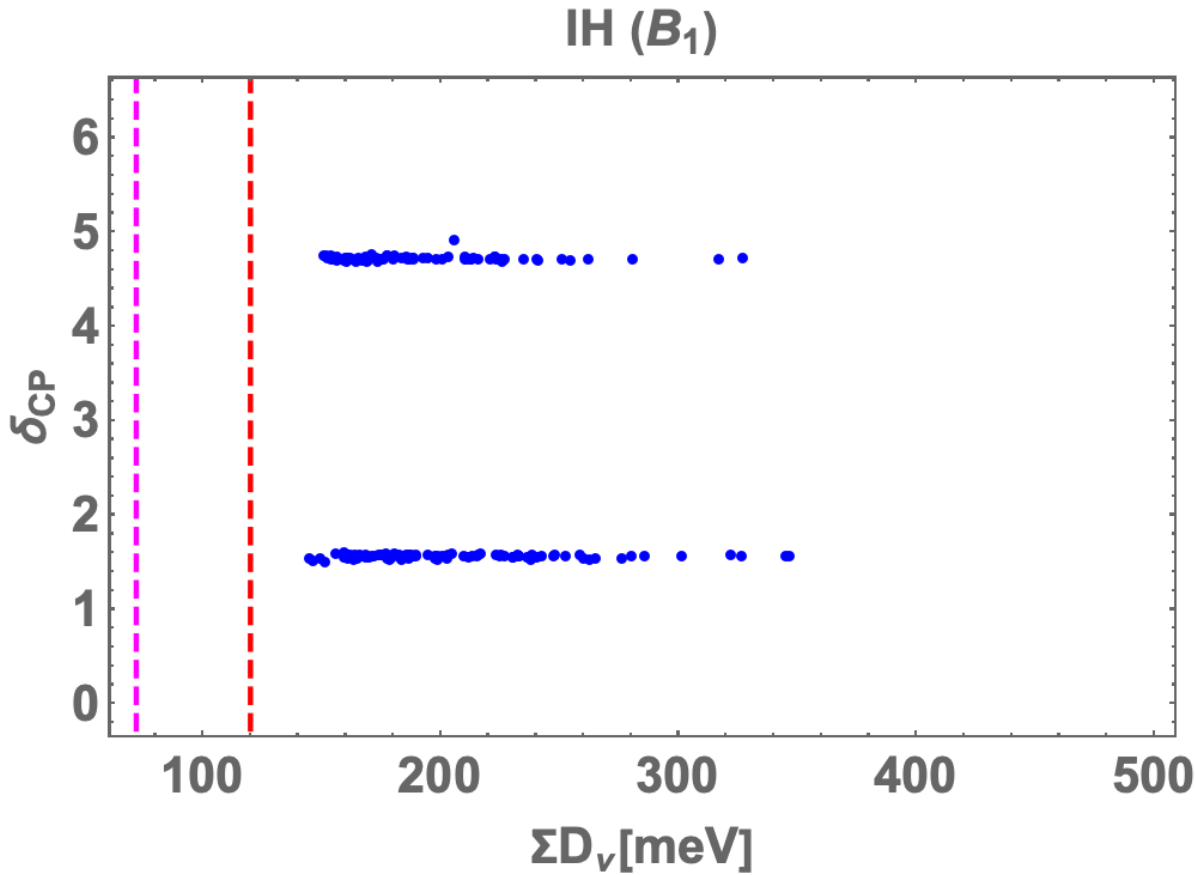} 
\caption{Numerical analyses in case of $B_1$.
All the legends are the same as Figs.~\ref{fig:tau=i_2}. }
  \label{fig:b1_2}
\end{center}\end{figure}
%%%%%%%%%%%%%%%%%%%   

%%%
In Fig.~\ref{fig:b1_2}, we show predictions in the case of $B_1$ where 
the up and down figures represent the NH and the IH cases, respectively. 
The left ones demonstrate neutrinoless double beta decay $\langle m_{ee}\rangle$ in terms of the lightest neutrino masses.
The dashed red horizontal lines are bounds from the current KamLAND-Zen as the Fig.~\ref{fig:tau=i_2}.
The allowed regions are given by $40(40)~{\rm meV}\lesssim D_{\nu_1}(\langle m_{ee}\rangle )\lesssim 120(120)~{\rm meV}$ for NH
and  $35(60)~{\rm meV}\lesssim D_{\nu_3}(\langle m_{ee}\rangle )\lesssim 110(120)~{\rm meV}$ for IH.
The center ones shows Majorana phases $\alpha$ and $\beta$ where 
the figures imply that both cases are localized at nearby $\alpha\simeq 0$ and $\beta\simeq \pi$.
The right ones represent the Dirac CP phase in terms of the sum of neutrino masses where
the vertical red (magenta) dashed line shows the upper bound of $\sum m_\nu$ the same as the Fig.~\ref{fig:tau=i_2}.
The allowed regions are given by $140(150)~{\rm meV}\lesssim  \sum D_\nu \lesssim 480(350)~{\rm meV}$ for NH(IH)
and  $\delta_{\rm CP}\simeq \pi/2, 3\pi/2$ for both cases.

\subsection{$B'_1$-like model}
$B'_1$-like model is realized by assigning $[\bm{1}]=\{1',1'',1\}$.
Similar way to the case of main text, the charged-lepton mass matrix is given by 
\begin{align}
m_\ell = \frac{v_d}{\sqrt2}
 \left(\begin{array}{ccc} f_1 & f_2 & f_3 \\
 f_3  & f_1 & f_2 \\
 f_2 & f_3 & f_1 \end{array} \right)
   %%%
   \left(\begin{array}{ccc} 0 & 0 & 1 \\
0 & 1 & 0 \\
1 & 0 & 0 \end{array} \right)
  %%%
   \left(\begin{array}{ccc} a_e & 0 & 0 \\
0 & b_e & 0 \\
0 & 0 & c_e \end{array} \right),
\label{eq:cgd-lep-b1p}
\end{align}
where $(f_1,f_2,f_3)$ is $A_4$ triplet with modular $+2$ that is defined as $(Y_1(\tau),Y_2(\tau),Y_3(\tau))$ in Appendix B.
The neutrino mass matrix is given by
\begin{align}
m_\nu = \frac{v_{\Delta_u}}{\sqrt2}
%\frac{v_{\Delta_u} |d_\nu|}{\sqrt2}
 \left(\begin{array}{ccc} a_\nu & b_\nu  & 0 \\
b_\nu  & 0 & c_\nu  \\
0 & c_\nu  &d_\nu \end{array} \right),
\label{eq:cgd-lep2-b1p}
\end{align}
where $a_\nu,b_\nu,c_\nu, d_\nu$ implicitly include $Y_1^{(4)}$ or $Y_{1'}^{(4)}$. 
%$a_\nu,b_\nu,c_\nu, d_\nu$ respectively include $Y_1^{(4)},Y_{1'}^{(4)},Y_{1}^{(4)},Y_{1'}^{(4)}$. 

 %%%%%%%%%%%%%%%%%%%
\begin{figure}[tb]
\begin{center}
\includegraphics[width=50.0mm]{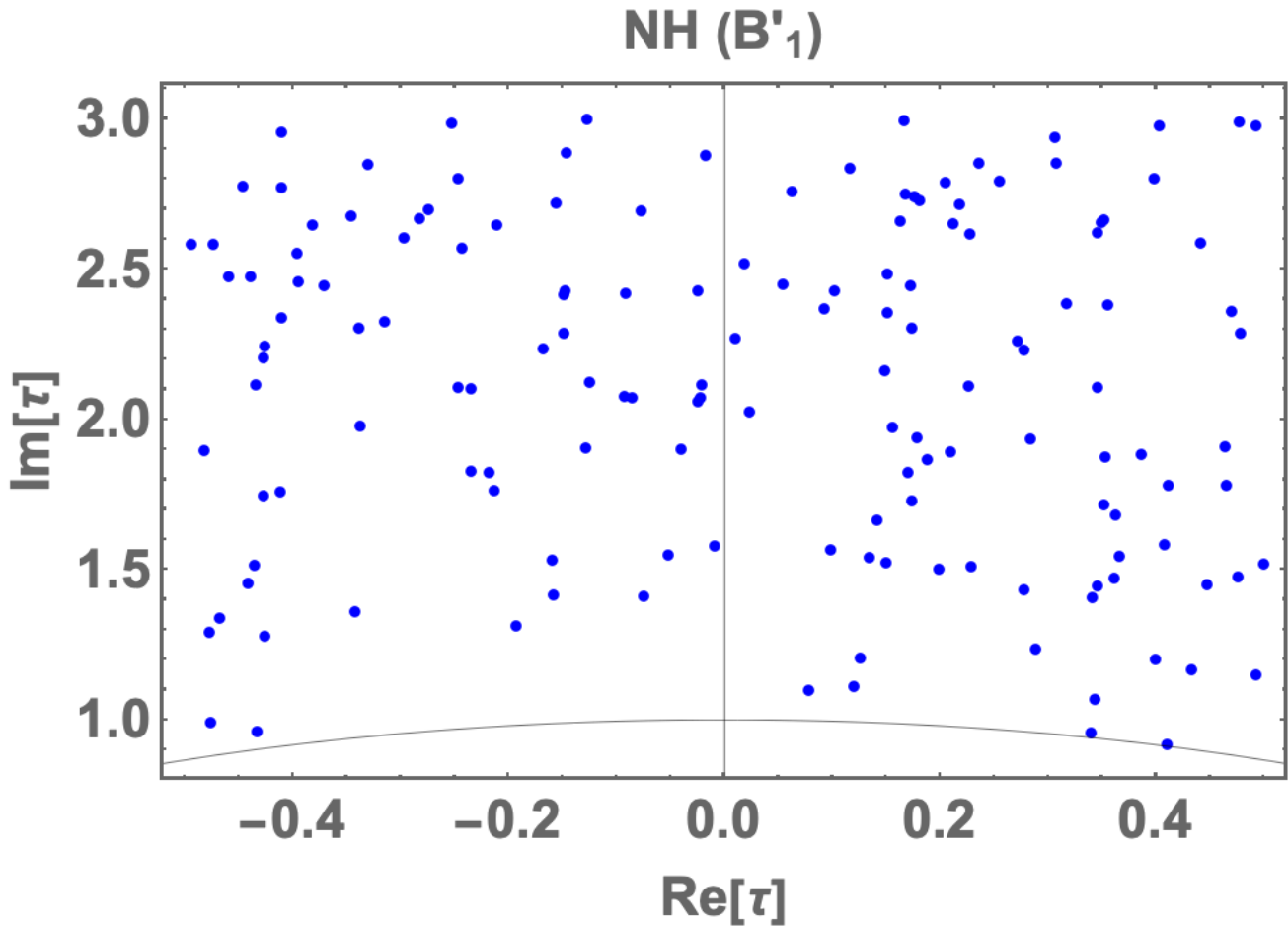} \quad
%%%
\includegraphics[width=50.0mm]{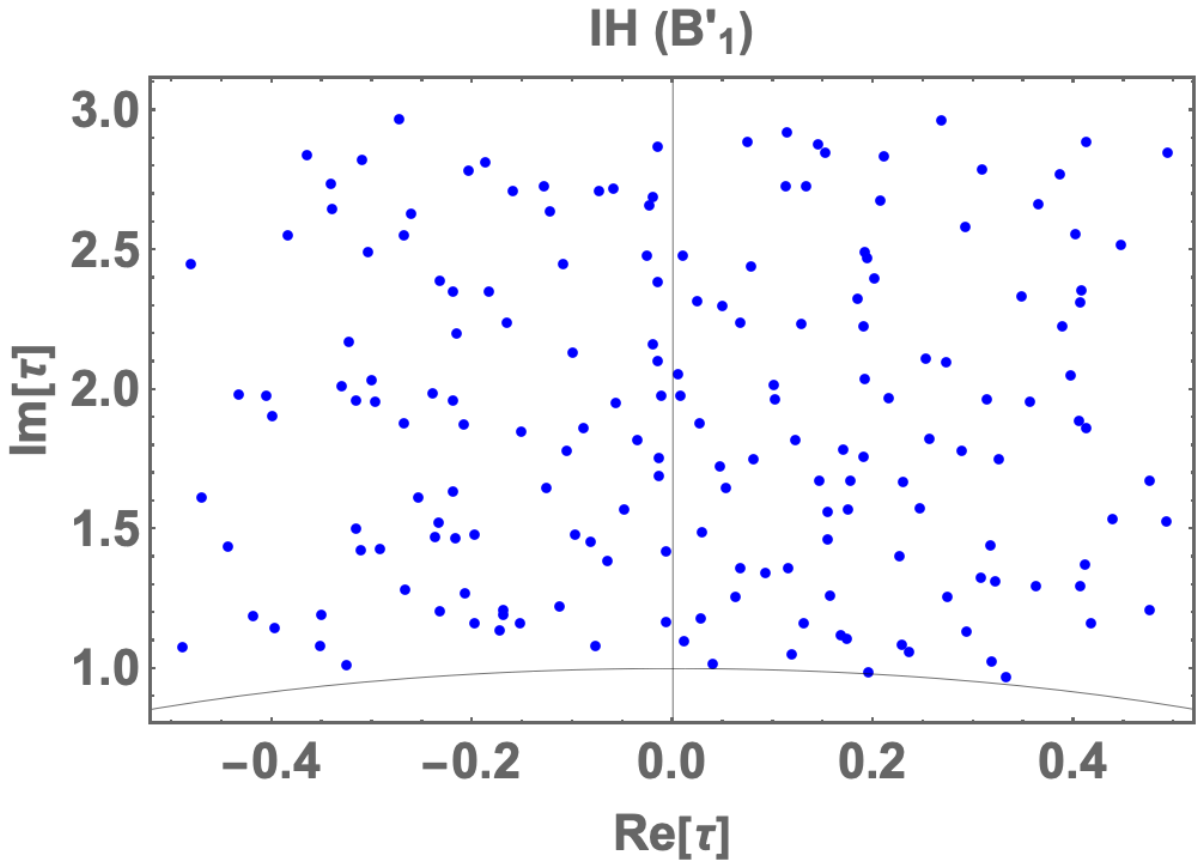} \quad
\caption{Numerical analysis of $\tau$ in case of $B'_1$-like model, where the legends of figures are the same as Fig.~\ref{fig:tau=i_1}.}
  \label{fig:b1p_1}
\end{center}\end{figure}
%%%%%%%%%%%%%%%%%%%   
%
In Fig.~\ref{fig:b1p_1}, we figure out the allowed range of $\tau$
where the left figure represents the case of NH and the right one is for IH.
They tell us that all ranges are allowed in the fundamental region up to ${\rm Im}[\tau]=3$.

 %%%%%%%%%%%%%%%%%%%
\begin{figure}[tb]
\begin{center}
\includegraphics[width=50.0mm]{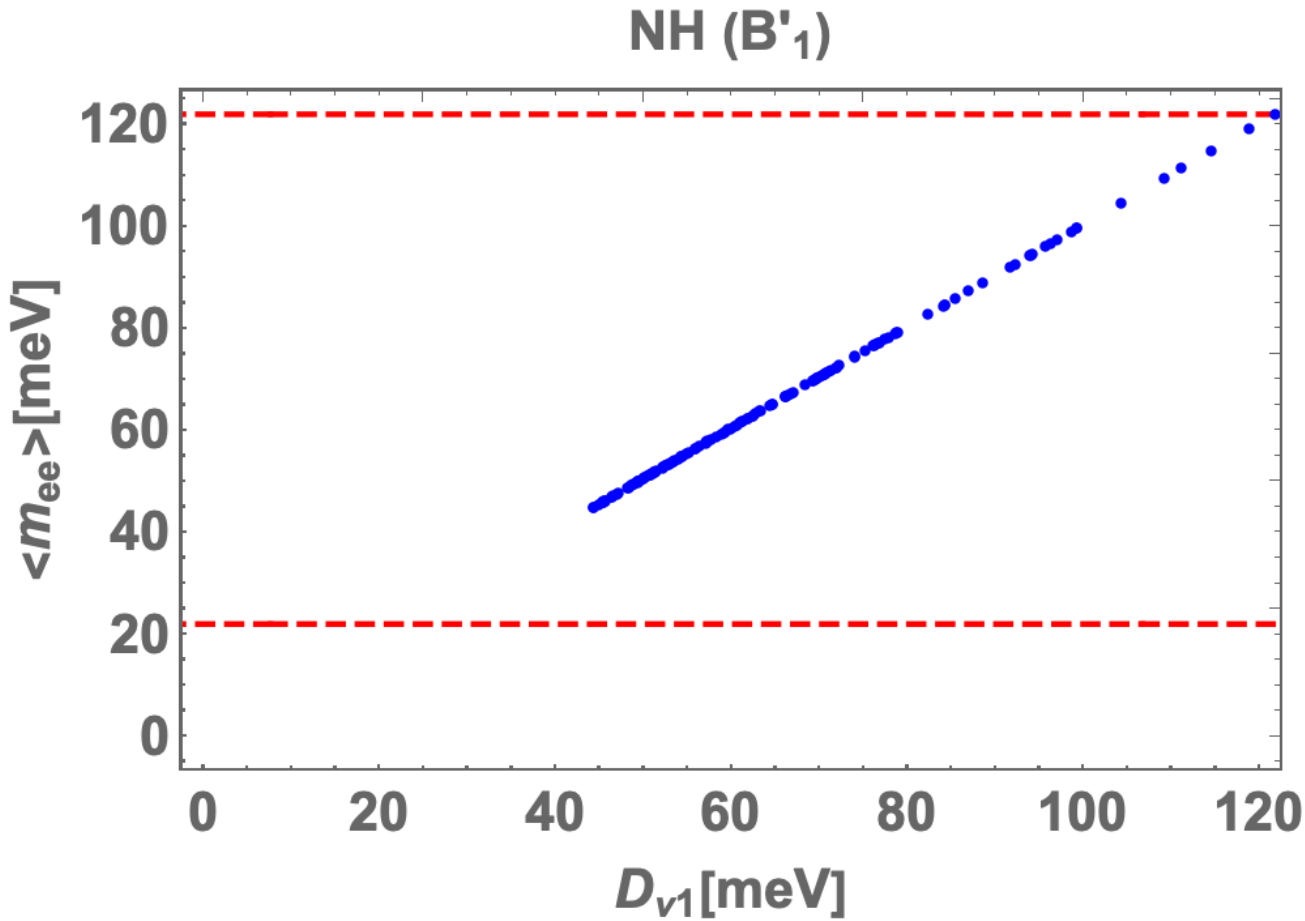} \quad
\includegraphics[width=50.0mm]{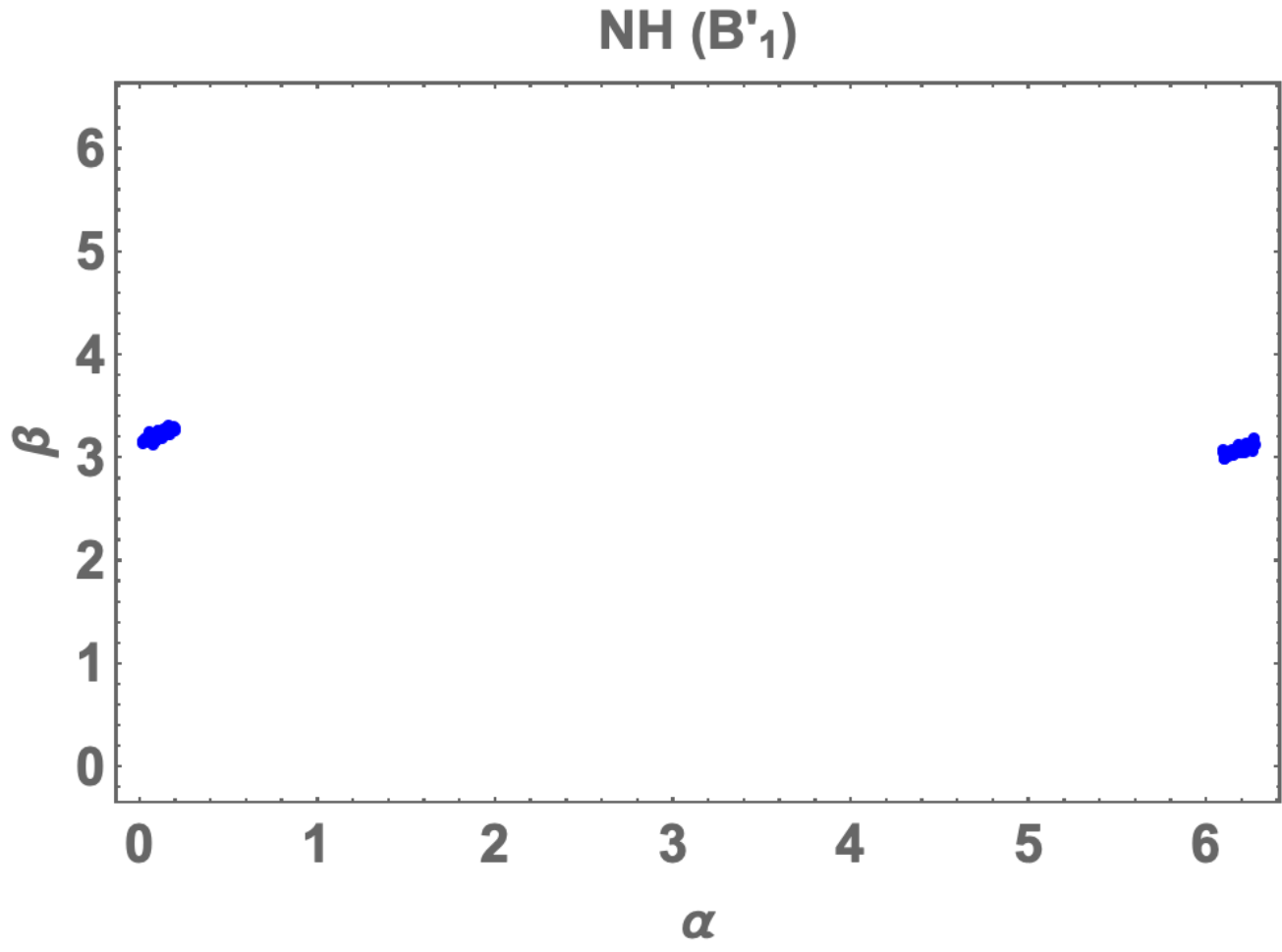}  \quad
\includegraphics[width=50.0mm]{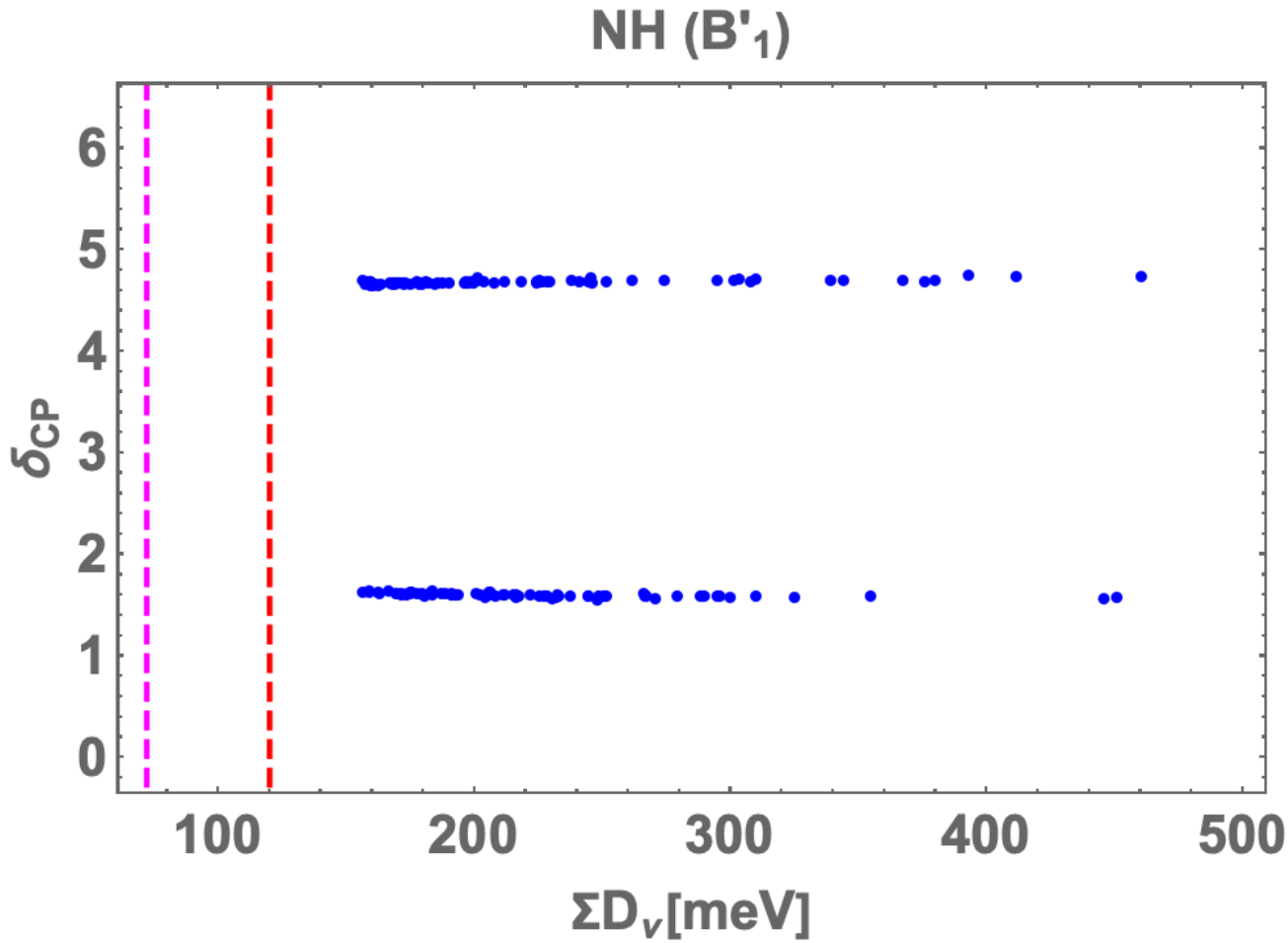} \\
%%%
\includegraphics[width=50.0mm]{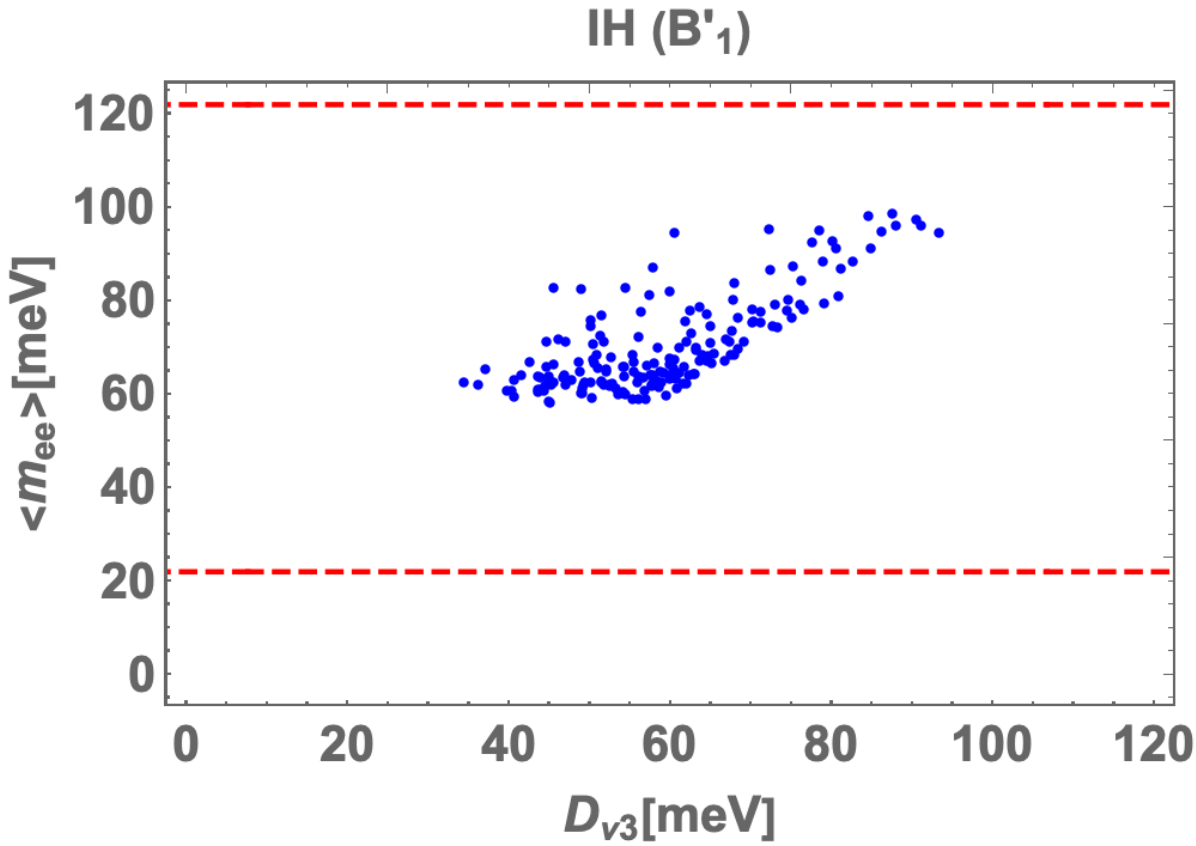} \quad
\includegraphics[width=50.0mm]{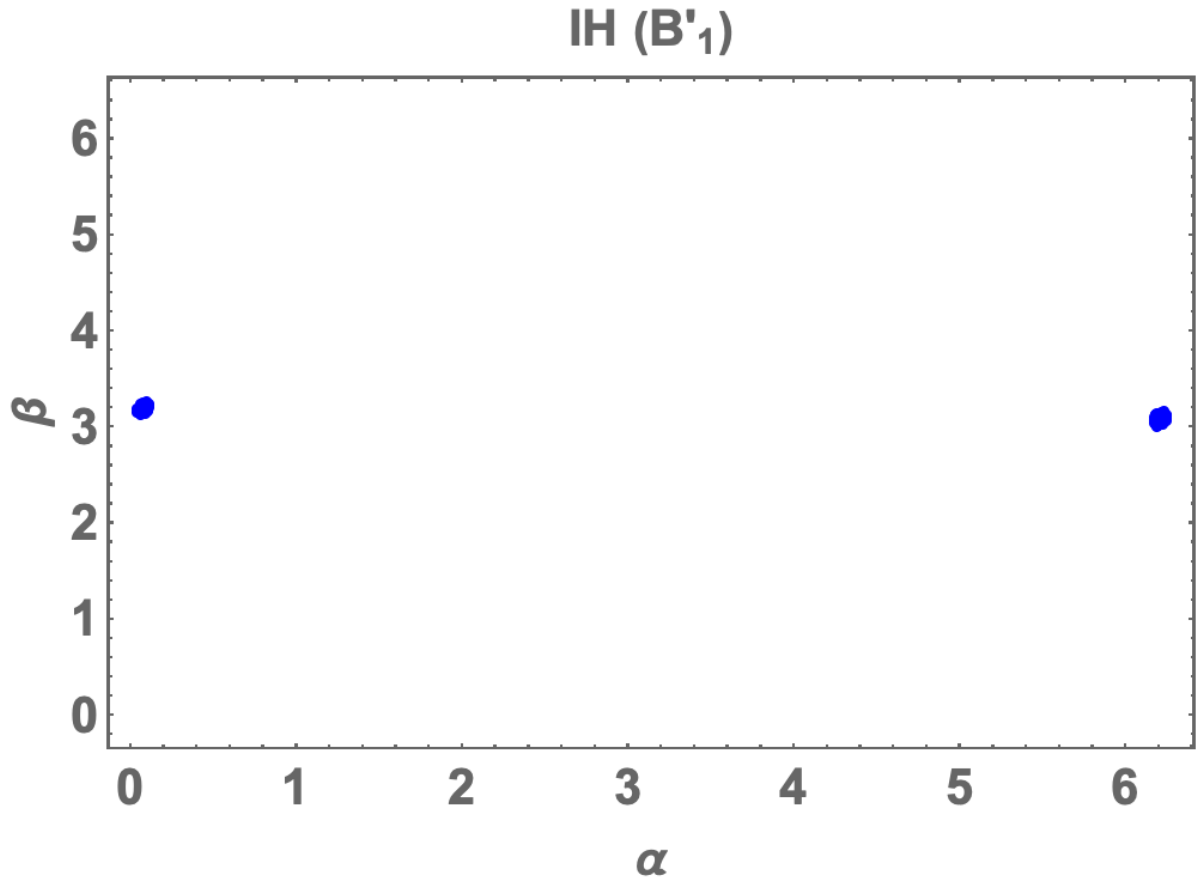}  \quad
\includegraphics[width=50.0mm]{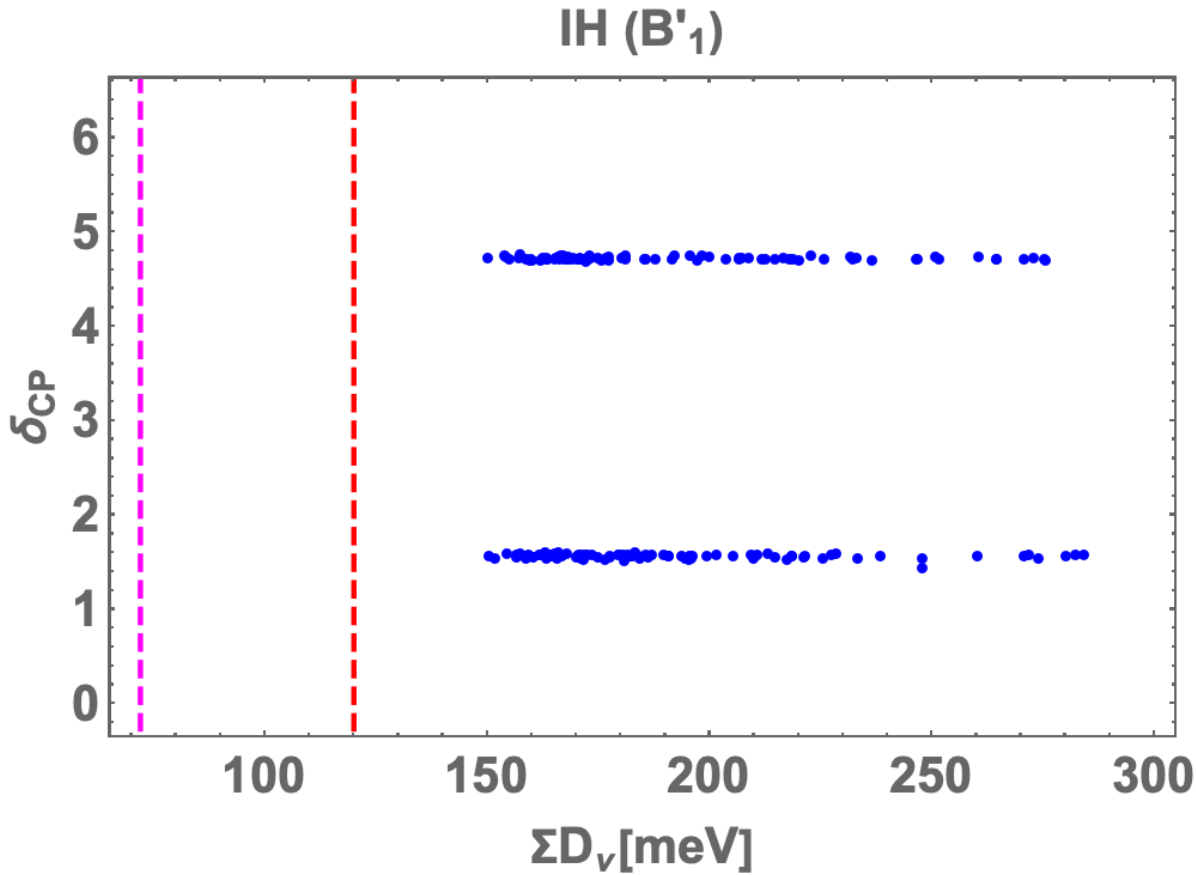} 
\caption{Numerical analyses in case of $B'_1$.
All the legends are the same as Figs.~\ref{fig:tau=i_2}. }
  \label{fig:b1p_2}
\end{center}\end{figure}
%%%%%%%%%%%%%%%%%%%   

%%%
In Fig.~\ref{fig:b1p_2}, we show predictions in the case of $B_1$ where 
the up and down figures represent the NH and the IH cases, respectively. 
The left ones demonstrate neutrinoless double beta decay $\langle m_{ee}\rangle$ in terms of the lightest neutrino masses.
The dashed red horizontal lines are bounds from the current KamLAND-Zen as the Fig.~\ref{fig:tau=i_2}.
The allowed regions are given by $45(40)~{\rm meV}\lesssim D_{\nu_1}(\langle m_{ee}\rangle )\lesssim 120(120)~{\rm meV}$ for NH
and  $30(55)~{\rm meV}\lesssim D_{\nu_3}(\langle m_{ee}\rangle )\lesssim 95(100)~{\rm meV}$ for IH.
The center ones shows Majorana phases $\alpha$ and $\beta$ where 
the figures imply that both cases are localized at nearby $\alpha\simeq 0$ and $\beta\simeq \pi$.
The right ones represent the Dirac CP phase in terms of the sum of neutrino masses where
the vertical red (magenta) dashed line shows the upper bound of $\sum m_\nu$ the same as the Fig.~\ref{fig:tau=i_2}.
The allowed regions are given by $145(150){\rm meV}\lesssim \sum D_\nu \lesssim 450(290){\rm meV}$ for NH(IH)
and  $\delta_{\rm CP}\simeq \pi/2, 3\pi/2$ for both cases.

\subsection{$B_2$-like model}
$B_2$-like model is realized by assigning $[\bm{1}]=\{1,1',1''\}$.
Similar way to the case of main text, the charged-lepton mass matrix is given by 
\begin{align}
m_\ell = \frac{v_d}{\sqrt2}
 \left(\begin{array}{ccc} f_1 & f_2 & f_3 \\
 f_3  & f_1 & f_2 \\
 f_2 & f_3 & f_1 \end{array} \right)
  %%%
   \left(\begin{array}{ccc} 
1 & 0 & 0 \\
0 & 0 & 1 \\
0 & 1 & 0 \end{array} \right)
 %%%
   \left(\begin{array}{ccc} a_e & 0 & 0 \\
0 & b_e & 0 \\
0 & 0 & c_e \end{array} \right).
\label{eq:cgd-lep-b2}
\end{align}
The neutrino mass matrix is given by
\begin{align}
m_\nu = \frac{v_{\Delta_u}}{\sqrt2}
%\frac{v_{\Delta_u} |d_\nu|}{\sqrt2}
 \left(\begin{array}{ccc} d_\nu & 0  & b_\nu \\
0  & a_\nu & c_\nu  \\
b_\nu & c_\nu  &0 \end{array} \right),
\label{eq:cgd-lep2-b1}
\end{align}
where $a_\nu,b_\nu,c_\nu, d_\nu$ implicitly include $Y_1^{(4)}$ or $Y_{1'}^{(4)}$. 
%$a_\nu,b_\nu,c_\nu, d_\nu$ respectively include $Y_1^{(4)},Y_{1'}^{(4)},Y_{1}^{(4)},Y_{1'}^{(4)}$. 

 %%%%%%%%%%%%%%%%%%%
\begin{figure}[tb]
\begin{center}
\includegraphics[width=50.0mm]{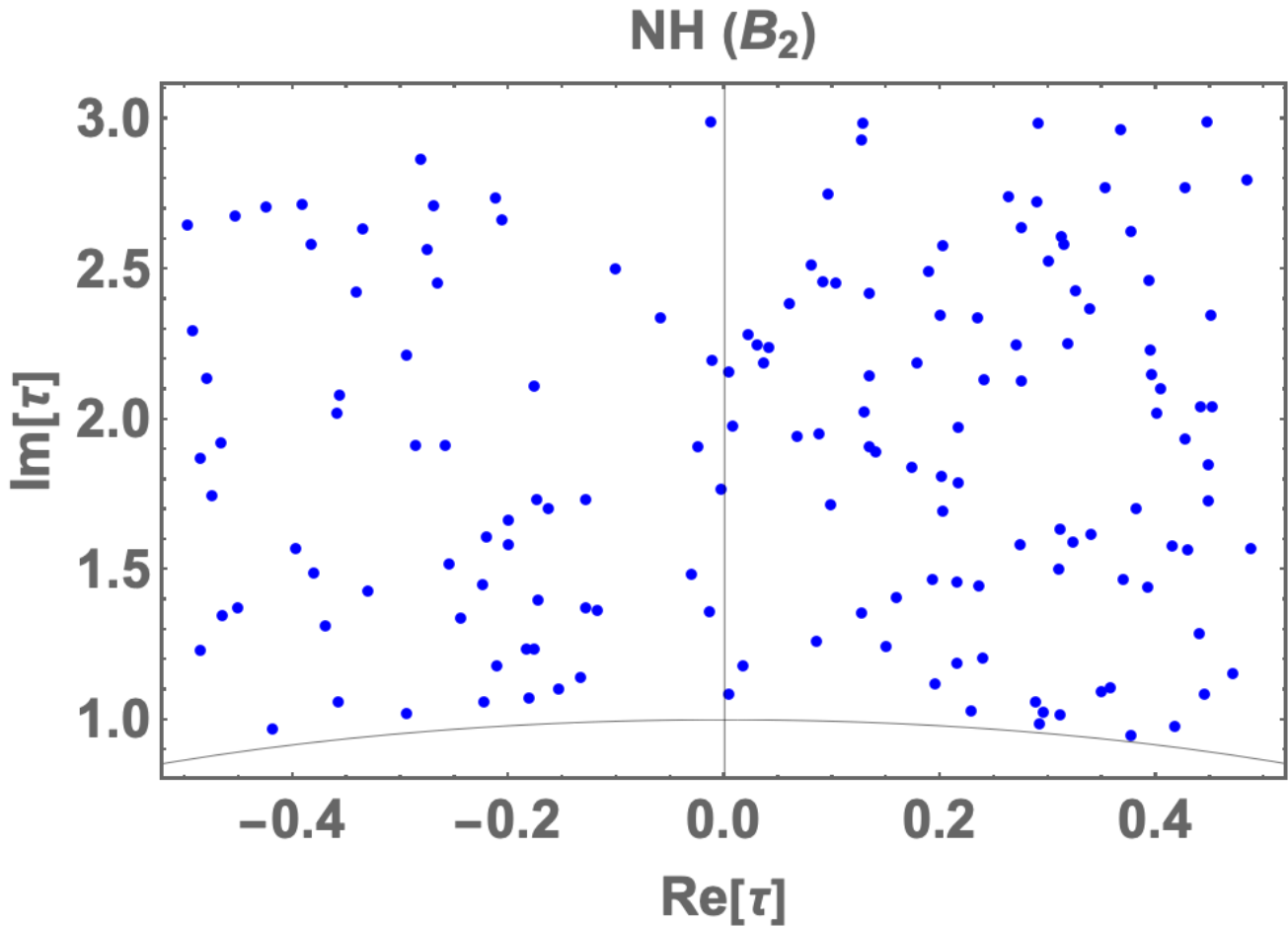} \quad
%%%
%\includegraphics[width=50.0mm]{figs/tau_ih-b2.pdf} \quad
\caption{Numerical analysis of $\tau$ in case of $B_2$-like model, where the legends of figures are the same as Fig.~\ref{fig:tau=i_1}.}
  \label{fig:b2_1}
\end{center}\end{figure}
%%%%%%%%%%%%%%%%%%%   
%
In Fig.~\ref{fig:b2_1}, we figure out the allowed range of $\tau$
where the left figure represents the case of NH and the right one is for IH.
They tell us that all ranges are allowed in the fundamental region up to ${\rm Im}[\tau]=3$.

 %%%%%%%%%%%%%%%%%%%
\begin{figure}[tb]
\begin{center}
\includegraphics[width=50.0mm]{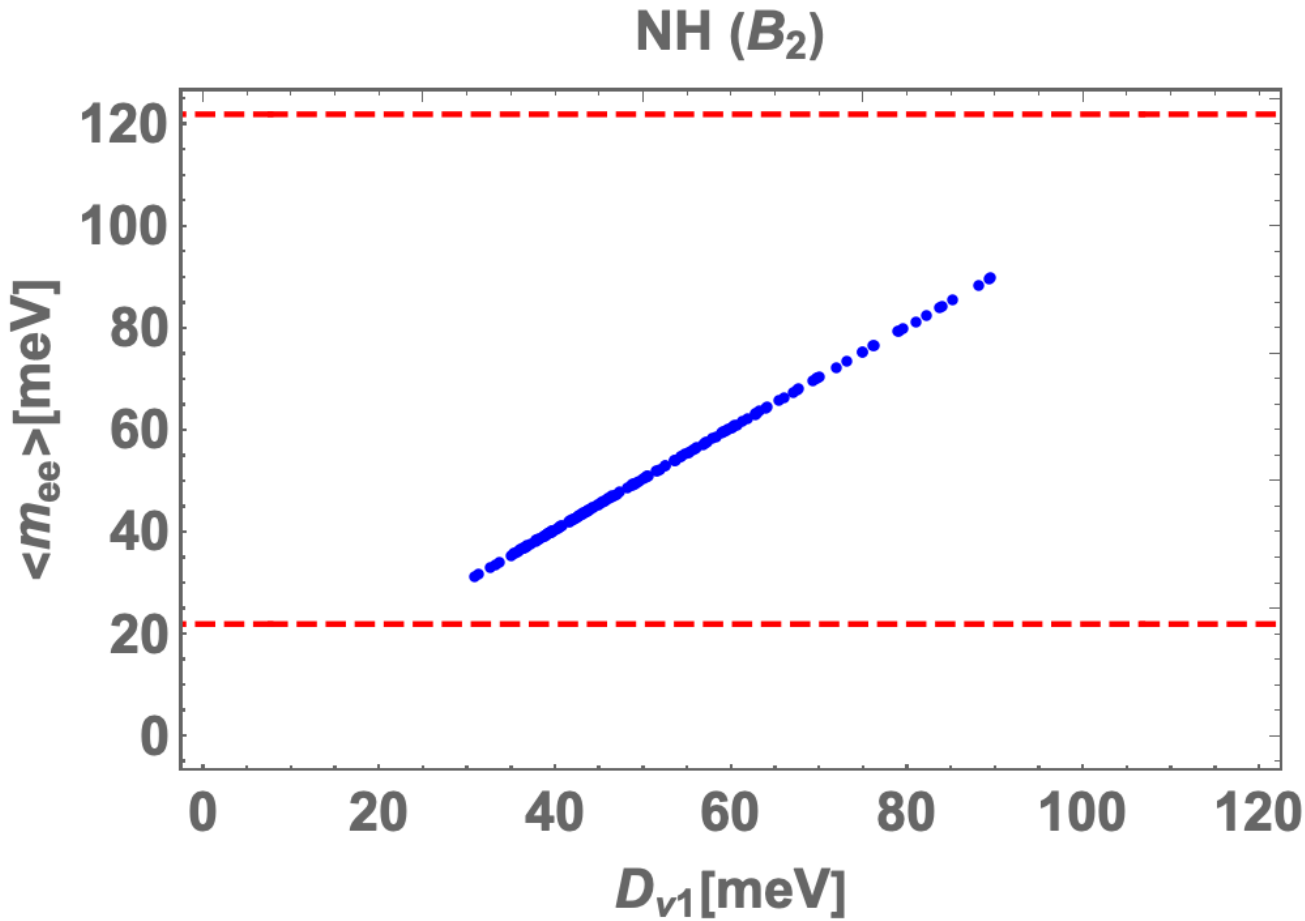} \quad
\includegraphics[width=50.0mm]{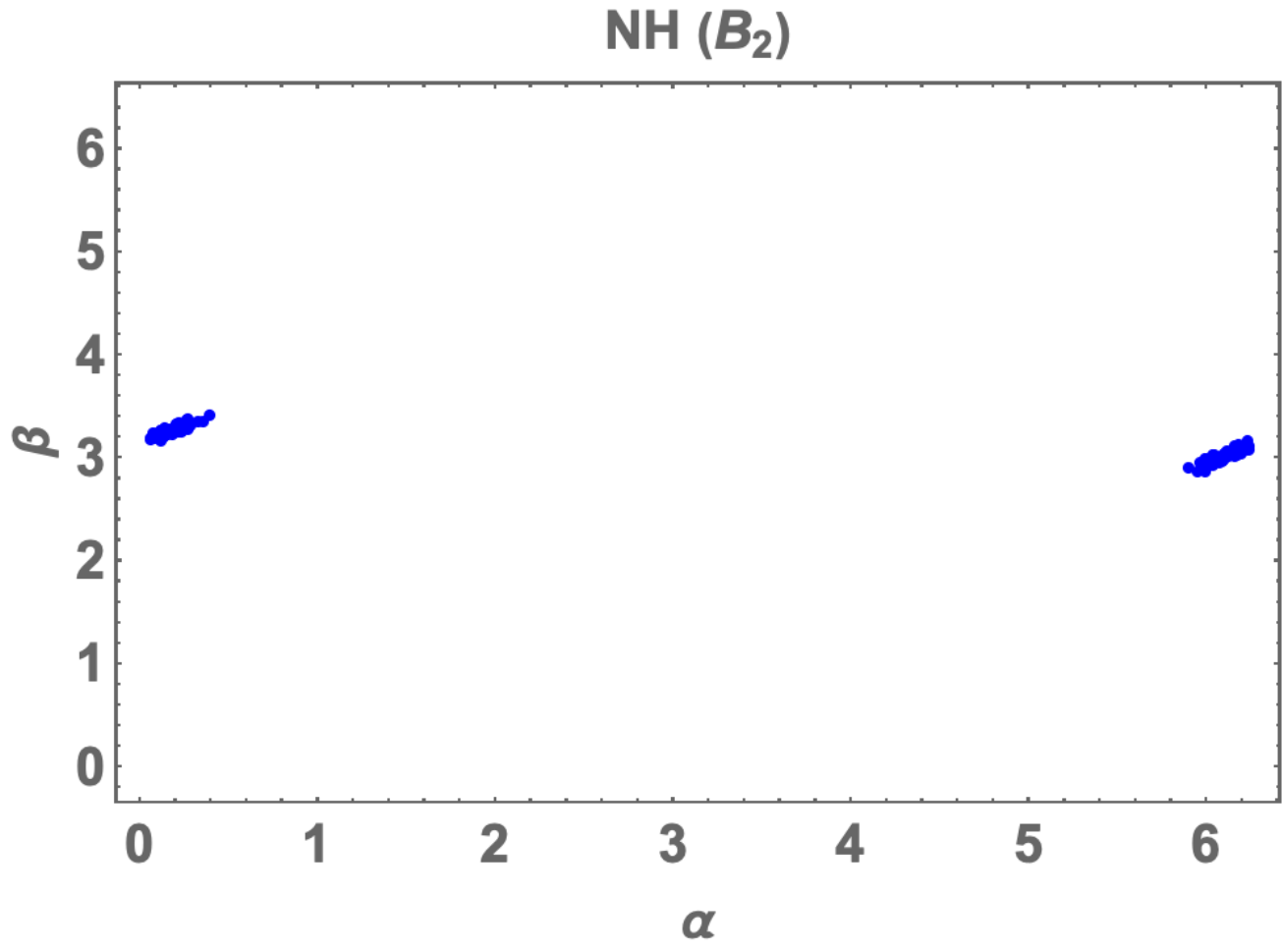}  \quad
\includegraphics[width=50.0mm]{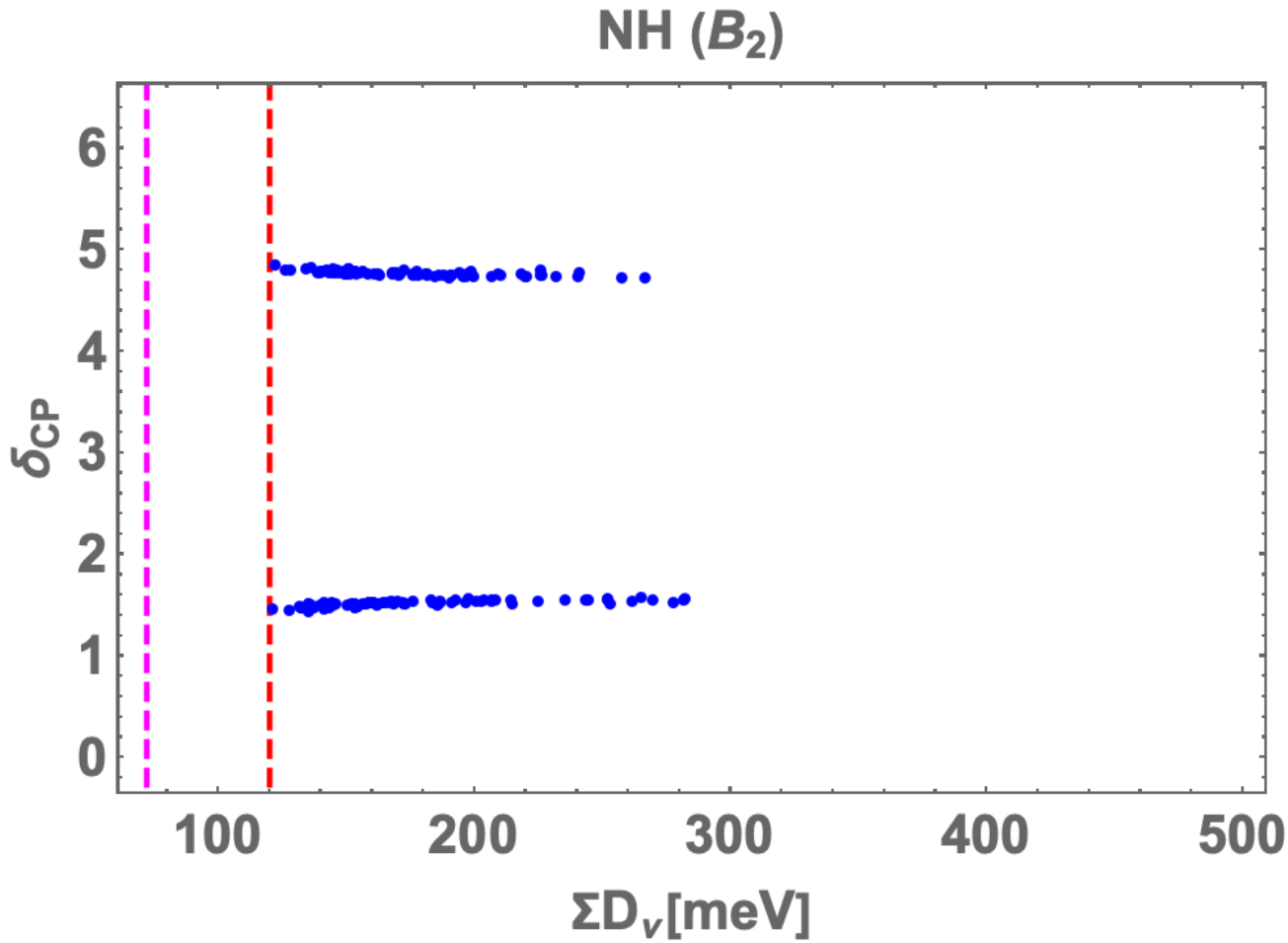} \\
%%%
%\includegraphics[width=50.0mm]{figs/m3-mee_ih-b2.pdf} \quad
%\includegraphics[width=50.0mm]{figs/majos_ih-b2.pdf}  \quad
%\includegraphics[width=50.0mm]{figs/sum-dcp_ih-b2.pdf} 
\caption{Numerical analyses in case of $B_2$.
All the legends are the same as Figs.~\ref{fig:tau=i_2}. }
  \label{fig:b2_2}
\end{center}\end{figure}
%%%%%%%%%%%%%%%%%%%   
%%%
In Fig.~\ref{fig:b2_2}, we show predictions in the case of $B_2$ where 
the only NH case has solutions. 
The left one demonstrates neutrinoless double beta decay $\langle m_{ee}\rangle$ in terms of the lightest neutrino masses.
The dashed red horizontal lines are bounds from the current KamLAND-Zen as the Fig.~\ref{fig:tau=i_2}.
The allowed regions are given by $30(30)~{\rm meV}\lesssim D_{\nu_1}(\langle m_{ee}\rangle )\lesssim 90(90)~{\rm meV}$.
The center one shows Majorana phases $\alpha$ and $\beta$ where 
the figure implies that they are localized at nearby $\alpha\simeq 0$ and $\beta\simeq \pi$.
The right one represents the Dirac CP phase in terms of the sum of neutrino masses where
the vertical red (magenta) dashed line shows the upper bound of $\sum m_\nu$ the same as the Fig.~\ref{fig:tau=i_2}.
The allowed regions are given by $121~{\rm meV}\lesssim \sum D_\nu \lesssim 280~{\rm meV}$
and  $\delta_{\rm CP}\simeq \pi/2, 3\pi/2$.

\subsection{$B'_2$-like model}
$B'_2$-like model is realized by assigning $[\bm{1}]=\{1',1,1''\}$.
Similar way to the case of main text, the charged-lepton mass matrix is given by 
\begin{align}
m_\ell = \frac{v_d}{\sqrt2}
 \left(\begin{array}{ccc} f_1 & f_2 & f_3 \\
 f_3  & f_1 & f_2 \\
 f_2 & f_3 & f_1 \end{array} \right)
  %%%
   \left(\begin{array}{ccc} 
0 & 1 & 0 \\
0 & 0 & 1 \\
1 & 0 & 0 \end{array} \right)
 %%%
   \left(\begin{array}{ccc} a_e & 0 & 0 \\
0 & b_e & 0 \\
0 & 0 & c_e \end{array} \right).
\label{eq:cgd-lep-b2p}
\end{align}
The neutrino mass matrix is given by
\begin{align}
m_\nu = \frac{v_{\Delta_u}}{\sqrt2}
%\frac{v_{\Delta_u} |d_\nu|}{\sqrt2}
 \left(\begin{array}{ccc} d_\nu & 0  & b_\nu \\
0  & a_\nu & c_\nu  \\
b_\nu & c_\nu  &0 \end{array} \right),
\label{eq:cgd-lep2-b2p}
\end{align}
where $a_\nu,b_\nu,c_\nu, d_\nu$ implicitly include $Y_1^{(4)}$ or $Y_{1'}^{(4)}$. 
%$a_\nu,b_\nu,c_\nu, d_\nu$ respectively include $Y_1^{(4)},Y_{1'}^{(4)},Y_{1}^{(4)},Y_{1'}^{(4)}$. 

 %%%%%%%%%%%%%%%%%%%
\begin{figure}[tb]
\begin{center}
\includegraphics[width=50.0mm]{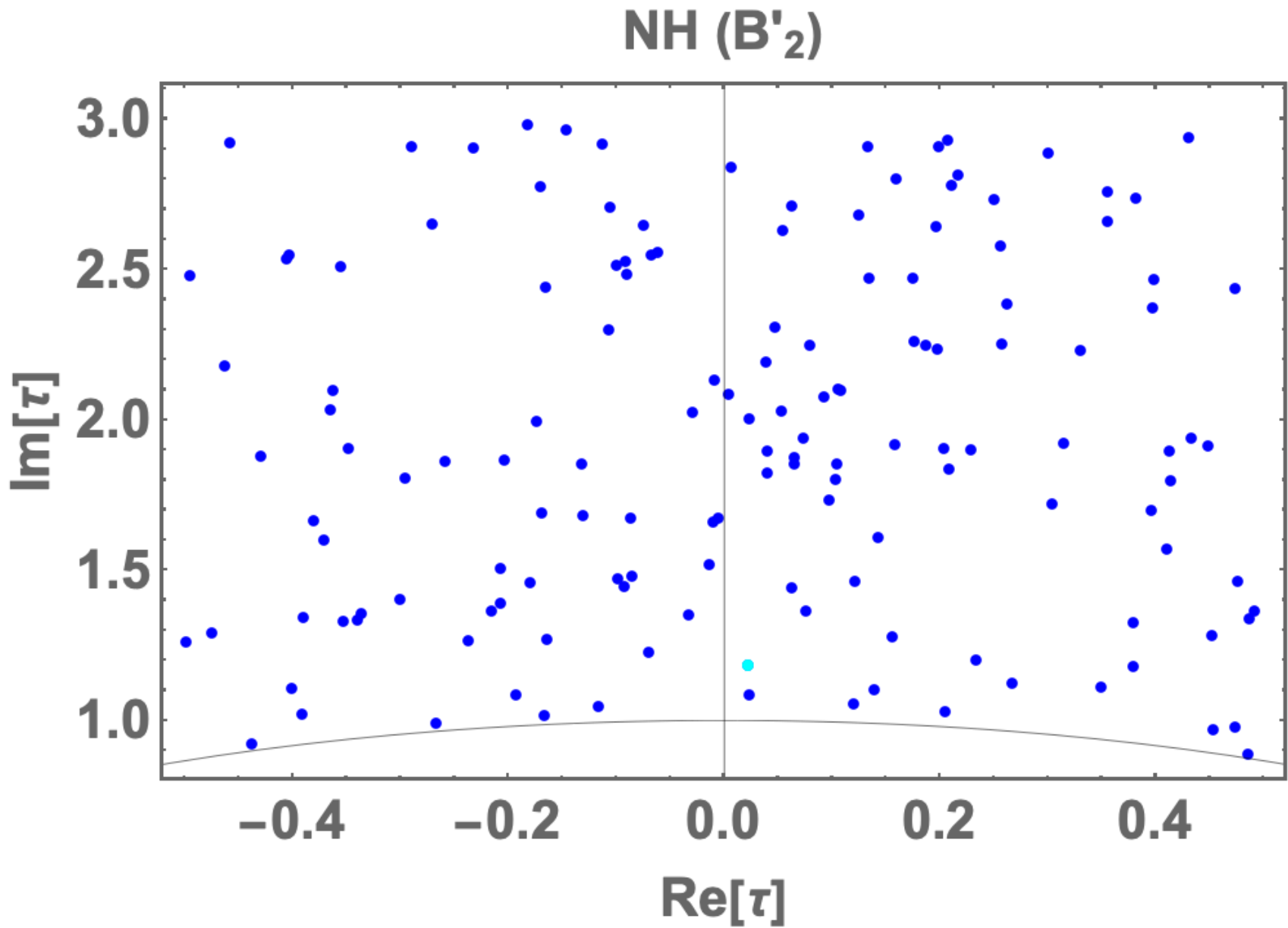} \quad
%%%
%\includegraphics[width=50.0mm]{figs/tau_ih-b2.pdf} \quad
\caption{Numerical analysis of $\tau$ in case of $B'_2$-like model, where the legends of figures are the same as Fig.~\ref{fig:tau=i_1}.}
  \label{fig:b2p_1}
\end{center}\end{figure}
%%%%%%%%%%%%%%%%%%%   
%
In Fig.~\ref{fig:b2p_1}, we figure out the allowed range of $\tau$
where the left figure represents the case of NH and the right one is for IH.
They tell us that all ranges are allowed in the fundamental region up to ${\rm Im}[\tau]=3$.

 %%%%%%%%%%%%%%%%%%%
\begin{figure}[tb]
\begin{center}
\includegraphics[width=50.0mm]{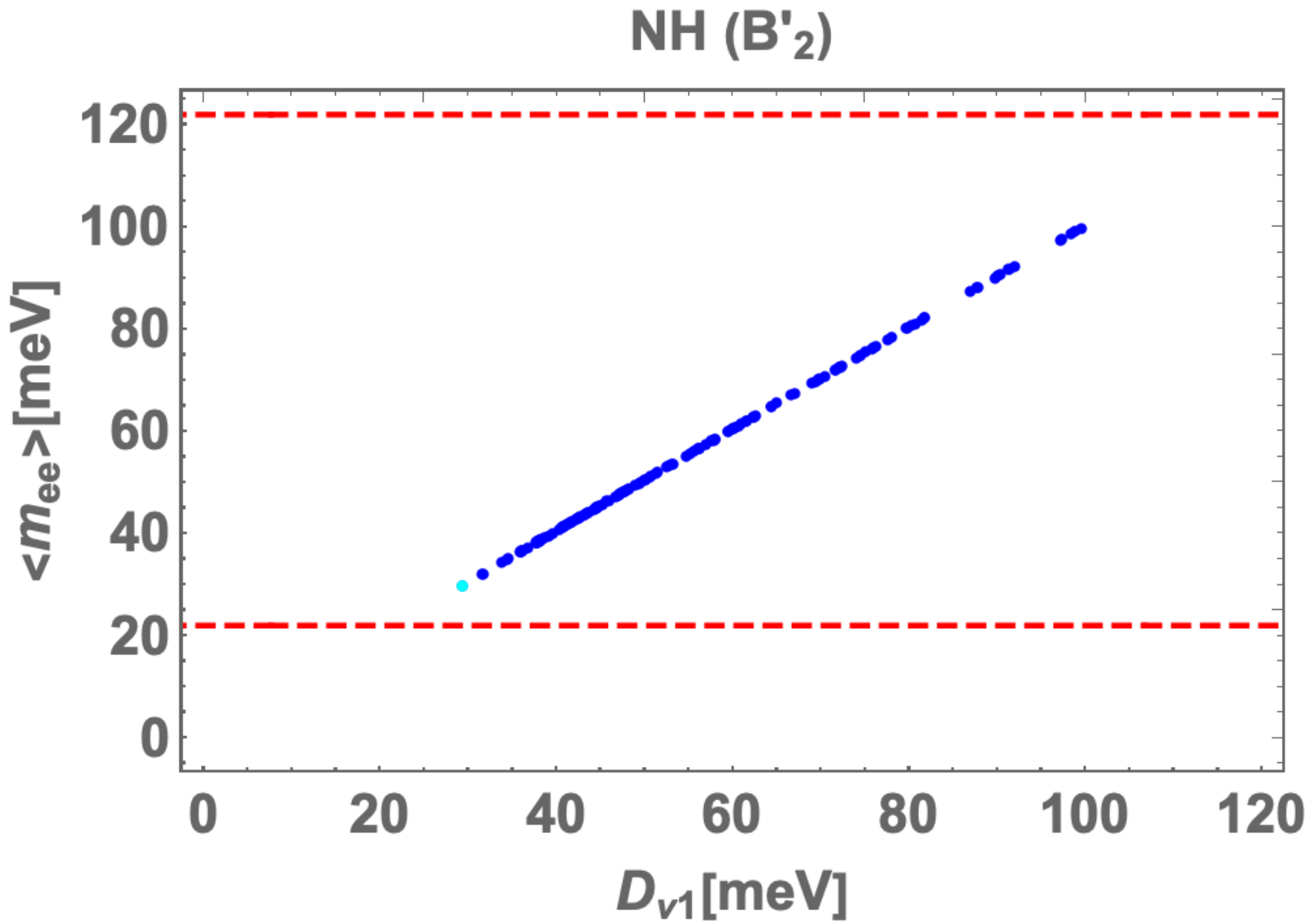} \quad
\includegraphics[width=50.0mm]{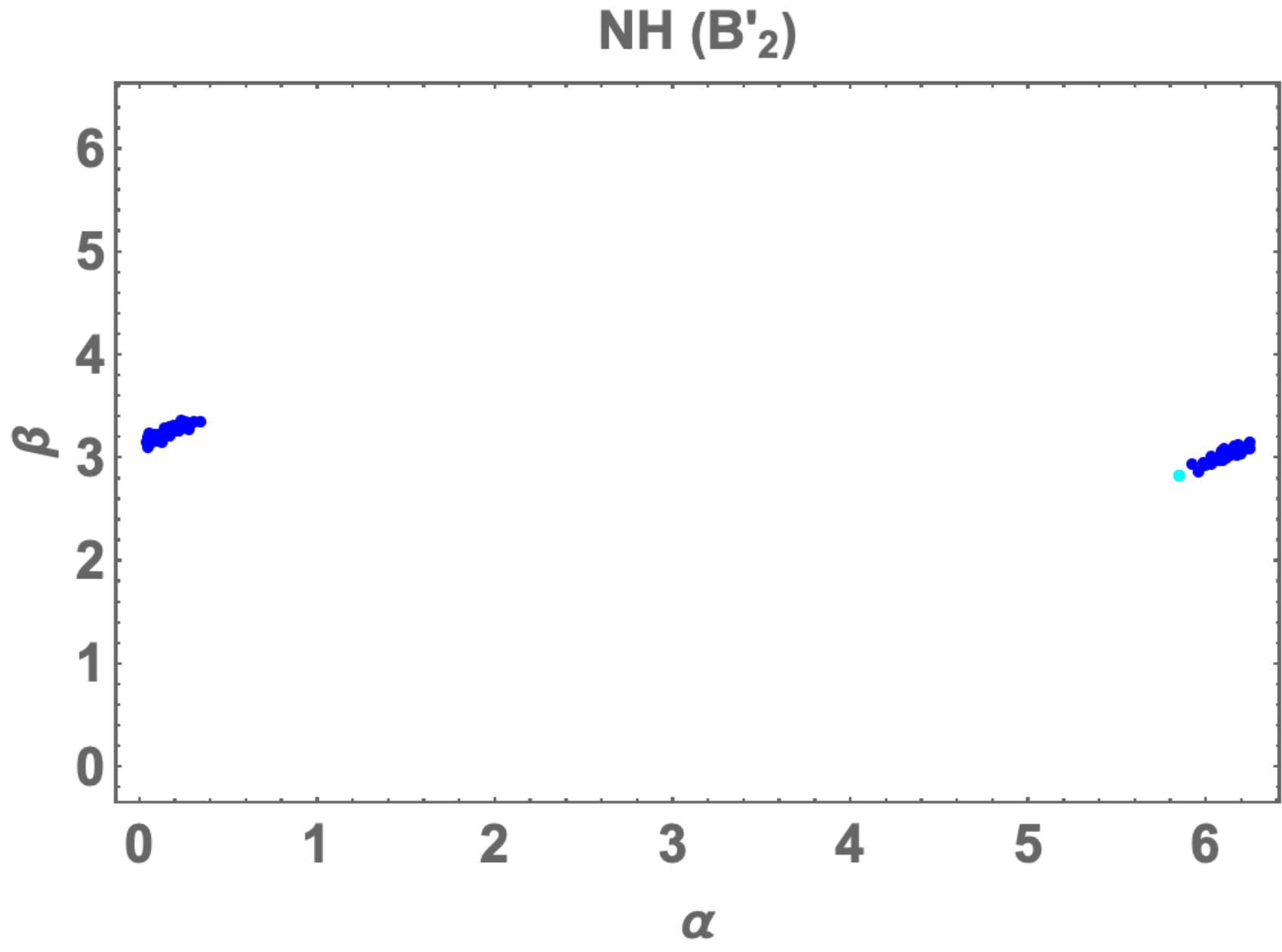}  \quad
\includegraphics[width=50.0mm]{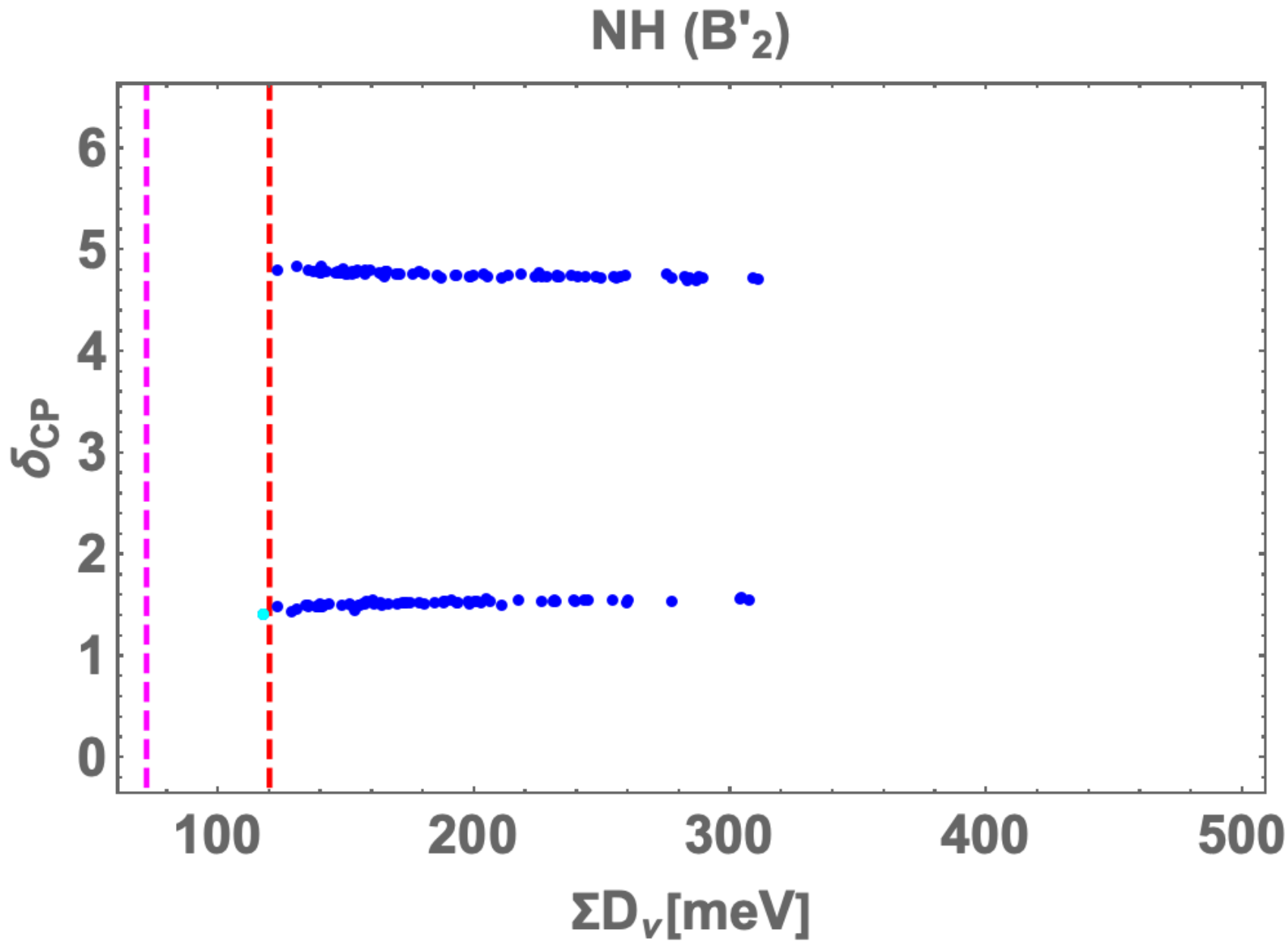} \\
%%%
%\includegraphics[width=50.0mm]{figs/m3-mee_ih-b2.pdf} \quad
%\includegraphics[width=50.0mm]{figs/majos_ih-b2.pdf}  \quad
%\includegraphics[width=50.0mm]{figs/sum-dcp_ih-b2.pdf} 
\caption{Numerical analyses in case of $B'_2$.
All the legends are the same as Figs.~\ref{fig:tau=i_2}. }
  \label{fig:b2p_2}
\end{center}\end{figure}
%%%%%%%%%%%%%%%%%%%   
%%%
In Fig.~\ref{fig:b2p_2}, we show predictions in the case of $B'_2$ where 
the only NH case has solutions. 
The left one demonstrates neutrinoless double beta decay $\langle m_{ee}\rangle$ in terms of the lightest neutrino masses.
The dashed red horizontal lines are bounds from the current KamLAND-Zen as the Fig.~\ref{fig:tau=i_2}.
The allowed regions are given by $30(30)~{\rm meV}\lesssim D_{\nu_1}(\langle m_{ee}\rangle )\lesssim 100(100)~{\rm meV}$.
The center one shows Majorana phases $\alpha$ and $\beta$ where 
the figure implies that they are localized at nearby $\alpha\simeq 0$ and $\beta\simeq \pi$.
The right one represents the Dirac CP phase in terms of the sum of neutrino masses where
the vertical red (magenta) dashed line shows the upper bound of $\sum m_\nu$ the same as the Fig.~\ref{fig:tau=i_2}.
The allowed regions are given by $119~{\rm meV}\lesssim \sum D_\nu \lesssim 320~{\rm meV}$
and  $\delta_{\rm CP}\simeq \pi/2, 3\pi/2$.
Note here that only this model marginally satisfies the cosmological lower bound.

\subsection{The other models}
If one assigns $[\bm{1}]=\{1'',1,1'\}$, one gets $E_3$-like model.
Then, the charged-lepton and neutrino mass matrices are respectively given by 
\begin{align}
m_\ell &= \frac{v_d}{\sqrt2}
 \left(\begin{array}{ccc} f_1 & f_2 & f_3 \\
 f_3  & f_1 & f_2 \\
 f_2 & f_3 & f_1 \end{array} \right)
  %%%
   \left(\begin{array}{ccc} 
0 & 1 & 0 \\
1 & 0 & 0 \\
0 & 0 & 1 \end{array} \right)
 %%%
   \left(\begin{array}{ccc} a_e & 0 & 0 \\
0 & b_e & 0 \\
0 & 0 & c_e \end{array} \right),\\
m_\nu & = \frac{v_{\Delta_u}}{\sqrt2}
%\frac{v_{\Delta_u} |d_\nu|}{\sqrt2}
 \left(\begin{array}{ccc} 
 0  & b_\nu & c_\nu \\
b_\nu  & a_\nu & 0  \\
c_\nu & 0  & d_\nu \end{array} \right),
\label{eq:cgd-lep2-b2p}
\end{align}
where $a_\nu,b_\nu,c_\nu, d_\nu$ implicitly include $Y_1^{(4)}$ or $Y_{1'}^{(4)}$. 
%$a_\nu,b_\nu,c_\nu, d_\nu$ respectively include $Y_1^{(4)},Y_{1'}^{(4)},Y_{1}^{(4)},Y_{1'}^{(4)}$. 
\\

When one assigns $[\bm{1}]=\{1'',1',1\}$, one also gets $E_3$-like model.
Then, the charged-lepton and neutrino mass matrix are respectively given by 
\begin{align}
m_\ell &= \frac{v_d}{\sqrt2}
 \left(\begin{array}{ccc} f_1 & f_2 & f_3 \\
 f_3  & f_1 & f_2 \\
 f_2 & f_3 & f_1 \end{array} \right)
  %%%
   \left(\begin{array}{ccc} 
0 & 0 & 1 \\
1 & 0 & 0 \\
0 & 1 & 0 \end{array} \right)
 %%%
   \left(\begin{array}{ccc} a_e & 0 & 0 \\
0 & b_e & 0 \\
0 & 0 & c_e \end{array} \right),\\
m_\nu & = \frac{v_{\Delta_u}}{\sqrt2}
%\frac{v_{\Delta_u} |d_\nu|}{\sqrt2}
 \left(\begin{array}{ccc} 
 0  & b_\nu & c_\nu \\
b_\nu  & a_\nu & 0  \\
c_\nu & 0  & d_\nu \end{array} \right),
\label{eq:cgd-lep2-b2p}
\end{align}
where $a_\nu,b_\nu,c_\nu, d_\nu$ implicitly include $Y_1^{(4)}$ or $Y_{1'}^{(4)}$.

Although the mass structures of charged-leptons are nontrivial,
we confirmed that the above two textures do not satisfy the current neutrino oscillation data.
If two of the left-handed leptons are assigned by the same singlet under the $A_4$, the lightest mass of charged-leptons
is zero due to the rank 2. Thus, these models are not valid in our framework.

\section{Modular forms  at nearby  fixed points }

%%%%%%%%%%%%%%%%%%%%%%%%%%%%%%%%%%%%%%%%%%%%%%%%%%%%%%%%%%%
%\subsection{Modular forms  at nearby  fixed points }
%%%%%%%%%%%%%%%%%%%%%%%%%%%%%%%%%%%%%%%%%%%%%%%%%%%%%%%%%%%%%%%%% 

Here, we show each of behavior at modular forms at nearby $\tau=i,\ \omega,\ i\infty$~\cite{Okada:2020ukr}.

\subsection{Modular forms at nearby   $\tau=i$}
\label{apdxA}    
In case of behavior at modular forms at nearby $\tau=i$,
We obtain  approximate linear forms  of  $Y_1(\tau)$, $Y_2(\tau)$ and $Y_3(\tau)$ by performing Taylor expansion of modular forms around $\tau=i$:
\begin{align}
\begin{aligned}
\tau=i +\epsilon \ , \qquad {\rm with } \qquad \epsilon=\epsilon_R+i\,\epsilon_I
\ , 
\end{aligned}
\label{epsilonS}
\end{align}
where $|\epsilon|\ll 1$ is expected.
Then we approximately obtain
\begin{align}
\begin{aligned}
\frac{Y_2(\tau)}{Y_1(\tau)}\simeq (1+\epsilon_1)\, (1-\sqrt{3}) \, , \quad 
\frac{Y_3(\tau)}{Y_1(\tau)}\simeq (1+\epsilon_2)\, (-2+\sqrt{3}) \, ,
\quad \epsilon_1=\frac{1}{2} \epsilon_2=2.05\,i\,\epsilon\,.
\end{aligned}
\label{epS12}
\end{align}
%%%%%%%%%%%%%%%%%%%%%%%%%%%%%%%%%%%%%%%%%%%%%%%
Applying Eq.~(\ref{epS12}), we find approximation forms of $ Y_{{\bm 3},{\bm 3}'}^{(k)}$ as follows:
\begin{align}
&  \begin{aligned}
\frac{Y_1^{(6)}(\tau)}{3Y_1^3(\tau)} \simeq 2\sqrt{3}-3+
\left (2\sqrt{3}-\frac{10}{3}\right )(\epsilon_1+\epsilon_2) \, ,  \quad
\end{aligned} \nonumber \\
&  \begin{aligned}
\frac{Y_2^{(6)}(\tau)}{3Y_1^3(\tau)} \simeq 5\sqrt{3}-9+\left (\frac{31}{\sqrt{3}}-\frac{55}{3}\right )\epsilon_1+
\left (\frac{16}{\sqrt{3}}-\frac{28}{3} \right )\epsilon_2\, ,
\end{aligned}
\nonumber\\
&  \begin{aligned}
\frac{Y_3^{(6)}(\tau)}{3Y_1^3(\tau)} \simeq 12-7\sqrt{3}+\left (\frac{38}{3}-\frac{22}{\sqrt{3}}
\right )\epsilon_1+
\left (\frac{74}{3}- \frac{43}{\sqrt{3}}\right )\epsilon_2\, ,
\end{aligned}
\nonumber\\
&  \begin{aligned}
\frac{Y_1^{'(6)}(\tau)}{3Y_1^3(\tau)} \simeq 7\sqrt{3}-12+
\left (2\sqrt{3}-\frac{10}{3}\right )\epsilon_1+
\left (17\sqrt{3}-\frac{88}{3}\right )\epsilon_2 \, , \end{aligned} \nonumber\\
&\begin{aligned}
\frac{Y_2^{'(6)}(\tau)}{3Y_1^3(\tau)}  \simeq 3-2\sqrt{3}+\left (\frac{2}{3}- \frac{2}{\sqrt{3}}\right )
\epsilon_1+
\left (\frac{14}{3}-\frac{8}{\sqrt{3}}- \right )\epsilon_2\, ,
\end{aligned}
\nonumber\\
&  \begin{aligned}
\frac{Y_3^{'(6)}(\tau)}{3Y_1^3(\tau)}  \simeq9-5\sqrt{3}+\left (\frac{35}{3}-\frac{19}{\sqrt{3}}
\right )\epsilon_1+
\left (\frac{38}{3}- \frac{22}{\sqrt{3}}\right )\epsilon_2\, ,
\end{aligned}
\label{epS666}
\end{align}
%where the last one denotes for the $A_4$ singlet.
These  forms are  agreement with exact numerical values within  $0.1\,\%$
for $|\epsilon|\leq 0.05$.

%%%%%%%%%%%%%%%%%%%%%%%%%%%%%%%%%%%%%%%%%%%%%%%%%%%%%%%%%%%%%%%%%%%%%%
%%%%%%%%%%%%%%%%%%%%%%%%%%%%%%%%%%%%%%%%%%%%%%%%%%%%%%%%%%%%%%%%%%%%%%
%%%%%%%%%%%%%%%%%%%%%%%%%%%%%%%%%%%%%%%%%%%%%%%%%%%%%%%%%%%%%%%%%%%%%%
\subsection{Modular forms at nearby  $\tau=\omega$}\label{apdxB}
In case of the behavior of modular forms at nearby $\tau=\omega$, we obtain the following forms by performing the same way as the case of $\tau=i$:
\begin{align}
\begin{aligned}
\tau= \omega+\epsilon \, , \qquad {\rm with } \qquad \epsilon=\epsilon_R+i\,\epsilon_I
\ , 
\end{aligned}
\label{epST}
\end{align}
where $|\epsilon|\ll 1$ is supposed.
Then, we approximately obtain
\begin{align}
\begin{aligned}
\frac{Y_2(\tau)}{Y_1(\tau)}\simeq \omega\,(1+\,\epsilon_1) \, , \quad 
\frac{Y_3(\tau)}{Y_1(\tau)}\simeq -\frac{1}{2}\omega^2 \, 
(1+\, \epsilon_2)\,  ,
\quad \epsilon_1=\frac{1}{2} \epsilon_2=2.1\,i\,\epsilon\,,
\end{aligned}
\label{epST12}
\end{align}
where $|\epsilon|\ll 1$.

%%%%%%%%%%%%%%%%%%%%%%%%%%%%%%%%%%%%%%%%%%%%%%%
Applying Eq.~(\ref{epST12}), we find approximation forms of $ Y_{{\bm 3},{\bm 3}'}^{(k)}$ as follows:
\begin{align}
&  \begin{aligned}
\frac{Y_1^{(6)}(\tau)}{Y_1^3(\tau)} \simeq -(\epsilon_1+\epsilon_2)\, ,  \quad
\end{aligned} \nonumber \\
&  \begin{aligned}
\frac{Y_2^{(6)}(\tau)}{Y_1^3(\tau)} \simeq
-\omega\, (\epsilon_1+\epsilon_2)\,  ,
\end{aligned}
\nonumber\\
&  \begin{aligned}
\frac{Y_3^{(6)}(\tau)}{Y_1^3(\tau)} \simeq
\frac{1}{2}\,\omega^2\, (\epsilon_1+\epsilon_2) ,
\end{aligned}
\nonumber\\
&  \begin{aligned}
\frac{Y_1^{'(6)}(\tau)}{Y_1^3(\tau)} \simeq -\frac{9}{8}
\left (1+\frac{8}{9}\epsilon_1+\frac{11}{9}\epsilon_2\right )\, , \end{aligned} \nonumber\\
&\begin{aligned}
\frac{Y_2^{'(6)}(\tau)}{Y_1^3(\tau)}  \simeq
\frac{9}{4}\,\omega\,
\left (1+\frac{8}{9}\epsilon_1+\frac{2}{9}\epsilon_2\right ) \, ,
\end{aligned}
\nonumber\\
&  \begin{aligned}
\frac{Y_3^{'(6)}(\tau)}{Y_1^3(\tau)}  \simeq
\frac{9}{4}\,\omega^2\,
\left (1+\frac{17}{9}\epsilon_1+\frac{2}{9}\epsilon_2\right )\, .
\end{aligned}
\label{epST666}
\end{align}
These approximate  forms are  agreement with exact numerical values within  $1\,\%$
 for $|\epsilon|\leq 0.05$.

%%%%%%%%%%%%%%%%%%%%%%%%%%%%%%%%%%%%%%%%%%%%%%%%%%%%%%%%%%%%%%%%%%%%%%
\subsection{Modular forms  towards  $\tau=i\infty$}
\label{apdxC}

In case of the behavior of modular forms at large value of imaginary part of $\tau$,
we approximately find modular forms  as follows:
\begin{align}
Y_1(\tau)\simeq 1+ 12 {\bf p}\,\epsilon\,,\quad  
Y_2(\tau)\simeq -6 {\bf p}^{\frac{1}{3}}\,\epsilon^{\frac{1}{3}}\,,
\quad  Y_3(\tau)\simeq -18 {\bf p}^{\frac{2}{3}}\,\epsilon^{\frac{2}{3}}\,, \quad
{\bf p}=e^{2\pi i\, {\rm Re}\, \tau}\,,\quad 
{\bf \epsilon}=e^{-2\pi\, {\rm Im}\, \tau}\, .
\label{epT12}
\end{align}
% where $q=\exp{(2\pi i\tau)}$ is suppressed. Taking leading terms of  Eq.\,(\ref{Y(2)}), 

%%%%%%%%%%%%%%%%%%%%%%%%%%%%%%%%%%%%%%%%%%%%%%%%%%%%%

%%%%%%%%%%%%%%%%%%%%%%%%%%%%%%%%%%%%%%%%%%%%%%%%%%%%%%%%
%%%%%%%%%%%%%%%%%%%  weight 4 and 6  %%%%%%%%%%%%%%%%%%%
%%%%%%%%%%%%%%%%%%%%%%%%%%%%%%%%%%%%%%%%%%%%%%%%%%%%%%%% 
Applying Eq.~(\ref{epST12}), we find approximation forms of $ Y_{3,3'}^{(k)}$ as follows:
\begin{align}
&  \begin{aligned}
Y_1^{(6)}(\tau)
\simeq 1+252\, {\bf p} \, \epsilon \, , \qquad 
Y_2^{(6)}(\tau) \simeq -6\, {\bf p}^{\frac{1}{3}}\,\epsilon^{\frac{1}{3}}\,,
\qquad 
Y_3^{(6)}(\tau) \simeq -18\, {\bf p}^{\frac{2}{3}}\,\epsilon^{\frac{2}{3}}\,,
\end{aligned}
\nonumber\\
&  \begin{aligned}
Y_1^{'(6)}(\tau)
\simeq 216\,{\bf p} \, \epsilon \, , \qquad\quad\  
Y_2^{'(6)}(\tau) \simeq -12\, {\bf p}^{\frac{1}{3}}\,\epsilon^{\frac{1}{3}}\,,
\quad \ \
Y_3^{'(6)}(\tau) \simeq 72\, {\bf p}^{\frac{2}{3}}\,\epsilon^{\frac{2}{3}}\,.
\end{aligned}
\label{epT666}
\end{align}

% Ref Style
% Including title
%\bibliographystyle{utphys}
\bibliography{ctma4.bib}
\end{document}